\documentclass[12pt]{article}
\usepackage{amsmath,amssymb}
\usepackage{graphicx}
\usepackage[pdfencoding=auto]{hyperref}
\usepackage{float}
\usepackage{caption,subcaption}
\begin{document}

\title{A perturbative production of massive Z bosons and
fermion-antifermion pairs from the vacuum in the de Sitter Universe}
\author{Diana Dumitrele, Mihaela-Andreea B\u aloi
and Cosmin Crucean \\ \thanks{diana.dumitrele97@e-uvt.ro, mihaela.baloi88@e-uvt.ro, cosmin.crucean@e-uvt.ro}\\
{\small \it Faculty of Physics, West University of Timi\c soara,}\\
{\small \it V. Parvan Avenue 4 RO-300223 Timi\c soara,  Romania}}

\begin{abstract}
In this paper we study the problem of neutral electro-weak interactions in a de Sitter geometry. We develop the formalism of reduction for the Proca field with the help of the solutions for the interacting fields and by using perturbative methods we obtain the definition of the transition amplitudes in the first order of perturbation theory. As an application to our formalism we study the generation of massive fermions and Z bosons from vacuum in the expanding de Sitter universe. Our results are the generalization to the curved geometry of the Weinberg-Salam electro-weak theory for the case of Z boson interaction with leptons. The probability is found to be a quantity that depends on the Hubble parameter and we prove that such perturbative processes are possible only for large expansion regime of the early Universe. The total probability and rate of transition are obtained for the case of large expansion and we use the dimensional regularization for extract finite results from the momenta integrals. In the Minkowski limit we obtain that the probability of particle generation from vacuum is vanishing recovering the well known result that forbids particle production in flat space-time due to the momentum-energy conservation.
\end{abstract}

\maketitle
\newpage

\section{Introduction}
In Minkowski field theory the massive vector fields are described by the Proca equation \cite{PR1}, which represents one of the fundamental field equations alongside the Dirac equation and Maxwell equation. The Proca equation in curved geometries was studied in \cite{2,WT}, where the problem of fundamental solutions and propagators was approached. It is a well established fact that the massive neutral boson $Z$ and massive charged $W^{\pm}$ bosons were generated in the early universe \cite{w1}, but the mechanisms that were involved are a matter of further investigations. Among these mechanisms one is the perturbative approach which studies the particle generation for fields interactions \cite{24,25,26,27,28,29,30,31,32,b1,b2,33,34,35}. This is because the computations related to this phenomenon need to consider quantum fields in the presence of large expansion conditions of the early universe \cite{LL,LL1,23,24,25,26,27,28,29,30,31,32,b1,b2,33,34,35}. This can be properly done by using perturbative methods suitable for studying  processes that generate particle production in the presence of strong gravitational fields \cite{17,18,24,cc} because the translational invariance with respect to time is lost in a non-stationary background. The first steps for considering the above methods were done in \cite{2,24,cc}, by constructing the theory of free fields and fields interactions in a de Sitter geometry, and adopting the scenario of the fields minimally coupled with gravity \cite{24,cc}. This can be done by considering the Weinberg-Salam theory \cite{PR1,w1,12,3,4,5,6,7,8,9,10,11,cr} of electro-weak interactions in de Sitter geometry and extending the definitions of the transition amplitudes from Minkowski theory to the curved space-time. In this paper we will study for the first time the problem of generating from vacuum the triplet Z boson and electron-positron pair in the inflation of the early universe. In the presenent study we restrict ourselves to consider only the interactions between neutral massive Z bosons and massive fermions. The theory of electro-weak interactions that implies the Z bosons and fermions can be developed by taking into consideration in the lagrangean density the interaction term with the usual electrodynamic coupling $-J^\mu A_\mu$ where the current represents the neutral current. It is known that the phenomenon of particle production must be considered along with the inflations but until now there have been no definitive conclusions related to the mechanisms of particle generation. We mention that the nonperturbative approach of the problem of massive bosons generation is less studied in literature, but among a few results we mention here the ones obtained in \cite{37}. The general framework related to the idea that the space expansion produces particles was discussed in the fundamental papers \cite{32,33,34,35}, where the authors prove that the phenomenon of particle generation is possible only in early universe. In this paper we propose a perturbative mechanism that could explain the generation of massive Z bosons  by computing the perturbative first order amplitude corresponding to the de Sitter electro-weak theory \cite{cc}. The perturbative approach to the problem of particle generation was proposed in \cite{17,18}, where it was proven that this mechanism could play an important role in the matter-antimatter generation in early universe.

The paper is organized as follows: in the second section we present a short review about the theory of the free Proca Field and the free Dirac field in de Sitter geometry, the third section is dedicated to the equations of interaction between fields and in the forth section we present the reduction formalism for the Proca field as the basis for defining the amplitudes of electro-weak theory in de Sitter geometry. The fifth section is dedicated to the study of the amplitudes and probabilities corresponding to the process of spontaneous generation from vacuum of the triplet Z boson and electron-positron pair. In this section we obtain the total probability and we use a combined dimensional regularization and Pauli-Villars regularization for computing the integrals over the final momenta. In the six section we present the computation for the rate of transition and we discuss the limit of large expansion factor and we give an estimation of the density number of Z bosons, while in the seven section we present our conclusions.

\section{Free fields}
For our study we consider the de Sitter metric element written in conformal form \cite{1}:
\begin{equation}\label{metr}
ds^2=dt^2-e^{2\omega t}d\vec{x}^2=\frac{1}{(\omega t_{c})^2}(dt_{c}^2-d\vec{x}^2),
\end{equation}
where the conformal time is related to the proper time by $t_{c}=\frac{-e^{-\omega t}}{\omega}$ and $\omega$ is the Hubble parameter ($\omega>0$).

In this paper we propose for the first time a perturbative approach for explaining the generation of massive Z bosons and massive fermions in early universe. We work in a de Sitter geometry and use the chart with the conformal time $t_{c}\in(-\infty,0)$, which covers the expanding portion of de Sitter space-time. For the line element (\ref{metr}), in the Cartesian gauge, we have the nonvanishing tetrad components,
\begin{equation}
e^{0}_{\widehat{0}}=-\omega t_c  ;\,\,\,e^{i}_{\widehat{j}}=-\delta^{i}_{\widehat{j}}\,\omega t_c.
\end{equation}
The theory of the free Dirac field in de Sitter geometry was studied in \cite{22} where the fundamental spinor solutions in the momentum-helicity basis were determined. The Dirac equation in de Sitter space-time in Cartesian gauge was studied in \cite{22}, and reads as:
\begin{equation}
\Big( -i\omega t_c( \gamma^0 \partial_{t_c} + \gamma^i \partial_i) + \frac{3i \omega}{2} \gamma^0 -m\Big)\psi(x) =0
\end{equation}
The fundamental solutions of the Dirac equation in de Sitter geometry are the $U,V$ spinors with a defined momentum and helicity \cite{22}:
\begin{eqnarray}\label{sol}
&&U_{\vec{p},\sigma}(t,\vec{x}\,)=\frac{\sqrt{\pi
p/\omega}}{(2\pi)^{3/2}}\left (\begin{array}{c}
\frac{1}{2}e^{\pi k/2}H^{(1)}_{\nu_{-}}(\frac{p}{\omega} e^{-\omega t})\xi_{\sigma}(\vec{p}\,)\\
\sigma e^{-\pi k/2}H^{(1)}_{\nu_{+}}(\frac{p}{\omega} e^{-\omega
t})\xi_{\sigma}(\vec{p}\,)
\end{array}\right)e^{i\vec{p}\cdot\vec{x}-2\omega t};\nonumber\\
&&V_{\vec{p},\sigma}(t,\vec{x}\,)=\frac{\sqrt{\pi
p/\omega}}{(2\pi)^{3/2}} \left(
\begin{array}{c}
-\sigma\,e^{-\pi k/2}H^{(2)}_{\nu_{-}}(\frac{p}{\omega} e^{-\omega t})\,
\eta_{\sigma}(\vec{p}\,)\\
\frac{1}{2}\,e^{\pi k/2}H^{(2)}_{\nu_{+}}(\frac{p}{\omega} e^{-\omega t}) \,\eta_{\sigma}(\vec{p}\,)
\end{array}\right)
e^{-i\vec{p}\cdot\vec{x}-2\omega t},
\end{eqnarray}
where $H^{(1)}_{\nu}(z), H^{(2)}_{\nu}(z)$ are Hankel functions of the first and second kind and $K=\frac{m}{\omega},\nu_{\pm}=\frac{1}{2}\pm iK$.  The helicity spinors satisfy the relation:
\begin{equation}\label{pa}
\vec{\sigma}\vec{p}\,\xi_{\sigma}(\vec{p}\,)=2p\sigma\xi_{\sigma}(\vec{p}\,)
\end{equation}
with $\sigma=\pm1/2$, where $\vec{\sigma}$ are the Pauli matrices, and $p=\mid\vec{p}\mid$ is the modulus of the momentum vector, while $\eta_{\sigma}(\vec{p}\,)= i\sigma_2
[\xi_{\sigma}(\vec{p}\,)]^{*}$.
The meaning of positive/negative frequencies can be established if we use the asymptotic expansion of the Hankel functions at large $z$ \cite{21}:
\begin{equation}
H^{(1,\,2)}_{\nu}(z)\simeq\left(\frac{2}{\pi z}\right)^{1/2}e^{\pm i(z-\nu\pi/2-\pi/4)},
\end{equation}
and we obtain that for $t\rightarrow-\infty$, the modes defined above behave as a positive/negative frequency modes with respect to conformal time $t_{c}=-e^{-\omega t}/\omega$:
\begin{equation}
U_{\vec{p},\lambda}(t,\vec{x}\,)\sim e^{-ipt_{c}};\,\,V_{\vec{p},\lambda}(t,\vec{x}\,)\sim e^{ipt_{c}}.
\end{equation}\label{bd}
This is the behavior in the infinite past for positive/negative frequency modes which defines the Bunch-Davies vacuum \cite{17}.

In the case of the Proca free field the equations for the spatial and temporal components of the vector potential were obtained in \cite{2}
\begin{eqnarray}\label{a111}
  \partial_{t_c} (\partial_i A_i) - \Delta A_0 + \frac{M_Z^2}{\omega^2t_c^2} A_0 =0,\\
  \partial_{t_c}^2 A_k - \Delta A_k - \partial_k (\partial_{t_c} A_0) + \partial_k (\partial_i A_i) + \frac{M_Z^2}{\omega^2t_c^2} A_k = 0.
\end{eqnarray}
The above equations and the Lorentz condition \cite{2}
\begin{equation}
 \partial_{t_c}A_0-\frac{2}{t_c}A_0 -\partial_i A_i=0
\end{equation}
determine the normalised solutions for the free Proca field in de Sitter geometry. The spatial part of the solution is \cite{2}:
\begin{eqnarray}\label{sol1}
\vec{f}_{\vec{P},\lambda}(x)=\left\{
\begin{array}{cll}
\frac{i\sqrt{\pi}\omega Pe^{-\pi k/2}}{2M_Z(2\pi)^{3/2}}\left[(\frac{1}{2}+ik)\frac{\sqrt{-t_{c}}}{P}
H^{(1)}_{ik}\left(-Pt_{c}\right)-(-t_{c})^{3/2}H^{(1)}_{1+ik}\left(-Pt_{c}\right)\right]e^{i\vec{P}\vec{x}}\vec{\epsilon}\,(\vec{n}_{P},\lambda)&{\rm ;}&\lambda=0\\
\frac{\sqrt{\pi}e^{-\pi k/2}}{2(2\pi)^{3/2}}\sqrt{-t_{c}}H^{(1)}_{ik}\left(-P t_{c}\right)e^{i\vec{P}\vec{x}}\vec{\epsilon}\,(\vec{n}_{P},\lambda)
&{\rm ;}&\lambda=\pm 1.
\end{array}\right.
\nonumber\\
\end{eqnarray}
The temporal component of the solution for the Proca equation is \cite{2}:
\begin{eqnarray}
f_{0\vec{P},\lambda}(x)=\left\{
\begin{array}{cll}
\frac{\sqrt{\pi}\omega P e^{-\pi k/2}}{2M_Z(2\pi)^{3/2}}(-t_{c})^{3/2}H^{(1)}_{ik}\left(-Pt_{c}\right)e^{i\vec{P}\vec{x}}&{\rm for}&\lambda=0\\
0
&{\rm for}&\lambda=\pm 1.
\end{array}\right.
\end{eqnarray}

In the above equations for the solutions of the Proca equation $\vec{n}_{P}=\vec{P}/\mathcal{P}$ and $\vec{\epsilon}\,(\vec{n}_{P},\lambda)$ are the polarization vectors. For $\lambda=\pm 1$ these vectors are transversal on the momentum
$\vec{P}\cdot\vec{\epsilon}\,(\vec{n}_{P},\lambda=\pm1)=0$ and for $\lambda=0$ the polarization vectors are longitudinal on the momentum
$\vec{P}\cdot\vec{\epsilon}\,(\vec{n}_{P},\lambda=0)=P$, since $\vec{\epsilon}\,(\vec{n}_{P},\lambda=0)=\vec{n}_{P}$. The mass of the Z boson is denoted by $M_Z$, while the parameter $k=\sqrt{\left(\frac{M_Z}{\omega}\right)^2-\frac{1}{4}}$ depends on the ratio $\frac{M_Z}{\omega}$, provided that  $\frac{M_Z}{\omega}>\frac{1}{2}$.

The above free field solutions will be the basis of our study for writing down the amplitudes and probabilities corresponding to the processes of particle generation from vacuum in an expanding background.

\section{Neutral current interaction in de Sitter space-time}

The perturbative QED was developed in \cite{24} by taking into account that the theory of quantum fields with spin can be correctly constructed only in orthogonal (non-holonomic) local frames where the half-integer spins have sense. In \cite{24} it was proposed that the generators of the covariant representation are the differential operators produced by the Killing vectors which are associated to isometries according to the  generalized Carter and McLenagan formula \cite{CML}. They form an algebra of conserved observables that commute with the operators of the field equations. With this method the quantum states were globally defined on the entire manifold, and these states are independent on the local coordinates, so the vacuum state is unique and stable \cite{24}. In this approach the quantum states are prepared and measured by the same global apparatus which consists of the largest freely generated algebra that includes the conserved operators \cite{24}. This method does not exclude the cosmological particle creation which may be observed  by using  local detectors \cite{GH,DW}.

For developing the theory of electro-weak interactions we will adopt the same methods as those used in \cite{24}, and
we assume that the electro-weak transitions are measured by the same global apparatus which prepares all the quantum states, and this includes the $in$ and $out$ asymptotic free fields which remain  minimally coupled to gravity. This apparatus complies with an asymptotic prescription of frequencies separation that assures the uniqueness and stability of both the vacuum states of the free Dirac and Proca fields \cite{2,22}. This means that it cannot record particle creation in the absence of the electromagnetic interaction, as the local detectors \cite{GH,DW}.

Let us comment on the problem related to the definition of the S-matrix in de Sitter geometry. It is well known that in de Sitter geometry the only time-like Killing vector is not time-like everywhere and this problem generated concerns related to the possibility of defining the energy operator correctly, the in and out fields and the scattering operator \cite{38,39}. Here we consider that the S matrix can be defined since the time-like Killing vector remains time-like everywhere inside the light cone where an observer can perform physical measurements \cite{24,41}. In this approach the energy operator can be defined, but it does not commute with the momentum operator \cite{22,24}. Another result worth mentioning here was obtained in \cite{40}, where an S-matrix formalism was analysed for weakly coupled field theories in the global de Sitter manifold. A remarkable result is that from the S-matrix obtained in \cite{40} one can recover the usual S-matrix from Minkowski space in the flat-space limit, a result that was also reported in \cite{24}.

Let us assume that the theory of interactions between the vector neutral fields and fermion fields in de Sitter space-time can be developed by using the prescription of minimal coupling by introducing the interaction term with the prescription $\mathcal{L}_{int}=-(j_{neutral})^{\mu}A_{\mu}$, because the Z bosons have no electric charge and the particle coincides with the antiparticle, and mediates the  neutral  current interactions $(j_{neutral})\,^{\mu}$, which is given by:
\begin{eqnarray}\label{l1}
(j_{neutral})\,^{\mu}&=&\overline{\psi}_{\nu_e}\,\gamma^{\hat\alpha}e_{\hat\alpha}^{\mu}\left(\frac{1-\gamma^{5}}{2}\right)\psi_{\nu_e}-\cos(2\theta_{W})\overline{\psi}_e\,\gamma^{\hat\alpha}e_{\hat\alpha}^{\mu}\left(\frac{1-\gamma^{5}}{2}\right)\psi_e
\nonumber\\&&+2\sin^2(\theta_{W})\overline{\psi}_e\,\gamma^{\hat\alpha}e_{\hat\alpha}^{\mu}\left(\frac{1+\gamma^{5}}{2}\right)\psi_e.
\end{eqnarray}
where $e_0$ is the electric charge, $\theta_{W}$ is the Weinberg angle and, $\psi_{\nu_{e}}$ designates the neutrino-antineutrino field, $A_{\alpha}(Z)$ designates the Z boson field and $\psi_e$ designates the electron-positron field. Since it is well known that the neutrinos are only left-handed we use the left projector $\frac{1-\gamma^{5}}{2}$, with the specification that for the electrons and positrons we also have the right-handed part, and the corresponding right projector is $\frac{1+\gamma^{5}}{2}$.
In our present paper we study only the interaction of Z bosons with the massive fermions and the first term from the neutral current will be not taken into account in our study. The theory of fields interaction is developed in the conformal chart $\{t_c,\vec x\}$ with the Cartesian coordinates. We mention that a detailed analysis of the interaction between Z boson and neutrinos that include the definition of the transition amplitudes can be found in \cite{cc}.

The action can be expressed with the tetrad gauge invariant Lagrangian density that gives the coupling between $Z$ bosons and massive fermions, written with point independent Dirac matrices $\gamma^{\hat\alpha}$ and the tetrad fields $e_{\hat\alpha}^{\mu}$ and the Lagrangian densities for Proca field and Dirac fields as:
\begin{equation}
  S = \int d^4x \sqrt{-g} \bigg( \mathcal{L}_{Dirac} + \mathcal{L}_{Proca} + \mathcal{L}_{int} \bigg),
\end{equation}
which can be expanded as:
\begin{multline}\label{act}
  S = \int d^4x \sqrt{-g} \Bigg\{  \frac{i}{2} \big( \overline{\psi} \gamma^{\hat{\alpha}} D_{\hat{\alpha}} \psi - (\overline{D_{\hat{\alpha}} \psi}) \gamma^{\hat{\alpha}} \psi \big) -m \overline{\psi} \psi - \frac{1}{4}F_{\mu\nu} F^{\mu\nu} + \frac{M_Z^2}{2} A_\mu A^\mu \\
  + \bigg(\frac{e_0 \cos(2 \theta_W)}{\sin(2 \theta_W)} \bigg) \overline{\psi} \gamma^{\hat{\alpha}} e^\mu_{\hat{\alpha}} \bigg(\frac{1-\gamma^5}{2}\bigg) \psi A_\mu - e_0 \tan(\theta_W) \overline{\psi} \gamma^{\hat{\alpha}} e^\mu_{\hat{\alpha}} \bigg(\frac{1+\gamma^5}{2}\bigg) \psi A_\mu  \Bigg\}.
\end{multline}
where $m$ is the mass of the Dirac field and $M_Z$ is the mass of the Z boson field, $F_{\mu\nu}=\partial_\mu A_{\nu}-\partial_\nu A_{\mu}$ is the field strength. The notation $D_{\hat{\alpha}}=\partial_{\hat{\alpha}}+\Gamma_{\hat{\alpha}}$ represents the covariant derivatives in local frames, that depend on spin connections $\Gamma_{\hat{\alpha}}=\Gamma_{\hat{\alpha}\hat{\mu}\hat{\nu}}S^{\hat{\mu}\hat{\nu}}$, which are given in terms of basis generators $S^{\hat{\mu}\hat{\nu}}=i/4\{\gamma^{\hat{\mu}},\gamma^{\hat{\nu}}\}$ of the spinor representation of $SL(2,\textbf{C})$ group.
The Euler-Lagrange equations for fields will give the equations for interaction between the massive Dirac field and Z bosons
\begin{equation}\label{e1}
  \frac{\partial \mathcal{L}}{\partial \overline{\psi}} - \partial_\rho \bigg(\frac{\partial \mathcal{L}}{\partial (\partial_\rho \overline{\psi})} \bigg) = 0
\end{equation}

\begin{equation}\label{e2}
  \frac{\partial \mathcal{L}}{\partial A_\mu} - \partial_\rho \bigg(\frac{\partial \mathcal{L}}{\partial (\partial_\rho A_\mu)} \bigg) = 0.
\end{equation}
The equations of interactions are obtained by using equations (\ref{e1}), (\ref{e2}) and (\ref{act}). The Dirac equations will couple the potential $A_{\alpha}(x)$
\begin{equation}
  \big(i \gamma^{\hat{\alpha}} D_{\hat{\alpha}} - m \big) \psi(x) = \frac{e_0 \cos(2 \theta_W)}{\sin(2 \theta_W)} \gamma^{\hat{\mu}} e^\alpha_{\hat{\mu}} \bigg(\frac{1-\gamma^5}{2}\bigg) \psi(x) A_\alpha(x) - \tan(\theta_W) \gamma^{\hat{\mu}} e^\alpha_{\hat{\mu}} \bigg(\frac{1+\gamma^5}{2}\bigg) \psi(x) A_\alpha(x).
\end{equation}
In the de Sitter metric the above equation become:
\begin{eqnarray}\label{e21}
  \Big( -i\omega t_c( \gamma^0 \partial_{t_c} + \gamma^i \partial_i) + \frac{3i \omega}{2} \gamma^0 -m\Big)\psi(x) = \frac{e_0 \cos(2 \theta_W)}{\sin(2 \theta_W)} \gamma^{\hat{\mu}} e^\alpha_{\hat{\mu}} \bigg(\frac{1-\gamma^5}{2}\bigg) \psi(x) A_\alpha(x) \nonumber\\- \tan(\theta_W) \gamma^{\hat{\mu}} e^\alpha_{\hat{\mu}} \bigg(\frac{1+\gamma^5}{2}\bigg) \psi(x) A_\alpha(x).
\end{eqnarray}

The Proca equation will have the neutral current as the source and this can be read as:
\begin{equation}\label{fs}
  \partial_\rho \big( g^{\rho \alpha} g^{\mu \beta} \sqrt{-g(x)}F_{\alpha \beta}(x) \big) + \sqrt{-g} \, M_Z^2 A^\mu(x) = -\sqrt{-g(x)} \,e^\mu_{\hat{\alpha}} J^{\hat{\alpha}}(x)
\end{equation}
and we specify that the Lorentz condition is deduced in the case of the free Proca theory
\begin{equation}
\partial_{\mu}(\sqrt{-g}A^{\mu})=0
\end{equation}
which is maintained for the interacting fields.
Equation (\ref{fs}) can be expanded in terms of the temporal and spatial components of the potential vector $A^\mu$ and current $J^{\hat{\alpha}}$:
\begin{equation}\label{ai1}
  \partial_{t_c} (\partial_i A_i) - \Delta A_0 + \frac{M_Z^2}{\omega^2t_c^2} A_0 = -\sqrt{-g} J_0
\end{equation}

\begin{equation}\label{ai2}
  \partial_{t_c}^2 A_k - \Delta A_k - \partial_k (\partial_{t_c} A_0) + \partial_k (\partial_i A_i) + \frac{M_Z^2}{\omega^2t_c^2} A_k = \sqrt{-g} J_k
\end{equation}
Equations for the fields in interaction (\ref{ai1}), (\ref{e21}), are coupled nonlinear equations and the solutions can be written down by following the methods from the flat space case.
Then the system of coupled equations is replaced with one system of integral equations that contain information about the initial conditions. That means that one selects a Green function $G(x,y)$ corresponding to one initial condition, which helps us write the solutions for the interacting equations as follows:

\begin{equation}\label{grr}
  A^\mu (x)= \hat{A}^\mu(x) - \int d^4 y \sqrt{-g(y)} G^{\mu\alpha} (x,y) J_\alpha (y)
\end{equation}
where $\hat{A}^\mu(x)$ is a free field and $G^{\mu\alpha}$ is a Green function of the Proca equation. In the case of coupled Dirac equations the solution is:
\begin{multline}
  \psi(x) = \hat{\psi}(x) - e\int d^4 y \sqrt{-g(y)} S(x-y) \bigg[ \frac{\cos(2 \theta_W)}{\sin(2 \theta_W)} \gamma^{\hat{\mu}} e^\alpha_{\hat{\mu}} \bigg(\frac{1-\gamma^5}{2}\bigg) \psi(y) A_\alpha (y)  \\
  - \tan(\theta_W) \gamma^{\hat{\mu}} e^\alpha_{\hat{\mu}} \bigg(\frac{1+\gamma^5}{2}\bigg) \psi(y)  A_\alpha (y) \bigg]
\end{multline}
where $\hat{\psi}(x)$ is a free field and $S(x-y)$ is a Green function of the Dirac equation. The Green functions for the Dirac field satisfy the equation \cite{22,24}:
\begin{equation}\label{gdir}
E_D\,S(x-y)=-\frac{1}{\sqrt{-g(x)}}\,\delta^4(x-y).
\end{equation}
where the Dirac operator reads:
\begin{equation}
E_D=-i\omega t_c( \gamma^0 \partial_{t_c} + \gamma^i \partial_i) + \frac{3i \omega}{2} \gamma^0 -m.
\end{equation}
At the end of this section we mention the important results in regards to the propagator of the Dirac field in de Sitter space-time. First the propagator of the Dirac field on the de Sitter space-time in configuration representation was constructed by Candelas and Reine \cite{CRR}. The same propagator was obtained as a mode sum by Koskma and Prokopec in the general Friedmann–Lemaître–Robertson–Walker spacetimes of arbitrary dimensions \cite{PC}. A new integral representation in momentum space of the Feynman propagator for massive Dirac field was obtained by Cot\u aescu \cite{COT}. All these results complete the theory of Dirac field in de Sitter space-time and propose new methods related to the regularization and will allow one to study the second order scattering processes in this geometry.

\section{Reduction formalism and transition amplitudes}
In this section we establish the definition of the transition amplitudes of the theory of interaction between Z bosons and massive fermions. The solutions for the equation of interacting fields will be used for constructing the reduction of the fields from the $in/out$ sectors and then by using the perturbation theory we will define the first order transition amplitudes of the electro-weak theory.
\subsection{Transversal modes}
Taking into account that for the transversal modes $\lambda=\pm1$ we have only the spatial part of the solutions ($A_0=0$) the equation of interaction can be rewritten in the form:
\begin{equation}\label{ec1}
  \bigg[ \bigg(\partial_{t_c}^2 - \Delta + \frac{\mu^2}{t_c^2} \bigg)\delta_{ik}  + \partial_i \partial_k \bigg]A_k =- \sqrt{-g} J_i
\end{equation}
where we introduce the Proca operator for the transversal modes $E_{ik}(x)=\bigg(\partial_{t_c}^2 - \Delta + \frac{\mu^2}{t_c^2} \bigg)\delta_{ik}  + \partial_i \partial_k $.

Since we want to obtain the solutions for the coupled field equation we must work with the Green function for the Proca equation \cite{2,WT}. The propagator for massive vector fields on de Sitter background of arbitrary dimension was constructed in \cite{WT}. The propagator was proven to be de Sitter invariant and possesses the correct flat spacetime and massless limits \cite{WT}.
For obtaining useful equations with the Green functions of the Proca equation that help us verify that the solution proposed in equation (\ref{grr}) is an exact solution of the interacting equation (\ref{ec1}), we use the relation \cite{2,WT}:
\begin{equation}
  \eta^{\alpha \beta} \partial_\alpha \big[ \partial_\beta G_{\mu \nu} (x,y) - \partial_\mu G_{\beta \nu} (x,y) \big] + \frac{\mu^2}{t_c^2} G_{\mu \nu} (x,y) = \eta_{\mu \nu} \delta^4 (x-y)
\end{equation}
Taking the spatial values for the indices $\mu=i,\,\nu=j$ in the above equation and knowing that the temporal parts of the Green functions are vanishing, we deduce the following equation:
\begin{equation}\label{g1}
  \bigg[ \delta_{ik} \bigg( \partial_{t_c}^2 - \Delta + \frac{\mu^2}{t_c^2} \bigg) + \partial_k \partial_i \bigg] G_{kj} (x,y) = E_{ik}(x)G_{kj} (x,y)=-\delta_{ij} \delta^4 (x-y)
\end{equation}
The solution for the equation (\ref{g1}) is written as such that by using equation (\ref{ec1}) one can verify that is an exact solution:
\begin{equation}\label{sp}
  A_k (x) = \hat{A}_k (x) - \int d^4 y \sqrt{-g(y)} \, G_{kj} (x,y) J_j (y)
\end{equation}
where $\hat{A}_k (x)$ is a free field such that $E_{ik}(x)\hat{A}_k (x)=0$. The Proca operator applied on the above equation then gives:
\begin{equation}
  {E}_{ik}(x) A_k (x) = {E}_{ik}(x) \hat{A}_k (x) - \int d^4y \sqrt{-g(y)} \, {E}_{ik}(x) G_{kj} (x,y) J_j (y)=-\sqrt{-g}  J_i (x)
\end{equation}
which complete our proof.

For constructing a theory of interactions, we seek operators that create independent particle states with each particle propagating with its physical mass.
Let us start constructing these operators, using  (\ref{sp}) which offers us the possibility of constructing free fields, which are asymptotic equal (at $t\rightarrow\pm\infty$)
with solutions of equation (\ref{ec1}). The retarded Green functions $G_{ij\,\,R}(x,y)$ vanish at $t_c\rightarrow-\infty$ ($t\rightarrow-\infty$) while the advanced ones $G_{ij\,\,A}(x,y)$ vanish for $t_c\rightarrow 0$ ($t\rightarrow\infty$). Further the equation (\ref{sp}) can be expressed with the retarded and advanced functions as follows:
\begin{eqnarray}
 A_k (x) = \hat{A}_{k\,\,R/A} (x) - \int d^4 y \sqrt{-g(y)} \, G_{kj\,\,R/A} (x,y) J_j (y)
\end{eqnarray}
Now we can define the free fields, $\hat{A}_{k\,\,R/A}(x)$ which satisfy:
\begin{eqnarray}
\lim_{t\rightarrow\mp\infty}(A_k(x)-\hat{A}_{k\,\,R/A}(x))=0.
\end{eqnarray}
The free fields $\hat{A_{k\,\,R}}(x)$ and $\hat{A_{k\,\,A}}(x)$ have the mass $M_Z$ and are equal at $t\rightarrow \pm\infty$ with exact solutions of the coupled equations and represent the fields before and after the interaction. As in Minkowski the free fields $\hat{A_{k\,\,R}}(x)$ and $\hat{A_{k\,\,A}}(x)$ are defined up to a normalization constant noted with $\sqrt{z_3}$. This allows us to define the $in/out$ fields up to a normalization constant $\sqrt{z_3}$ with the help of retarded and advanced Green functions:
\begin{equation}
 \sqrt{z_3} A_k^{in/out}(x) = A_k (x) + \int d^4 y \sqrt{-g(y)}\, G_{kj\,\,R/A}(x,y) {E}_{ij}(y) A_i (y)
\end{equation}
The reduction formalism can be developed now by using the expansion of the field operator
\begin{equation}
  A^k (x) = \sum_{\lambda} \int d^3 p\, \Big( a(\vec{p},\lambda) f_{\vec{p},\lambda}^k (x) + b^+ (\vec{p},\lambda) {f_{\vec{p},\lambda}^k}^* (x) \Big)
\end{equation}
with the observation that from the above normalization we can define the creation and annihilation operators
\begin{eqnarray}
  a(\vec{p},\lambda) &=& \langle f_{\vec{p},\lambda}^k (x), A_k (x) \rangle \nonumber \\
   &=& i \int d^3 x {f_{\vec{p},\lambda}^k}^* (x) \overleftrightarrow{\partial_{t_c}} A_k (x).
\end{eqnarray}

Considering two states $|\alpha\rangle_{in}$ and $|\beta\rangle_{out}$, then the probability of transition from state $\alpha$ to state $\beta$ is defined as the scalar
product of the two states: ${}_{out}\langle \beta|\alpha\rangle_{in}$. These are the elements of the matrix for the scattering operator $S_{\beta\alpha}$, which can be used in applications. The scattering operator assures the stability of the vacuum state and one particle state, and in addition transforms any $out$ field in the equivalent $in$ field.

Then the reduction of the Proca particle from the $out$ state gives:
\begin{multline}
  {}_{out}\langle \beta 1 (\vec{p}\,', \lambda ') | \alpha 1 (\vec{p},\lambda) \rangle_{in} = {}_{out}\langle  \beta | \big( a_{out} (\vec{p}\,',\lambda') - a_{in} (\vec{p}\,',\lambda') \big) | \alpha 1 (\vec{p},\lambda) \rangle_{in} \\
  + {}_{out}\langle  \beta | a_{in} (\vec{p}\,',\lambda') a^+_{in} (\vec{p},\lambda) |  \alpha \rangle_{in}
\end{multline}
where the difference is
\begin{eqnarray}
  a_{out} (\vec{p}\,',\lambda') - a_{in} (\vec{p}\,',\lambda') &=& i \int d^3 x {f_{\vec{p},\lambda}^k}^* (x) \overleftrightarrow{\partial_{t_c}} \Big( {A_k}^{out} (x) - {A_k}^{in} (x) \Big)  \nonumber \\
   &=& i\int d^3 x \int d^4 y \sqrt{-g(y)}{f_{\vec{p},\lambda}^k}^* (x) \overleftrightarrow{\partial_{t_c}} G_{kj} (x,y) E_{ij} (y) A_i (y)
   \nonumber\\
\end{eqnarray}
where the functions $G_{kj} (x,y)$ are the total commutator functions defined as:
\begin{equation}
  G_{kj} (x,y) = i \sum_{\lambda} \int d^3 p \Big( f_{{\vec{p},\lambda}_k} (x) {f_{{\vec{p},\lambda}_j}}^* (y) - {f_{{\vec{p},\lambda}_k}}^* (x) f_{{\vec{p},\lambda}_j} (y) \Big).
\end{equation}
The calculation is completed by using the integral
\begin{equation}
  i \int d^3 x\, {f_{\vec{p},\lambda}^k}^* (x) \overleftrightarrow{\partial_{t_c}} G_{kj} (x,y) = i {f_{{\vec{p},\lambda}_j}}^* (y),
\end{equation}
which helps us write down the final result for the reduction of the Proca particle from the $out$ state
\begin{multline}
  {}_{out}\langle  \beta 1 (\vec{p}\,', \lambda ') | \alpha 1 (\vec{p},\lambda) \rangle_{in} = \delta_{\lambda \lambda'} \delta^3 (\vec{p}-\vec{p}\,') {}_{out}\langle \beta | \alpha \rangle_{in}  \\
  + i \int d^4 y \sqrt{-g(y)} \cdot {f_{{\vec{p}',\lambda'}_i}}^* (y) E_{ij}(x) {}_{out}\langle \beta | A _j (x) | \alpha, 1(\vec{p},\lambda) \rangle_{in},
\end{multline}
where the first term is dropped since it represents the case where the particle makes the transition $in-out$ without interacting with other particles. The reduction of the Proca particle from the $in$ state gives the final result
\begin{multline}
  {}_{out}\langle \beta 1 (\vec{p}\,', \lambda') | \alpha 1(\vec{p},\lambda) \rangle_{in} = \delta_{\lambda\lambda'} \delta^3 (\vec{p}-\vec{p}\,') {}_{out}\langle  \beta | \alpha \rangle_{in} \\
  + i \int d^4 y \sqrt{-g(y)} \cdot {}_{out}\langle \beta 1(\vec{p}\,',\lambda') | A_j (x) | \alpha \rangle_{in} \overleftarrow{E_{ij}}(x) f_{{\vec{p}\,',\lambda'}_i} (y)
\end{multline}
Then the generalized Green functions can be expressed in terms of free fields as:
\begin{equation}
\langle 0|\Psi(x_1)\overline\Psi(x_3)\mathcal{A}_k(x_3)...|0\rangle=\frac{\langle 0|\psi(x_1)\overline\psi(x_3)A_k(x_3)...\mathbf{S}|0\rangle}{\langle 0|\mathbf{S}|0\rangle},
\end{equation}
where the scattering operator is
\begin{eqnarray}\label{s}
&&\mathbf{S} =T  exp\left[-i\int d^4 x \sqrt{-g(x)}\,\mathcal{L}_{int}\right] \nonumber\\
&&=T  exp\left[-i\int d^4 x \sqrt{-g(x)}\left(\frac{e_0 \cos(2 \theta_W)}{\sin(2 \theta_W)} \overline{\psi} \gamma^{\hat{\alpha}} e^\mu_{\hat{\alpha}} \bigg(\frac{1-\gamma^5}{2}\bigg) \psi A_\mu - e_0 \tan(\theta_W) \overline{\psi} \gamma^{\hat{\alpha}} e^\mu_{\hat{\alpha}} \bigg(\frac{1+\gamma^5}{2}\bigg) \psi A_\mu\right) \right].
\nonumber\\
\end{eqnarray}
For example, if we consider a process of spontaneous emission from de Sitter vacuum of the triplet Z boson and the electron-positron pair the reduction procedure gives:
\begin{multline}
  {}_{out}\langle  1(\vec{P},\lambda), 1(\vec{p},\sigma), 1(\vec{p}\,',\sigma ') | 0 \rangle_{in} = \bigg( \frac{i}{\sqrt{z_3}} \bigg) \bigg( \frac{i}{\sqrt{z_2}} \bigg)^2 \int d^4 y_1 \sqrt{-g(y_1)} \int d^4 y_2 \sqrt{-g(y_2)} \\
  \times \int d^4 y_3 \sqrt{-g(y_3)} \overline{U}_{\vec{p},\sigma}(y_1) E_D (y_1) {f_{{\vec{p},\lambda}_i}}^* (y_2) E_{il} (y_2) \langle 0 | T[\psi(y_1) A_l (y_2) \overline{\psi} (y_3)] | 0 \rangle \overleftarrow{E_D} (y_3) V_{\vec{p}\,',\sigma'}(y_3)
\end{multline}
The Green functions from the equation can be expressed in terms of Feynmann propagators by considering all the T contractions from the Wick theorem.
Then by using the first order in perturbation theory with $\mathbf{S}^{(1)}$ and spatial index $\mu=i$, and by making the T contraction between the cinematical part and the dynamical part we obtain the definition of the transition amplitude in the first order of the perturbation theory for the transversal modes of the Proca field:
\begin{eqnarray}
  A_{i \rightarrow f} &=& -i \int d^4 x \sqrt{-g(x)} \int d^4 y_1 \sqrt{-g(y_1)} \int d^4 y_2 \sqrt{-g(y_2)} \int d^4 y_3 \sqrt{-g(y_3)} \bigg[ \overline{U}_{\vec{p},\sigma}(y_1)  \nonumber  \\
  && \times \frac{e_0 \cos(2 \theta_W)}{\sin(2 \theta_W)} E_D (y_1) S(y_1 - x) \gamma^{\hat{j}} e^k_{\hat{j}} \bigg(\frac{1-\gamma^5}{2}\bigg) {f_{{\vec{p},\lambda}_i}}^* (y_2) E_{il} (y_2) D_{lk} (y_2 , x) E_D (y_3)  \nonumber  \\
  && \times S(y_3 - x)  V_{\vec{p}\,',\sigma'}(y_3) - e_0 \tan(\theta_W) \overline{U}_{\vec{p},\sigma}(y_1) E_D (y_1) S(y_1 - x) \gamma^{\hat{j}} e^k_{\hat{j}} \bigg(\frac{1+\gamma^5}{2}\bigg) \nonumber \\
  && \times {f_{{\vec{p},\lambda}_i}}^* (y_2) E_{il} (y_2) \frac{1}{i} D_{lk} (y_2 , x) E_D (y_3) S(y_3 - x) V_{\vec{p}\,',\sigma'}(y_3)  \bigg].
\end{eqnarray}
We use the action of the field operators on propagators and solve the integrals using the properties of the delta Dirac functions, and obtain the final result
\begin{eqnarray}
 A_{i \rightarrow f} &=& \int d^4 x \sqrt{-g(x)} \bigg[ \frac{e_0 \cos(2 \theta_W)}{\sin(2 \theta_W)} \overline{U}_{\vec{p},\sigma}(x) \gamma^{\hat{j}} e^k_{\hat{j}} \bigg(\frac{1-\gamma^5}{2}\bigg) V_{\vec{p}\,',\sigma'}(x) {f_{{\vec{p},\lambda}\,_k}}^* (x)   \nonumber\\
 && - e_0 \tan(\theta_W) \overline{U}_{\vec{p},\sigma}(x) \gamma^{\hat{j}} e^k_{\hat{j}} \bigg(\frac{1+\gamma^5}{2}\bigg) V_{\vec{p}\,',\sigma'}(x) {f_{{\vec{p},\lambda}\,_k}}^* (x)  \bigg].
\end{eqnarray}
All of the first order processes can be computed by using the above formalism. The above definition of the transition amplitude holds in both charts $\{t,\vec x\}$ and $\{t_c,\vec x\}$. In addition in the limit $\omega \rightarrow 0$ the amplitude is reduced to the well known formula from the Minkowski theory.

The reduction formalism for the Dirac field in de Sitter space-time was developed in \cite{rc,24}, and here we used the final results for replacing the $in\ out$ Dirac fields in our amplitudes.

\subsection{Longitudinal modes}
In the case of longitudinal modes the temporal and spatial components of the four vector potential are combined in coupled equations according to equations (\ref{ai1}), (\ref{ai2}) and are useful for writing the equation with all components by rewriting the Proca operator with temporal and spatial components. In this setup the Proca equation can be written as:
\begin{equation}
   \bigg[ \delta^\beta_\mu \bigg(\partial_\rho \partial^\rho + \frac{\mu^2}{t_c^2} \bigg) - \partial^\beta \partial_\mu \bigg] A_\beta ={E}^\beta_\mu (x) A_\beta =  0.
\end{equation}

The Green function of the Proca field satisfies the equation:
\begin{equation}\label{gpro}
  \bigg[ \delta^\beta_\mu \bigg( \partial_\rho \partial^\rho + \frac{\mu^2}{t_c^2} \bigg) - \partial^\beta \partial_\mu \bigg] G_{\beta\nu} (x,y) = \eta_{\mu\nu} \delta^4 (x-y)
\end{equation}
from which we deduce the following relations:
\begin{equation}
  \bigg( -\Delta + \frac{\mu^2}{t_c^2} \bigg) G_{00} (x,y) + \partial_i \partial_{t_c} G_{i0} (x,y) = \delta^4 (x-y),
\end{equation}
\begin{equation}
  \bigg[ \delta_{ik} \bigg( \partial^2_{t_c} - \Delta + \frac{\mu^2}{t_c^2} \bigg) + \partial_k \partial_i \bigg] G_{kj} (x,y) - \partial_{t_c} \partial_i G_{0j} (x,y) = -\delta_{ij} \delta^4 (x-y)
\end{equation}
which combine the spatial and temporal components of the field.
Then one can observe that the equations of interaction (\ref{ai1}), (\ref{ai2}) can be combined in a single equation:
\begin{multline}\label{alo}
  {E}^\mu_\nu (x) A^\nu (x) = \frac{e_0 \cos(2 \theta_W)}{\sin(2 \theta_W)} \overline{\psi} \gamma^{\hat{\alpha}} e^\mu_{\hat{\alpha}} \bigg(\frac{1-\gamma^5}{2}\bigg) \psi \sqrt{-g(x)} \\
  - e_0 \tan(\theta_W) \overline{\psi}\gamma^{\hat{\alpha}} e^\mu_{\hat{\alpha}} \bigg(\frac{1+\gamma^5}{2}\bigg) \psi \sqrt{-g(x)} = \sqrt{-g(x)} J^\mu (x).
\end{multline}
with the solution written in terms of Green functions:
\begin{equation}
  A_\mu (x) = \hat{A_\mu} (x) + \int d^4 y \sqrt{-g(y)} G_{\mu\beta} (x,y) J^\beta (y),
\end{equation}
where $ \hat{A_\mu} (x)$ is a free field such that $ {E}^\mu_\nu (x) \hat{A_\mu} (x)=0$.
One can verify that this is an exact solution by applying the Proca operator ${E}^\mu_\nu (x)$
\begin{eqnarray}
  {E}^\mu_\nu (x) A_\mu (x) &=& {E}^\mu_\nu \hat{A_\mu} (x) + \int d^4 y \sqrt{-g(y)} {E}^\mu_\nu G_{\mu\beta} (x,y) J^\beta (y) \nonumber\\
   &=& \int d^4 y \sqrt{-g(y)} \cdot \eta_{\nu\beta} \delta^4 (x-y) J^\beta (y) \nonumber\\
   &=& \sqrt{-g(x)} J_\nu (x)
\end{eqnarray}
Then we can define the $in\ out$ fields in terms of the retarded and advanced Green functions up to a normalization constant denoted with $ \sqrt{z_3}$
\begin{equation}
 \sqrt{z_3} {A_\mu}^{in/out} (x) = A_\mu - \int d^4 y \sqrt{-g(y)}  {G_{\mu\beta}}_{R/A} (x,y) {E}^\beta_\alpha (y) A^\alpha (y).
\end{equation}
Since we take into account only the longitudinal components for $\lambda=0$ the field operator can be expanded as:
\begin{equation}
  A^\mu (x) =\int d^3 p\, \Big( a(\vec{p},\lambda) f_{\vec{p},\lambda=0}^\mu (x) + b^+ (\vec{p},\lambda) {f_{\vec{p},\lambda=0}^\mu}^* (x) \Big).
\end{equation}
The reduction formalism can be constructed by considering the transition between two states as in the previous section
\begin{multline}
  {}_{out}\langle \beta 1 (\vec{p}\,', 0) | \alpha 1 (\vec{p},0) \rangle_{in} = {}_{out}\langle \beta | \big( a_{out} (\vec{p}\,',0) - a_{in} (\vec{p}\,',0) \big) | \alpha 1 (\vec{p},0) \rangle_{in} \\
  + {}_{out}\langle \beta | a_{in} (\vec{p}\,',0) a^+_{in} (\vec{p},0) | \alpha \rangle_{in}
\end{multline}
and the definitions of the creation and annihilation operators in terms of quantities with four components. The difference between the operators gives:
\begin{eqnarray}
   a_{out} (\vec{p}\,',0) - a_{in} (\vec{p}\,',0) = i \int d^3 x {f_{\vec{p},0}^\mu}^* (x) \overleftrightarrow{\partial_{t_c}} \Big( {A_\mu}_{out} (x) - {A_\mu}_{in} (x) \Big)\nonumber\\
= -i \int d^3 x {f_{\vec{p},0}^\mu}^* (x) \overleftrightarrow{\partial_{t_c}} \int d^4 y \sqrt{-g(y)} \cdot \Big( {G_{\mu\beta}}_A (x,y) \\
  - {G_{\mu\beta}}_R (x,y) \big) {E}_\alpha^\beta (y) A^\alpha (y)
\end{eqnarray}
The commutator function is defined using the mode functions as
\begin{equation}
  G_{\mu\beta}(x,y) = {G_{\mu\beta}}_A(x,y) - {G_{\mu\beta}}_R(x,y) = i\int d^3 p \Big( {f_{\vec{p},0}}_\mu (x) {f_{\vec{p},0}}_\beta (y) - {f_{\vec{p},0}}_\mu^* (x) {f_{\vec{p},0}}_\beta (y) \Big)
\end{equation}
Then we compute the integral with the commutator function
\begin{equation}
  i\int d^3 x {f_{\vec{p},0}^\mu}^* (x) \overleftrightarrow{\partial_{t_c}} G_{\mu\beta} (x,y) = i \int d^3 p \delta^3 (\vec{p} - \vec{p}') {f_{\vec{p},0}}_\beta^* (y) = i{f_{\vec{p},0}}_\beta^* (y)
\end{equation}
and we obtain the final result:
\begin{equation}
  a_{out} (\vec{p}\,',0) - a_{in} (\vec{p}\,',0) = i \int d^3 y \sqrt{-g(y)} {f_{\vec{p},0}}_\beta (y) {E}^\beta_\alpha (y) A^\alpha (y)
\end{equation}
The final result for the reduction of the Proca particle from the $out$ state gives:
\begin{multline}
  {}_{out}\langle \beta 1 (\vec{p}\,', 0) | \alpha 1 (\vec{p},0) \rangle_{in} = \delta_{\lambda \lambda'} \delta^3 (\vec{p}-\vec{p}\,')\, {}_{out}\langle \beta | \alpha \rangle_{in}  \\
  + i \int d^4 y \sqrt{-g(y)} {f_{{\vec{p}',0}_\beta}}^* (y) {E}^\beta_\alpha (y) {}_{out}\langle \beta | A ^\alpha (x) | \alpha, 1(\vec{p},\lambda) \rangle_{in}
\end{multline}
with the specification that the reduction from the $in$ state can be obtained in a similar manner. Let us consider the amplitude for the same process of Z boson and electron-positron emission from vacuum, and the associated amplitude written with the reduction procedure for the modes with $\lambda=0$:
\begin{multline}
  {}_{out}\langle 1(\vec{P},\lambda), 1(\vec{p},\sigma), 1(\vec{p}\,',\sigma ') |  0 \rangle_{in} = \bigg( \frac{i}{\sqrt{z_3}} \bigg) \bigg( \frac{i}{\sqrt{z_2}} \bigg)^2 \int d^4 y_1 \sqrt{-g(y_1)} \int d^4 y_2 \sqrt{-g(y_2)} \\
  \times \int d^4 y_3 \sqrt{-g(y_3)} \overline{U}_{\vec{p},\sigma}(y_1) E_D (y_1) {f_{\vec{p},0}^\beta}^* (y_2) {E}^\mu_\beta (y_2) \langle 0 | T[\psi(y_1) A_\mu (y_2) \overline{\psi} (y_3) | 0 \rangle \overleftarrow{E_D} (y_3) V_{\vec{p}\,',\sigma'}(y_3)
\end{multline}
By using the term corresponding to the first order in perturbation theory from the expansion of the scattering operator (\ref{s}) and performing all possible T contractions we obtain:
\begin{eqnarray}
  A_{i \rightarrow f} &=& \int d^4 x \sqrt{-g(x)} \int d^4 y_1 \sqrt{-g(y_1)} \int d^4 y_2 \sqrt{-g(y_2)} \int d^4 y_3 \sqrt{-g(y_3)} \bigg[ \overline{U}_{\vec{p},\sigma}(y_1)  \nonumber  \\
  && \times \frac{e_0 \cos(2 \theta_W)}{\sin(2 \theta_W)} E_D (y_1) S(y_1 - x) \gamma^{\alpha} \bigg(\frac{1-\gamma^5}{2}\bigg) {f_{\vec{p},0}^\beta}^* (y_2) {E}^\mu_\beta (y_2) D_{\mu\alpha} (x,y_2)  \nonumber  \\
  && \times S(y_3 - x) \overleftarrow{E_D} (y_3) V_{\vec{p}\,',\sigma'}(y_3) - e_0 \tan(\theta_W) \overline{U}_{\vec{p},\sigma}(y_1) E_D (y_1) S(y_1 - x) \gamma^{\alpha} \bigg(\frac{1+\gamma^5}{2}\bigg) \nonumber \\
  && \times {f_{\vec{p},0}^\beta}^* (y_2) {E}^\mu_\beta (y_2) D_{\mu\alpha} (x,y_2) S(y_3 - x) \overleftarrow{E_D} (y_3) V_{\vec{p}\,',\sigma'}(y_3)  \bigg]
\end{eqnarray}
The final form of the transition amplitude contains both the spatial and temporal parts of the Proca field and is obtained by using the relations with Green functions (\ref{gdir}), (\ref{gpro}). We also note that on a curved background the definition of the transition amplitude is similar to the flat space case up to the measure of integration $\sqrt{-g(x)}$:
\begin{multline}
  A_{i \rightarrow f} = \int d^4 x \sqrt{-g(x)} \bigg[ \frac{e_0 \cos(2 \theta_W)}{\sin(2 \theta_W)} \overline{U}_{\vec{p},\sigma}(x) \gamma^{\hat{\nu}} e^\alpha_{\hat{\nu}} \bigg(\frac{1-\gamma^5}{2}\bigg) V_{\vec{p}\,',\sigma'}(x) {f_{{\vec{P},0}_\alpha}}^* (x)   \\
  - e_0 \tan(\theta_W) \overline{U}_{\vec{p},\sigma}(x) \gamma^{\hat{\nu}} e^\alpha_{\hat{\nu}} \bigg(\frac{1+\gamma^5}{2}\bigg) V_{\vec{p}\,',\sigma'}(x) {f_{{\vec{P},0}_\alpha}}^* (x)  \bigg].
\end{multline}
The amplitudes defined above are the starting point for developing the perturbative electro-weak interactions in de Sitter metric. This include the interesting phenomena related to particle production in early universe. Finally we must point out that the amplitude obtained in the cases
$\lambda=\pm1$ and $\lambda=0$  could be emerged in a single formula that contains the contributions of the temporal part and spatial part of the Proca field.

\section{Probability of transition}

In this section we compute the amplitude of the transition for the process of electron-positron and Z boson generation from de Sitter vacuum, using the solutions of the Proca equation corresponding to $\lambda=\pm1$ or $\lambda=0$. The first order transition amplitude corresponding to the process $vac\rightarrow Z+e^-+e^+$, which is the spontaneous generation from the de Sitter vacuum of a Z boson and an electron-positron pair is given by :
\begin{eqnarray}
  A_{Ze\bar{e}} = \int d^4 x \sqrt{-g} \Bigg\{ \bigg(\frac{e_0 \cos(2 \theta_W)}{\sin(2 \theta_W)} \bigg) \overline{U}_{\vec{p},\sigma}(x) \gamma^{\hat{\alpha}} e^\mu_{\hat{\alpha}} \bigg(\frac{1-\gamma^5}{2}\bigg) V_{\vec{p}\,',\sigma'}(x) f^*_{\mu\,\vec{P},\lambda}(x) \\
  - e_0 \tan(\theta_W) \overline{U}_{\vec{p},\sigma}(x) \gamma^{\hat{\alpha}} e^\mu_{\hat{\alpha}} \bigg(\frac{1+\gamma^5}{2}\bigg) V_{\vec{p}\,',\sigma'}(x) f^*_{\mu\,\vec{P}, \lambda}(x) \Bigg\}
\end{eqnarray}
where $e_0$ is the electric charge, $\theta_{W}$ is the Weinberg angle. We specify that for the electrons and positrons we have the left-handed part with the projector $\frac{1-\gamma^{5}}{2}$ and also the right-handed part and the corresponding right projector $\frac{1+\gamma^{5}}{2}$. The Z boson has no electric charge and the particle coincides with the antiparticle, and mediates the  neutral  current interactions.

\subsection{Amplitude of transition for $\lambda=\pm1$}
The modes with $\lambda=\pm1$ only have the spatial components non-vanishing, so the amplitude in this case is:
\begin{eqnarray}\label{a}
\mathcal{A}_{Ze\overline{e}}(\lambda=\pm1)&=&\int d^4x\sqrt{-g}\,\biggl\{\left(\frac{e_0\cos(2\theta_{W})}{\sin(2\theta_{W})}\right)\overline{U}_{\vec{p},\sigma}(x)\,\gamma^{\hat i}e_{\hat i}^{j}\left(\frac{1-\gamma^{5}}{2}\right)V_{\vec{p}\,',\sigma'}(x) f_{j\vec{P},\lambda=\pm1}^*(x)\nonumber\\
&&-(e_0\tan(\theta_{W}))\overline{U}_{\vec{p},\sigma}(x)\,\gamma^{\hat i}e_{\hat i}^{j}\left(\frac{1+\gamma^{5}}{2}\right)V_{\vec{p}\,',\sigma'}(x)f_{j\vec{P},\lambda=\pm1}^*(x)\biggl\}\nonumber\\
\end{eqnarray}

The amplitude of the process is computed using the equations (\ref{sol}), (\ref{sol1}) and the variable change $z=-t_c$ in the temporal integral, while the spatial integral gives the momentum conservation in the process and we introduce the sign function for the helicities:
\begin{eqnarray}\label{a1}
&&\mathcal{A}_{Ze\overline{e}}(\lambda=\pm1)=\frac{\pi^{3/2}\sqrt{pp'}\, e^{-\pi k/2}}{8(2\pi)^{3/2}}\,\delta^3(\vec{p}+\vec{p}\,'+\vec{P})\nonumber\\
&&\times\biggl\{ \frac{e_0\cos(2\theta_{W})}{\sin(2\theta_{W})}\,sgn(\sigma')
\int_0^{\infty} dz\left[
z^{3/2}H^{(2)}_{\nu_{+}}(pz)H^{(2)}_{\nu_{-}}(p'z)H^{(2)}_{-ik}\left(Pz\right)\right]\nonumber\\
&&-e_0\tan(\theta_{W})\,sgn(\sigma)\int_0^{\infty} dz\left[
z^{3/2}H^{(2)}_{\nu_{-}}(pz)H^{(2)}_{\nu_{+}}(p'z)H^{(2)}_{-ik}\left(Pz\right)\right] \biggl\}\nonumber\\
&&\times \xi_{\sigma}^+(\vec{p}\,)\,\vec{\sigma}\vec{\epsilon}\,^*\,(\vec{n}_{P},\lambda=\pm1)\eta_{\sigma'}(\vec{p}\,')
\end{eqnarray}
The delta Dirac function of momenta assures the momentum conservation in this process. By using the equations (\ref{a3}) from Apendix we arrive at the final result
\begin{eqnarray}\label{a1}
&&\mathcal{A}_{Ze\overline{e}}(\lambda=\pm1)=\frac{i^{-1/2}e_0}{(8\pi)}\,\delta^3(\vec{p}+\vec{p}\,'+\vec{\mathcal{P}})
\,\xi_{\sigma}^+(\vec{p}\,)\,\vec{\sigma}\vec{\epsilon}\,^*\,(\vec{n}_{\mathcal{P}},\lambda=\pm1)\eta_{\sigma'}(\vec{p}\,')\nonumber\\
&&\times\biggl\{\frac{\cos(2\theta_{W})}{\sin(2\theta_{W})}\,sgn(\sigma')A_1(\lambda=\pm1)
-\tan(\theta_{W})\,sgn(\sigma)A_2(\lambda=\pm1) \biggl\},
\end{eqnarray}
where the functions $A_1(\lambda=\pm1)\,,A_2(\lambda=\pm1)$ are defined as :
\begin{eqnarray}\label{aa3}
  A_1(\lambda=\pm1) &=& p'  B_{Kk} (pp'P) -  B_{1Kk} (pp'P)  \nonumber\\
  && -  B_{2Kk} (pp'P) + pB_{-Kk} (pp'P) , \\
  A_2(\lambda=\pm1) &=& p B_{Kk} (pp'P) - B_{1Kk} (pp'P)  \nonumber\\
  && -  B_{2Kk} (pp'P) + p' B_{-Kk} (pp'P)
\end{eqnarray}
and the $B$ functions are defined in terms of Appel hypergeometric functions and gamma Euler functions:
\begin{eqnarray}
  B_{Kk} (pp'P) = \frac{(pp')^{-iK} (iP)^{-\frac{5}{2}+2iK}e^{\pi K}}{\cosh^2(\pi K)\Gamma \big( \frac{1}{2} - iK \big) \Gamma \big( \frac{3}{2} - iK \big) } \Gamma \bigg(\frac{5-4iK+2ik}{4} \bigg) \Gamma \bigg(\frac{5-4iK-2ik}{4} \bigg) \nonumber\\
  \times F_4 \bigg( \frac{5-4iK+2ik}{4}, \frac{5-4iK-2ik}{4}, \frac{3}{2}-iK, \frac{1}{2}-iK; \Big(\frac{p}{P}\Big)^2, \Big(\frac{p'}{P}\Big)^2 \bigg);
\end{eqnarray}

\begin{eqnarray}
  B_{1Kk} (pp'P) = \frac{(p)^{-iK} (p')^{iK} (iP)^{-3/2}}{\cosh^2(\pi K)\Gamma \big(\frac{1}{2}-iK \big) \Gamma \big( \frac{1}{2} +iK\big)} \Gamma \bigg(\frac{3+2ik}{4} \bigg) \Gamma \bigg(\frac{3-2ik}{4} \bigg) \nonumber\\\
  \times F_4 \bigg( \frac{3+2ik}{4}, \frac{3-2ik}{4}, \frac{1}{2}-iK, \frac{1}{2}+iK; \Big(\frac{p}{P}\Big)^2, \Big(\frac{p'}{P}\Big)^2 \bigg);
\end{eqnarray}

\begin{eqnarray}
  B_{2Kk} (pp'P) = \frac{(p)^{1+iK} (p')^{1-iK} (iP)^{-\frac{7}{2}}}{\cosh^2(\pi K)\Gamma \big(\frac{3}{2}+iK \big) \Gamma \big( \frac{3}{2} - iK\big)} \Gamma \bigg(\frac{7+2ik}{4} \bigg) \Gamma \bigg(\frac{7-2ik}{4} \bigg) \nonumber\\\
  \times F_4 \bigg( \frac{7+2ik}{4}, \frac{7-2ik}{4}, \frac{3}{2}+iK, \frac{3}{2}-iK; \Big(\frac{p}{P}\Big)^2, \Big(\frac{p'}{P}\Big)^2 \bigg).
\end{eqnarray}
The Appel hypergeometric functions of double argument are less studied in literature but the double infinite sums that define these functions are always convergent when the two algebraic arguments are subunitary. In our case this is translated in the ratio of momenta less than one and we will use this observation in our further analysis. Our amplitude depends on both the ratios between the fermion mass per expansion factor $\frac{m}{\omega}$ and the Z boson mass per expansion factor $\frac{M_Z}{\omega}$ and these are the key parameters that give the amplitude dependence of the space expansion. From our mathematical results given in terms of Appel hypergeometric function we must obtain the physical significance of these amplitudes and how the probabilities change with the expansion parameter.

\begin{figure}[h!t]
\includegraphics[scale=0.35]{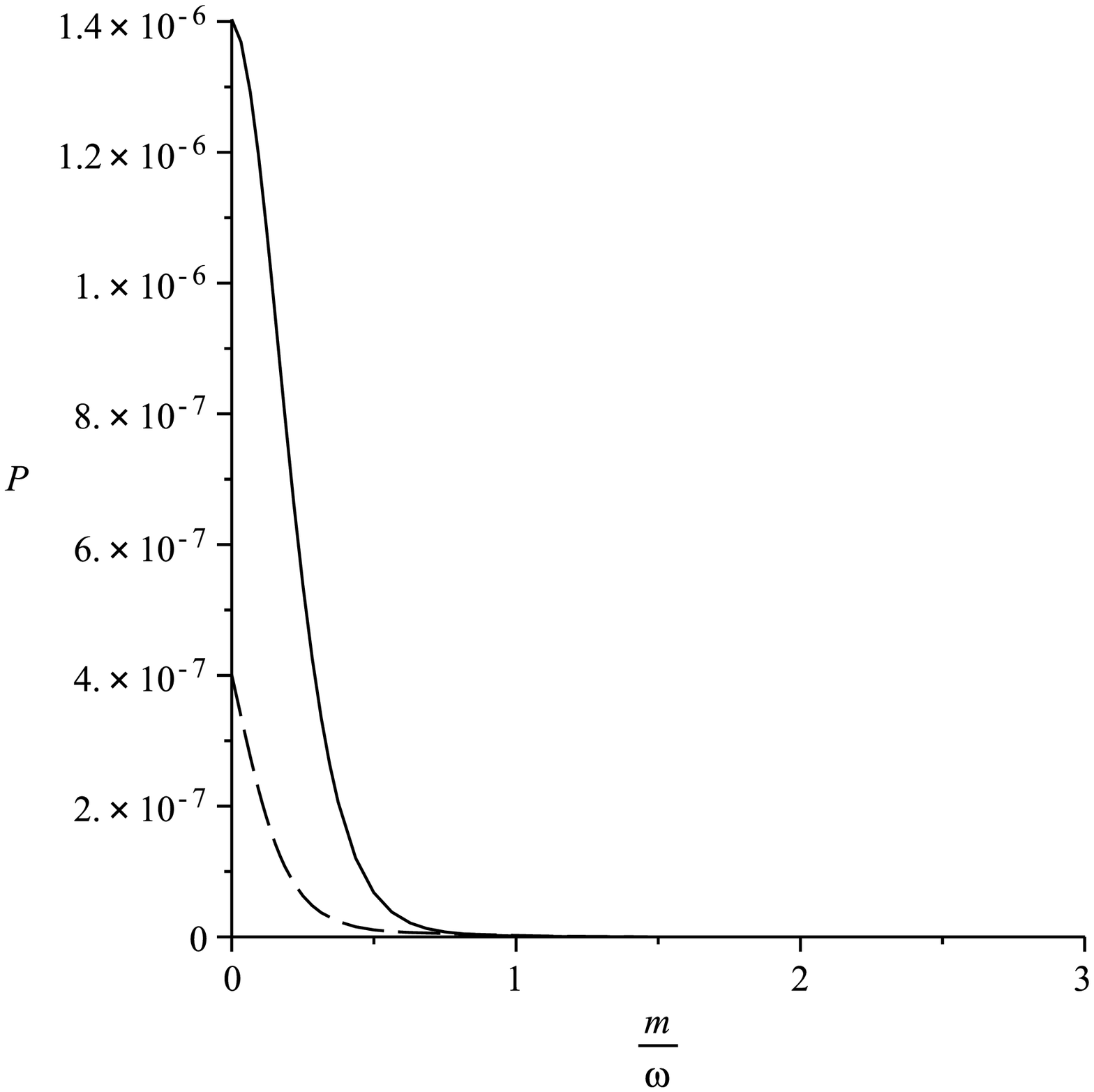}
\includegraphics[scale=0.35]{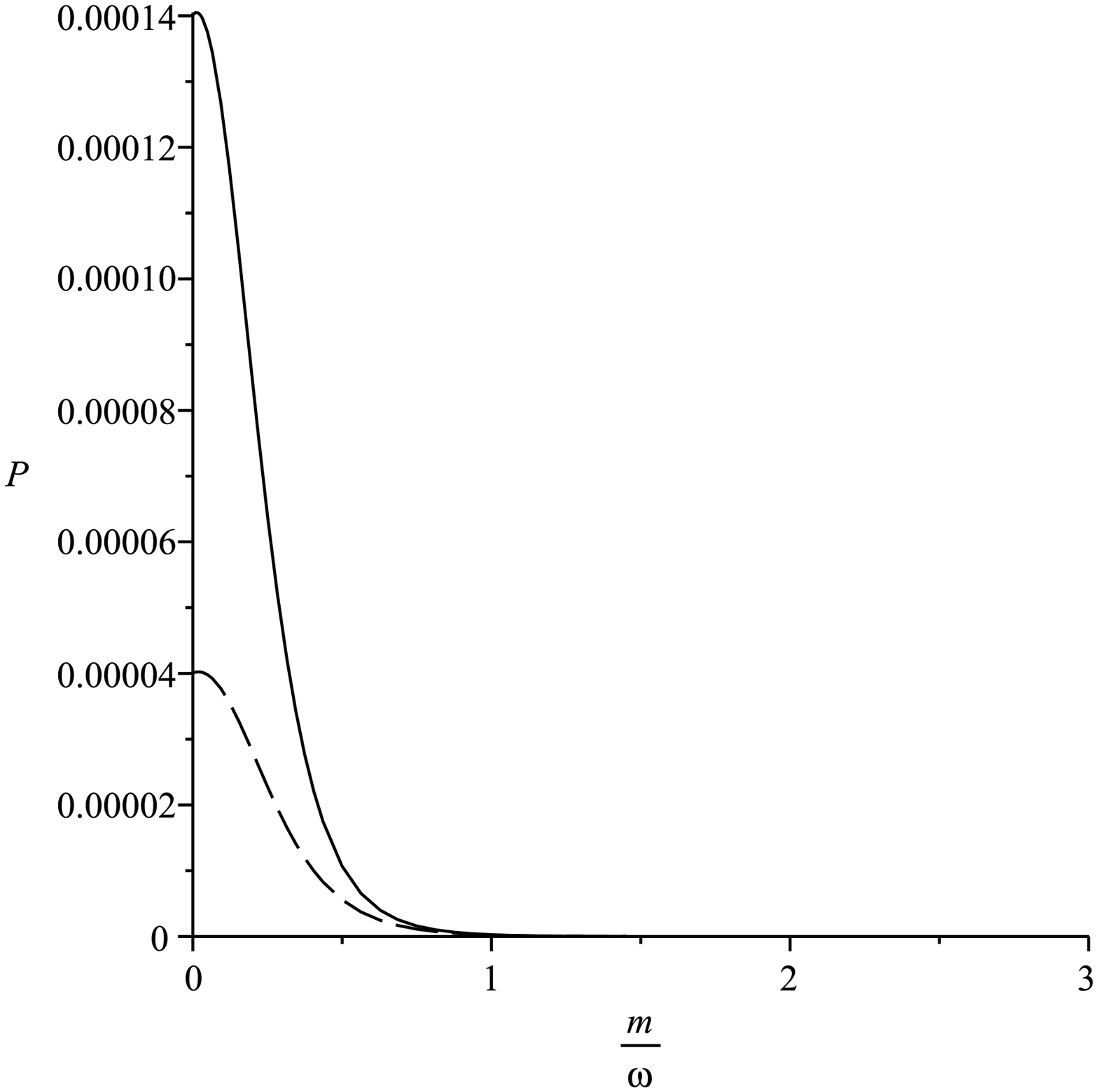}
\caption{Probability as a function of $m/\omega$ for $\lambda=\pm 1$ with $M_Z/\omega = 0.9$, $\sigma$ and $\sigma'$ having the same sign and opposite signs respectively. The dotted line is for $p/P = 0.1, p'/P=0.2$, and the solid line is for $p/P = 0.3, p'/P=0.4$ }
\centering
\label{f1}
\end{figure}

\begin{figure}[h!t]
\includegraphics[scale=0.35]{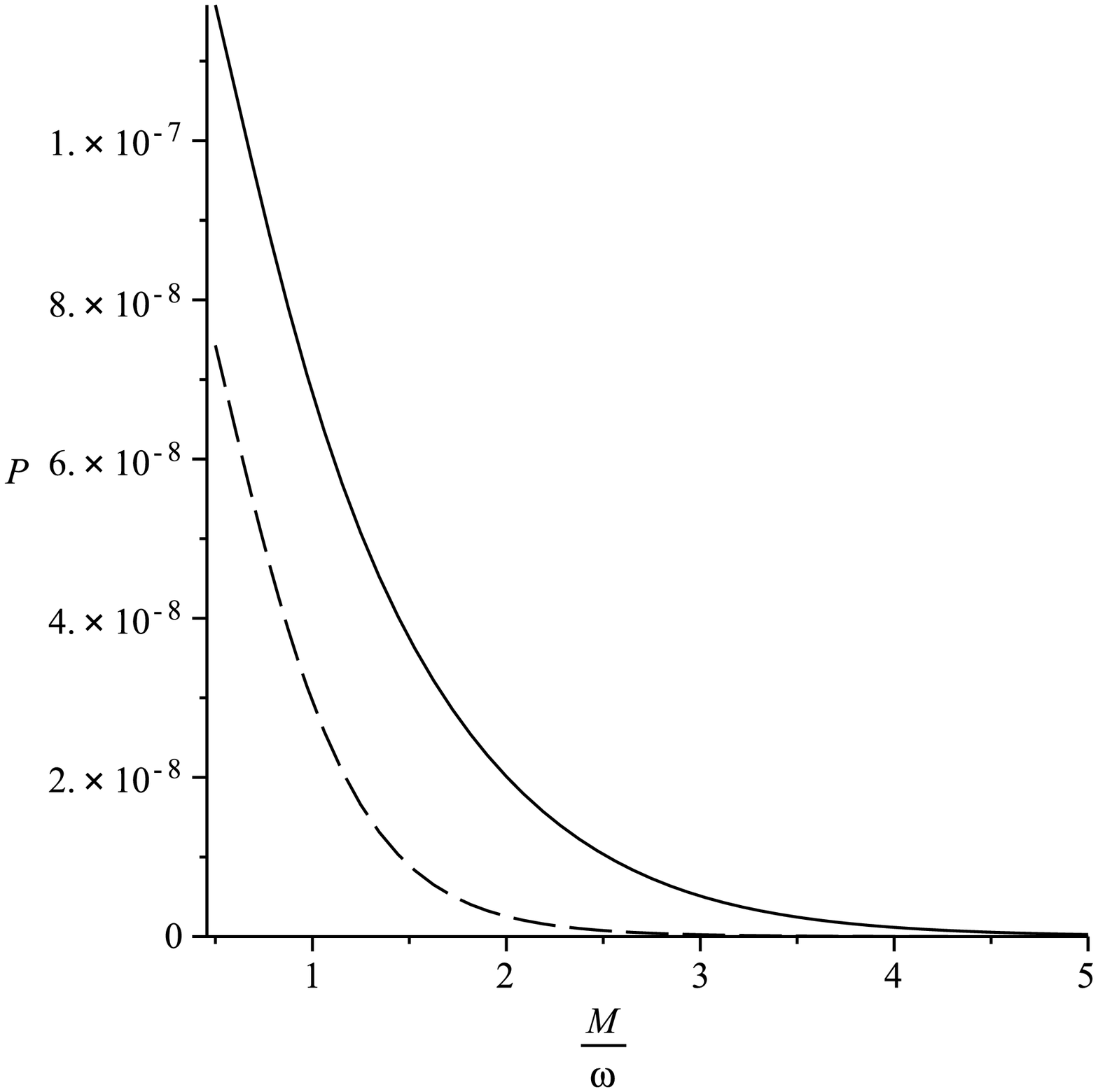}
\includegraphics[scale=0.35]{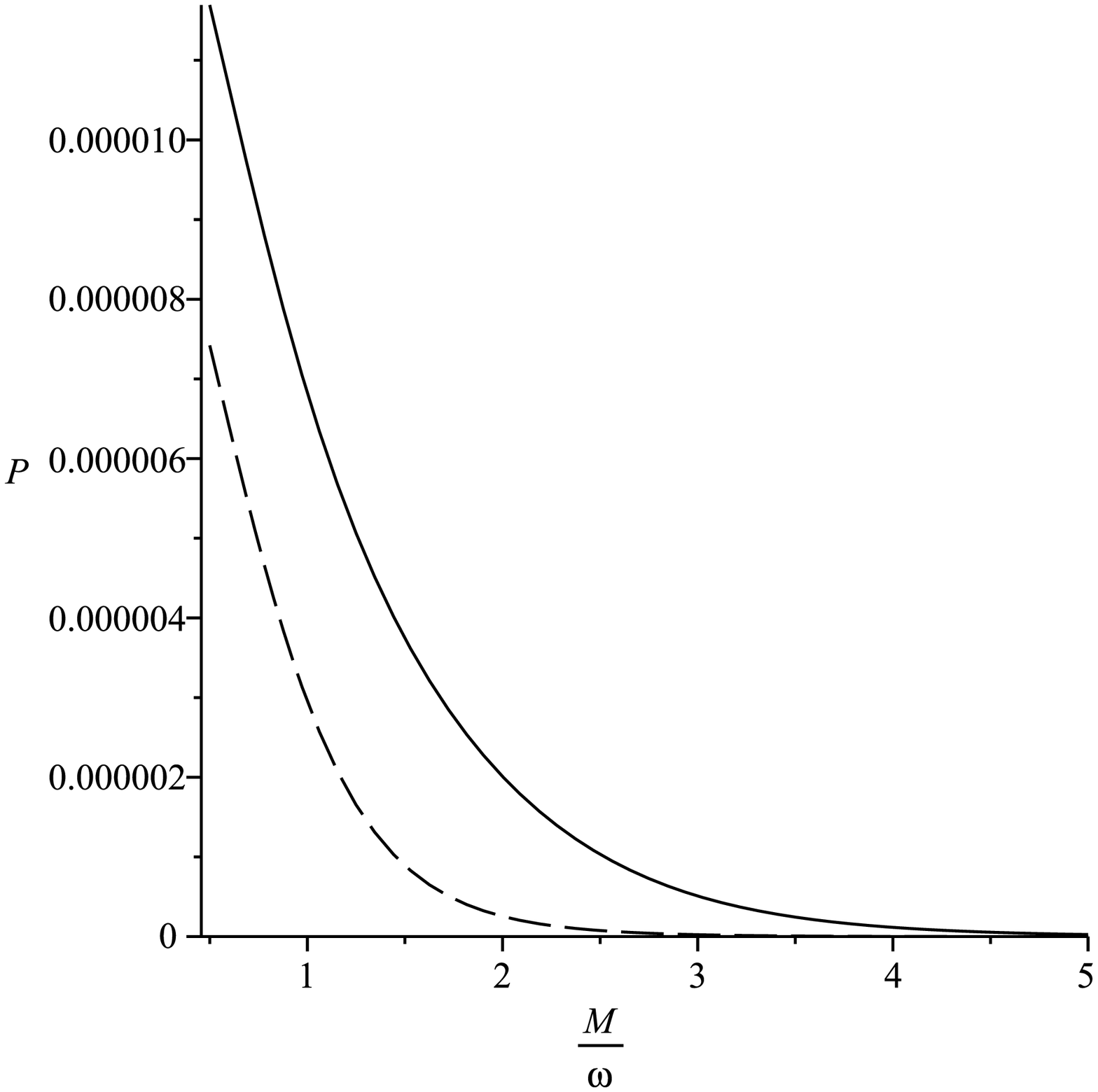}
\caption{Probability as a function of $M_Z/\omega$ for $\lambda=\pm 1$ with $m/\omega = 0.5$, $\sigma$ and $\sigma'$ having the same sign and opposite signs respectively. The dotted line is for $p/P = 0.1, p'/P=0.2$, and the solid line is for $p/P = 0.3, p'/P=0.4$ }
\centering
\label{f2}
\end{figure}

The probability of spontaneous generation of the triplet Z boson and electron-positron pair from vacuum is defined by taking the square modulus of the amplitude and sum after the helicities. Since the amplitude is proportional with the delta Dirac function depending on momentum $|\delta^3(\vec p\,)|^2=V\delta^3(\vec p\,)$ we will define the probability in volume unit \cite{cc}:
\begin{eqnarray}\label{prob}
  P(\lambda = \pm 1) &=& \frac{e^2}{64 \pi^2} \delta^3 (\vec{p}+\vec{p}\,' +\vec{P})\, \frac{1}{8}\sum_{\sigma\sigma'\lambda} \Bigg\{ \bigg(\frac{\cos(2 \theta_W)}{\sin(2 \theta_W)} \bigg)^2 |A_1|^2  + \tan^2 (\theta_W) |A_2|^2   \\
  &&- \bigg(\frac{\cos(2 \theta_W)}{\sin(2 \theta_W)} \bigg) \tan(\theta_W) sgn(\sigma') sgn(\sigma) \big[A_1 \cdot A_2^* + A_1^* \cdot A_2 \big] \Bigg\}\\
  &&\times|\xi_{\sigma}^+(\vec{p}\,)\,\vec{\sigma}\vec{\epsilon}\,^*\,(\vec{n}_{\mathcal{P}},\lambda=\pm1)\eta_{\sigma'}(\vec{p}\,')|^2.
\end{eqnarray}
For a correct analysis we write each Appel hypergeometric function of double arguments with its definition given in Appendix by equation (\ref{a4}), and expand it as an infinite sum \cite{21}. For example the functions from $B_{Kk}(p,p'P)$ could be written as
\begin{eqnarray}
&&F_4 \bigg( \frac{5-4iK+2ik}{4}, \frac{5-4iK-2ik}{4}, \frac{3}{2}-iK, \frac{1}{2}-iK; \Big(\frac{p}{P}\Big)^2, \Big(\frac{p'}{P}\Big)^2 \bigg)=\nonumber\\
&&\sum_{m,n=0}^{\infty}\frac{\Gamma\left(\frac{5-4iK+2ik}{4}+m+n\right)\Gamma\left(\frac{5-4iK-2ik}{4}+m+n\right)\Gamma\left(\frac{3}{2}-iK\right)\Gamma\left(\frac{1}{2}-iK\right)}{\Gamma\left(\frac{5-4iK+2ik}{4}\right)
\Gamma\left(\frac{5-4iK-2ik}{4}\right)\Gamma\left(\frac{3}{2}-iK+m\right)\Gamma\left(\frac{1}{2}-iK+n\right)m!n!}\Big(\frac{p}{P}\Big)^{2m} \Big(\frac{p'}{P}\Big)^{2n}
\nonumber\\
\end{eqnarray}
The above series is convergent for fixed values of the ratio $\frac{m}{\omega}\,,\frac{M_Z}{\omega}$ when the momenta ratio are less than one i.e. $\frac{p}{P}<1,\,\frac{p'}{P}<1$.
Our numerical calculations prove that the series converges very rapidly and keeps constant values up to infinity. For a complete analysis we will plot the probability in terms of the parameters $\frac{m}{\omega}\,,\frac{M_Z}{\omega}$ for fixed ratios between the momenta $\frac{p}{P}\,,\frac{p'}{P}$ and we mention that in our plots we use the above expansions for the Appel functions taking the sums from zero to infinity. Thus we recover the exact variation of the probability with the expansion parameter as follows in Figs.(\ref{f1}),(\ref{f2}).

Our graphs Figs.(\ref{f1}),(\ref{f2}) prove that the probability of spontaneous generation from vacuum of the triplet Z boson and electron-positron pair is nonvanishing only when the ratios between the masses and the expansion parameter have small values and is vanishing when these ratios become bigger. For $\frac{m}{\omega}\rightarrow \infty\,,\frac{M_Z}{\omega}\rightarrow \infty$ the probabilities vanish and we recover the Minkowski limit where this process is forbidden as a perturbative process by the energy conservation law \cite{12,19,20}.

\begin{figure}[h!t]
\includegraphics[scale=0.35]{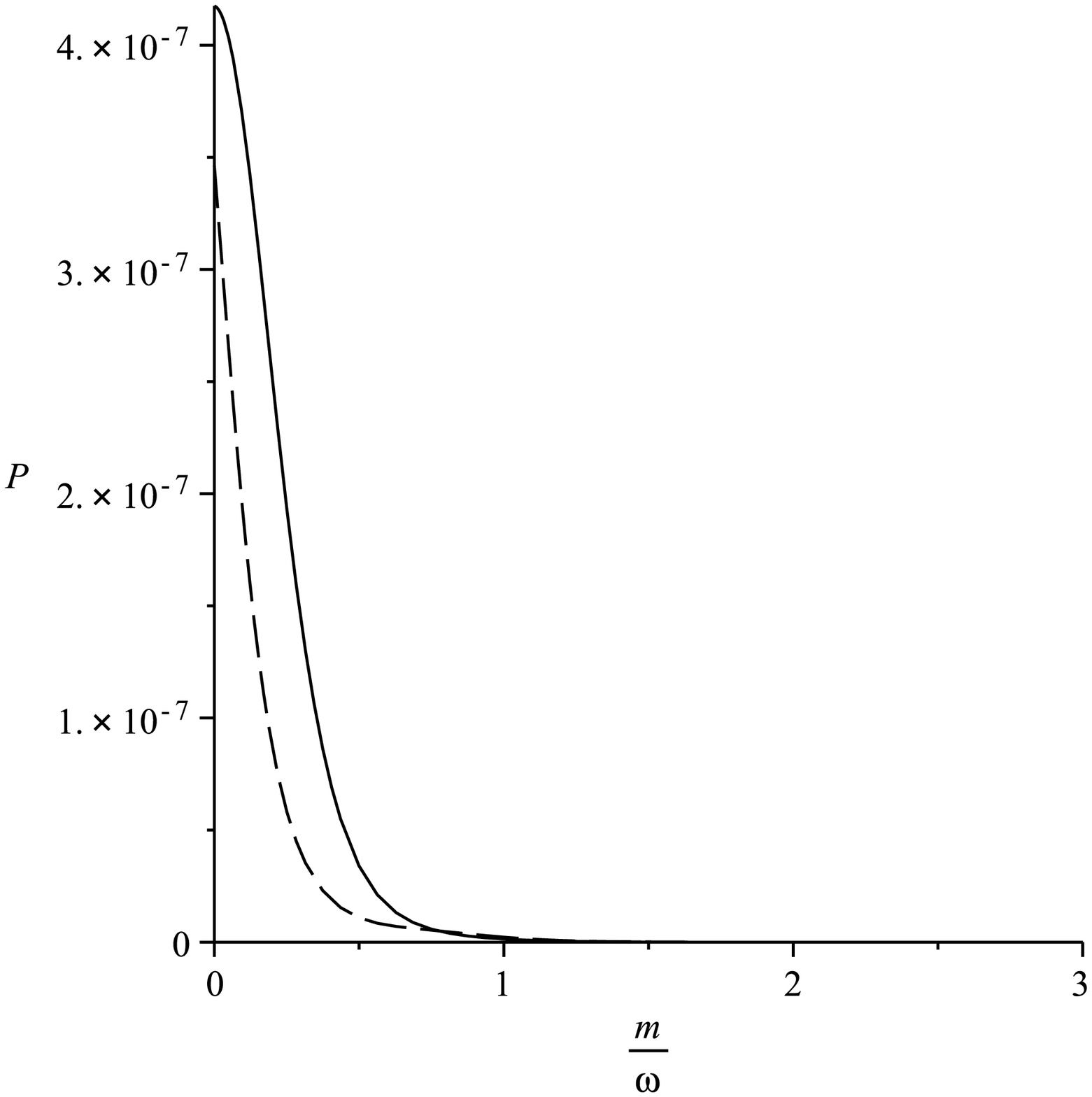}
\includegraphics[scale=0.35]{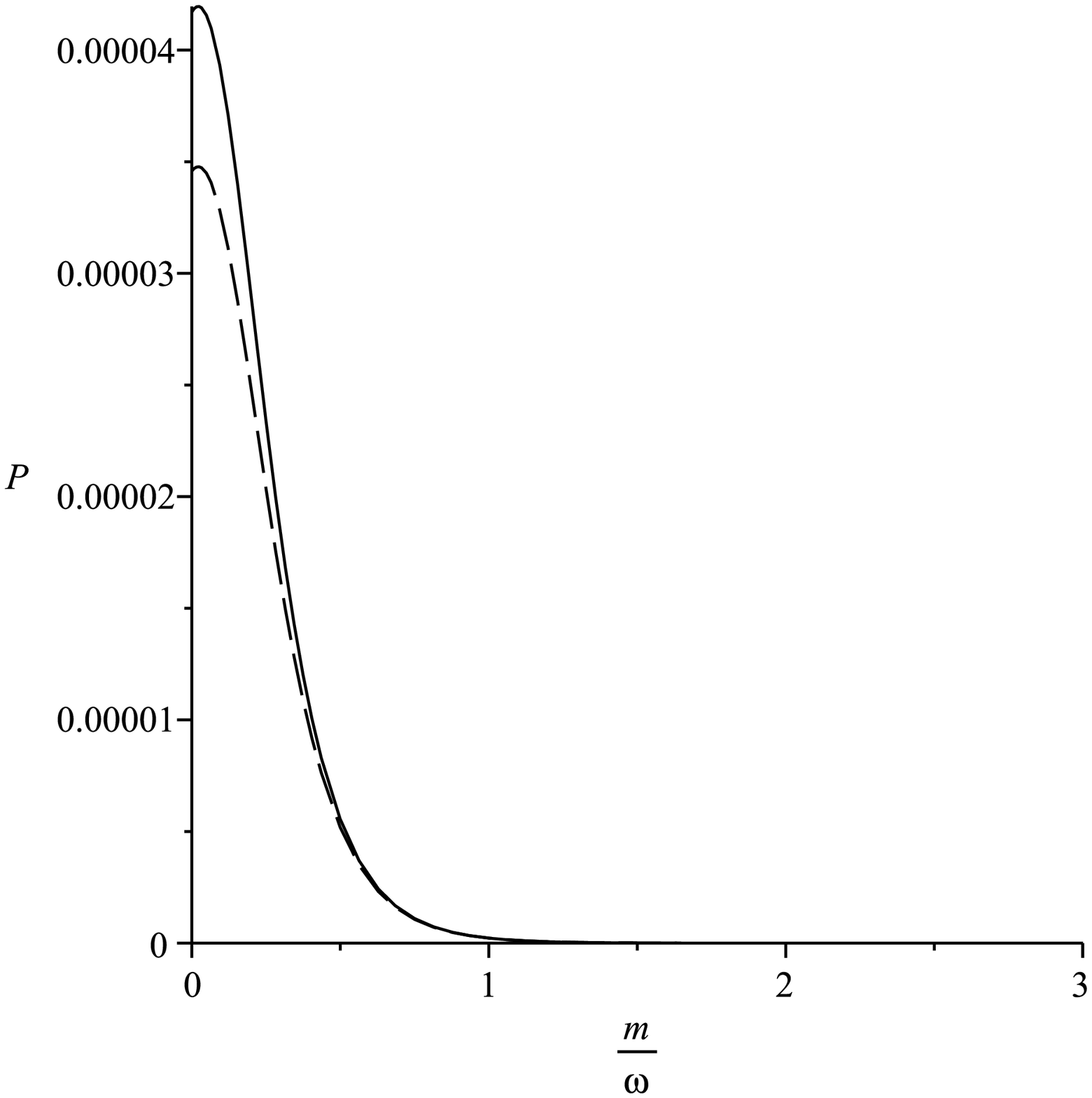}
\caption{Probability as a function of $m/\omega$ for $\lambda=\pm1 $, disregarding the $F_4$ functions, with $M_Z/\omega = 0.9$, $\sigma$ and $\sigma'$ having the same sign and opposite signs respectively. The dotted line is for $p/P = 0.1, p'/P=0.2$, and the solid line is for $p/P = 0.3, p'/P=0.4$ }
\centering
\label{f11}
\end{figure}

\begin{figure}[h!t]
\includegraphics[scale=0.35]{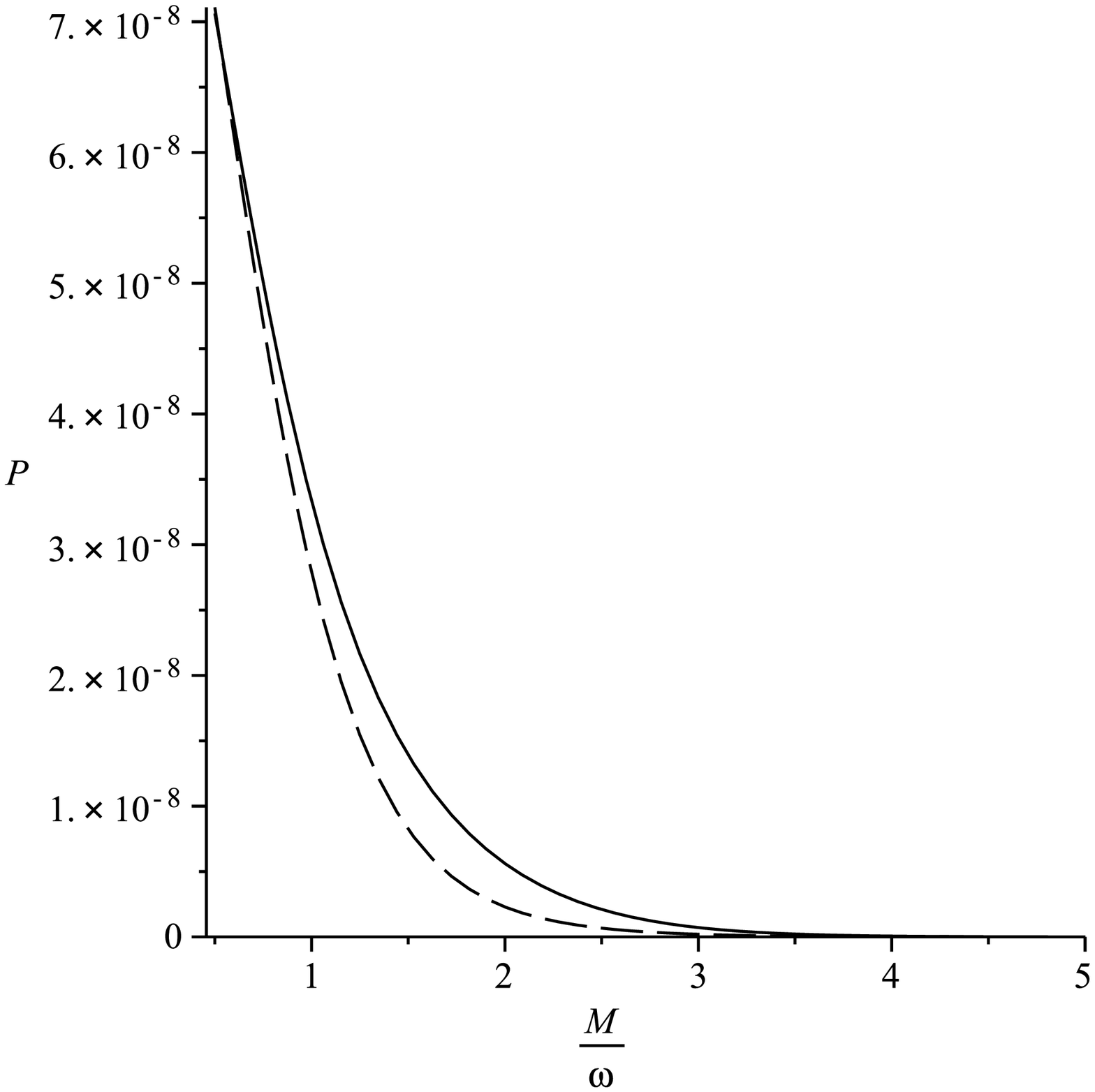}
\includegraphics[scale=0.35]{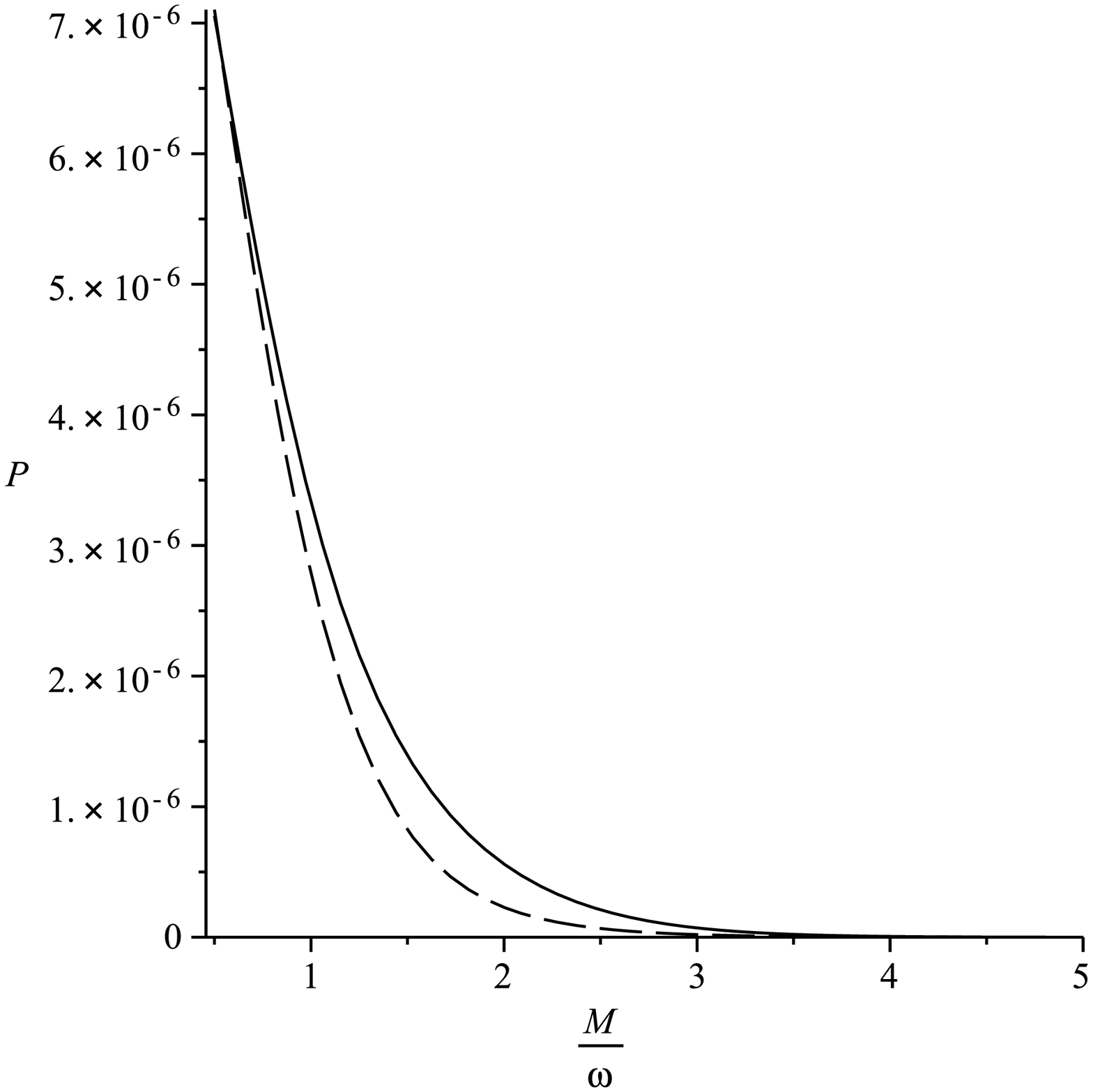}
\caption{Probability as a function of $M_Z/\omega$ for $\lambda=\pm1$, disregarding the $F_4$ functions, with $m/\omega = 0.5$, $\sigma$ and $\sigma'$ having the same sign and opposite signs respectively. The dotted line is for $p/P = 0.1, p'/P=0.2$, and the solid line is for $p/P = 0.3, p'/P=0.4$ }
\centering
\label{f22}
\end{figure}

Another issue regarding our study is related to the computation of the total probability obtained after solving the integrals after the final momenta in equation (\ref{prob}).
The functions that define the probabilities are written in terms of complicated Appel hypergeometric functions that have as algebraic argument ratios of momenta and for this reason it will be of interest to find an approximation which can replace our functions $B$ with a more suitable expression that allows us to obtain an analytical result.
Let us analyse the result for the probability. First a numerical and graphical analysis proves that the graphs of the probability in terms of parameters $K,k$ are preserved if we neglect the contributions of the Appel hypergeometric functions. This can be verified by an analytical and numerical study and for large values of the parameters $K,k$ the graphs for the probability remains the same, while for small values there are small variations of the probabilities with the specification that the profile of the curves are preserved as in Figs.(\ref{f11}),(\ref{f22}). In this situation it is useful to work with the simplified $B$ functions that can describe the behaviour of our amplitudes and probabilities in terms of any values of parameters $m/\omega,\,M_Z/\omega$ as the functions defined in equation (\ref{aa3}):
\begin{equation}
  B_{Kk} (pp'P) = \frac{(pp')^{-iK} P^{-\frac{5}{2}+2iK} e^{\pi K}}{\Gamma \big( \frac{1}{2} - iK \big) \Gamma \big( \frac{3}{2} - iK \big) } \Gamma \bigg(\frac{5-4iK+2ik}{4} \bigg) \Gamma \bigg(\frac{5-4iK-2ik}{4} \bigg)
\end{equation}

\begin{equation}
   B_{1Kk} (pp'P) = \frac{(p)^{-iK} (p')^{iK} P^{-3/2}}{\Gamma \big(\frac{1}{2}-iK \big) \Gamma \big( \frac{1}{2} +iK\big)} \Gamma \bigg(\frac{3+2ik}{4} \bigg) \Gamma \bigg(\frac{3-2ik}{4} \bigg)
\end{equation}

\begin{equation}
  B_{2Kk} (pp'P) = \frac{(p)^{1+iK} (p')^{1-iK} P^{-7/2}}{\Gamma \big(\frac{3}{2}+iK \big) \Gamma \big( \frac{3}{2} - iK\big)} \Gamma \bigg(\frac{7+2ik}{4} \bigg) \Gamma \bigg(\frac{7-2ik}{4} \bigg)
\end{equation}

\begin{equation}
  B_{-Kk} (pp'P) = \frac{(pp')^{iK} P^{-\frac{5}{2}-2iK} e^{-\pi K}}{\Gamma \big( \frac{1}{2} + iK \big) \Gamma \big( \frac{3}{2} + iK \big) } \Gamma \bigg(\frac{5+4iK-2ik}{4} \bigg) \Gamma \bigg(\frac{5+4iK+2ik}{4} \bigg)
\end{equation}
We specify that we present the above simplified functions for a future study related to the computation of total probability in the general case, which is a quantity dependent on the parameters $m/\omega,\,M_Z/\omega$. Even with the simplified functions defined above the calculation of the total probability is a complicated task since the resulted integrals will contain momenta at imaginary powers and this leads to integrals that are not well defined. For this reason it will be of interest to discuss the situation when the expansion parameter is much larger than the particle masses. The above $B$ functions become very simple in the limit where the expansion parameter is much larger than the particle mass $\omega>>m,M_Z$ and this represents the limit of large expansion that is interesting for computing the probabilities and density number of particles.

\,
\,
\,
\,
\,

Let us comment on the results related to helicity. In the case when the helicity of Z boson is $\lambda=1$ then for $\sigma=\sigma'=-1/2$ the helicity is conserved in the process and for the rest of the combinations of $\sigma,\sigma'$ the helicity conservation law is broken. The helicity not being conserved is due to the fact that the particles have mass like in Minkowski theory. Our graphical results prove that the processes which do not conserve helicity have higher probability to happen and these results are presented in the above right figures for $\sigma$ and $\sigma'$ having opposite signs.

\subsection{Minkowski limit}
In this section we will approach the Minkowski limit of our result for the amplitude and probability. This is the limit when the expansion parameter vanishes $\omega=0$, or in our notations the ratios between the masses of particles and the expansion parameter become infinite i.e $K=\frac{m}{\omega}\rightarrow \infty$,\,$\frac{M_Z}{\omega}\rightarrow \infty$. First the functions $B_{Kk} (pp'P),B_{1Kk} (pp'P),B_{2Kk} (pp'P)$ that define the amplitude could be approximated for large values of the parameters $K=\frac{m}{\omega} $,\,$\frac{M_Z}{\omega} $. In all hypergeometric Appel functions and gamma Euler functions we replace $ik=i\frac{M_Z}{\omega}$ for $\frac{M_Z}{\omega}>>1$, and in addition we use the Stirling formula for approximating the gamma Euler functions for large arguments
\begin{equation}
\Gamma(z)\simeq e^{-z}z^z\left(\frac{2\pi}{z}\right)^{1/2}.
\end{equation}
The behaviour of these functions for large $\frac{m}{\omega}>>1 $,\,$\frac{M_Z}{\omega}>>1$ will be mainly determined by the factors in front of the Appel functions and we obtain:
\begin{eqnarray}
  &&B_{K>>1k>>1} (pp'P) = \frac{(pp')^{-iK} P^{-\frac{5}{2} +2iK} e^{-(\frac{5-4iK}{2})} (1-i) \frac{\sqrt{2}}{2} }{e^{2\pi K} e^{-2+2iK} \big(\frac{1}{2}-iK \big)^{-iK} \big(\frac{3}{2}-iK\big)^{1-iK} \frac{1}{4} \sqrt{25-16K^2 - 40iK+4\frac{M^2}{\omega^2}}}  \nonumber\\
  &&\times \bigg(\frac{5-4iK+2i\frac{M}{\omega}}{4} \bigg)^{\frac{5-4iK+2i\frac{M}{\omega}}{4}} \bigg(\frac{5-4iK-2i\frac{M}{\omega}}{4} \bigg)^{\frac{5-4iK-2i\frac{M}{\omega}}{4}}\nonumber\\
  &&\times F_4 \bigg( \frac{5-4iK+2i\frac{M_Z}{\omega}}{4}, \frac{5-4iK-2i\frac{M_Z}{\omega}}{4}, \frac{3}{2}-iK, \frac{1}{2}-iK; \Big(\frac{p}{P}\Big)^2, \Big(\frac{p'}{P}\Big)^2 \bigg)
\end{eqnarray}

\begin{eqnarray}
  B_{1K>>1k>>1} (pp'P) &=& \frac{- \big(\frac{p'}{p} \big)^{iK} P^{-3/2} e^{-(\frac{3}{2})} (1+i) \sqrt{2} }{ e^{\pi K} \frac{1}{2}\sqrt{\left(\frac{M_Z}{\omega}\right)^2+\frac{9}{4}}} \bigg(\frac{3+2i\frac{M_Z}{\omega}}{4} \bigg)^{\frac{3+2i\frac{M_Z}{\omega}}{4}} \bigg(\frac{3-2i\frac{M_Z}{\omega}}{4} \bigg)^{\frac{3-2i\frac{M_Z}{\omega}}{4}}\nonumber\\
   &&\times F_4 \bigg( \frac{3+2i\frac{M_Z}{\omega}}{4}, \frac{3-2i\frac{M_Z}{\omega}}{4}, \frac{1}{2}-iK, \frac{1}{2}+iK; \Big(\frac{p}{P}\Big)^2, \Big(\frac{p'}{P}\Big)^2 \bigg)
\end{eqnarray}

\begin{eqnarray}
  B_{2K>>1k>>1} (pp'P) &=& \frac{pp'\big(\frac{p}{p'} \big)^{iK} P^{-7/2} e^{-(\frac{7}{2})}(1+i) \sqrt{2} }{ e^{\pi K} \frac{1}{4} \sqrt{\left(\frac{M_Z}{\omega}\right)^2 + \frac{49}{4}} } \bigg(\frac{7+2i\frac{M_Z}{\omega}}{4} \bigg)^{\frac{7+2i\frac{M_Z}{\omega}}{4}} \bigg(\frac{7-2i\frac{M_Z}{\omega}}{4} \bigg)^{\frac{7-2i\frac{M_Z}{\omega}}{4}}\nonumber\\
  &&\times F_4 \bigg( \frac{7+2i\frac{M_Z}{\omega}}{4}, \frac{7-2i\frac{M_Z}{\omega}}{4}, \frac{3}{2}+iK, \frac{3}{2}-iK; \Big(\frac{p}{P}\Big)^2, \Big(\frac{p'}{P}\Big)^2 \bigg)
\end{eqnarray}
From the above equations one can observe that in this limit the $B$ functions vanish as $e^{-\pi m/\omega}$ and $,\frac{1}{M_Z/\omega}$ multiplied by factors at imaginary powers, while the probabilities vanish as $e^{-2\pi m/\omega}$ and $\frac{1}{(M_Z/\omega)^2}$.

A graphical analysis with the above approximated functions proves that both the real and imaginary parts are very convergent in terms of parameters $m/\omega,\,M_Z/\omega$. We specify that in our graphs we take one of the ratios between the mass of the particle and the expansion factor to be greater than one, with the observation that the behaviour remains the same for subunitary values of $m/\omega,\,M_Z/\omega$. In the Minkowski limit the amplitude and probability are vanishing, and we recover the well known fact that in flat space-time the spontaneous particle generation from vacuum is forbidden by the energy-momentum conservation.

\begin{figure}[h!t]
\includegraphics[scale=0.35]{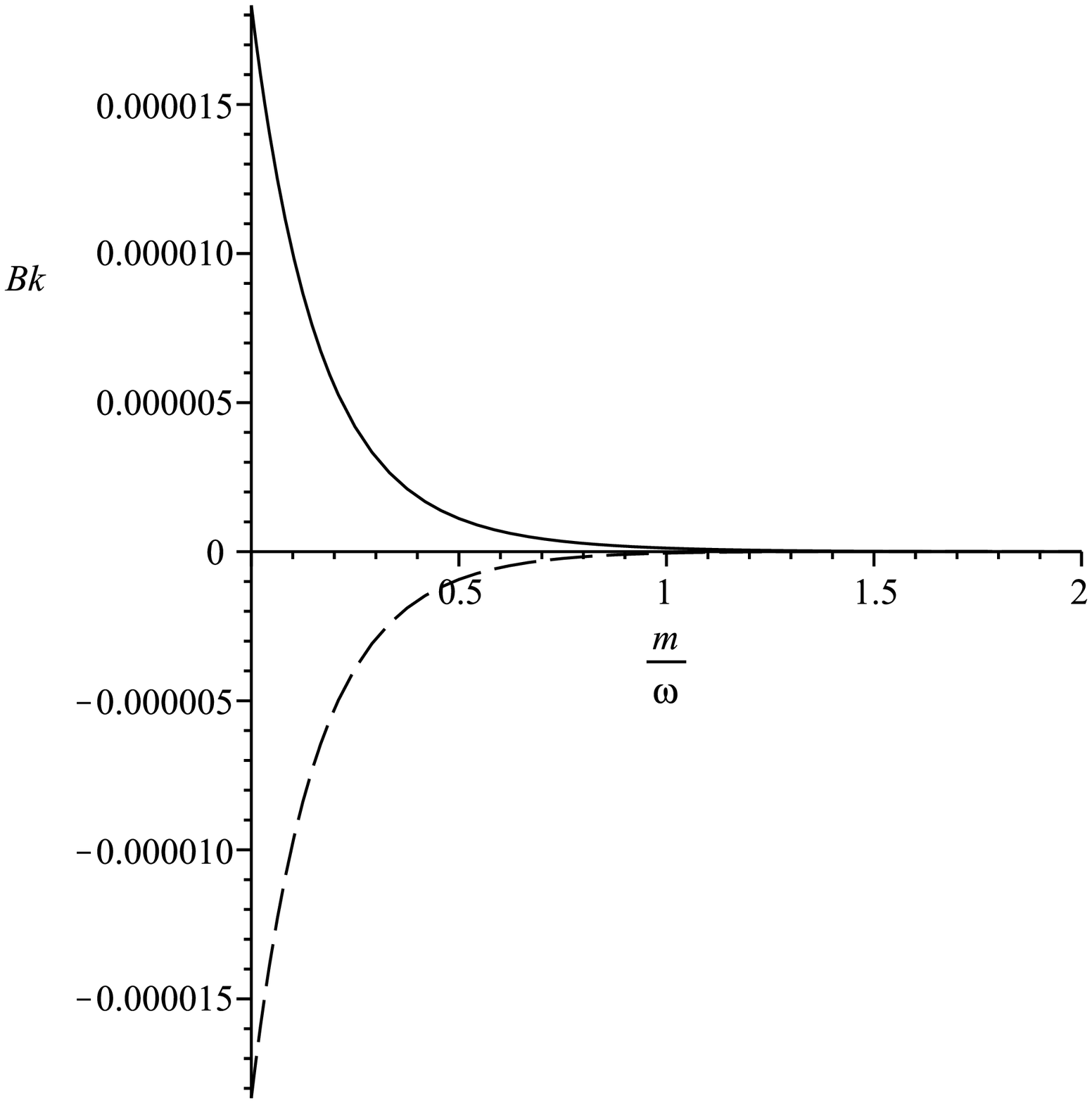}
\includegraphics[scale=0.35]{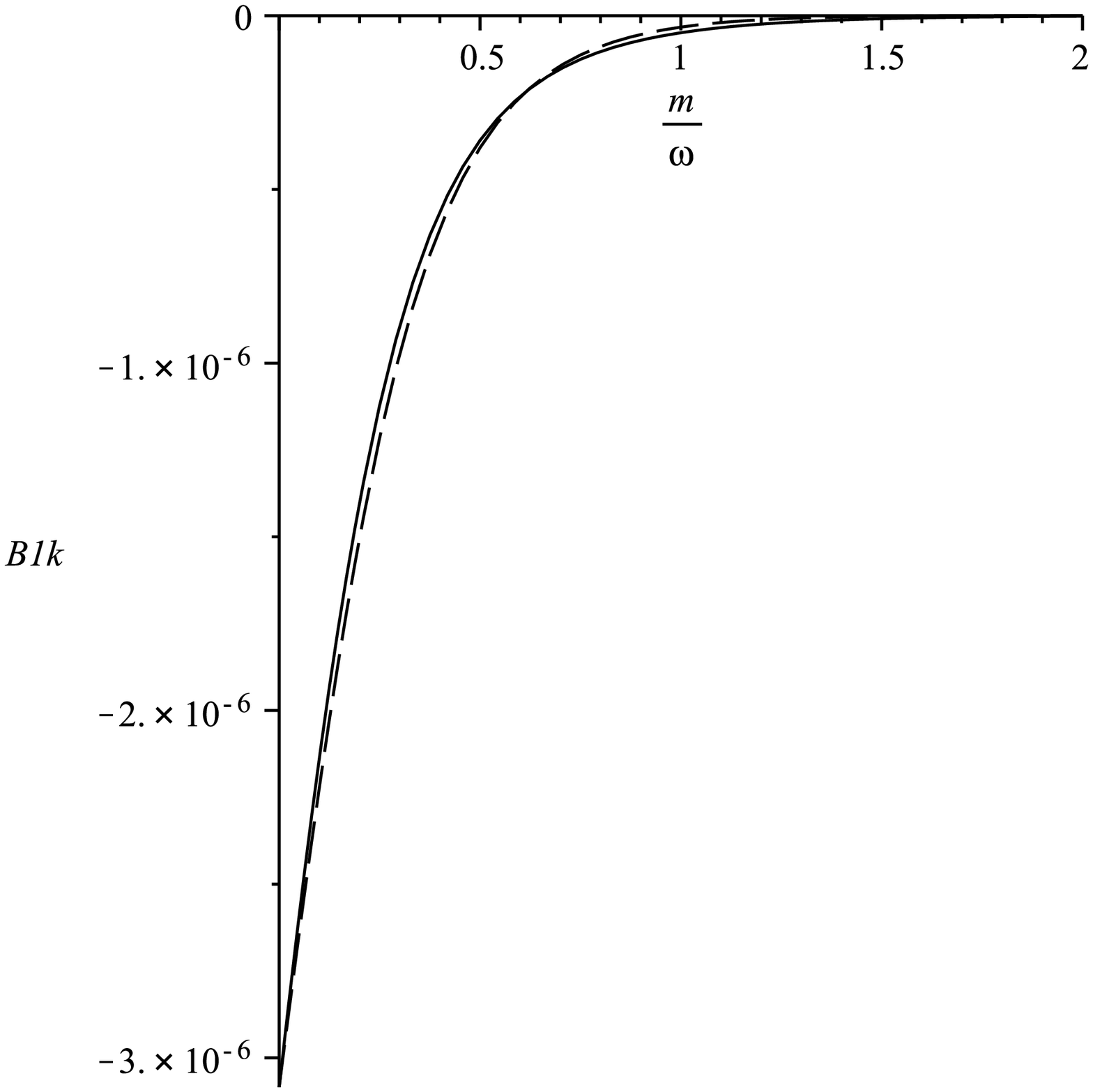}
\caption{The real and imaginary parts of the $B_{Kk}, B_{1Kk}$ as functions of $m/\omega$ with $M_Z/\omega = 10$, $p = 0.1, p'=0.2, P=1$. The dotted lines represent the imaginary parts, and the solid lines the real parts.}
\centering
\end{figure}
\begin{figure}[h!t]
\includegraphics[scale=0.35]{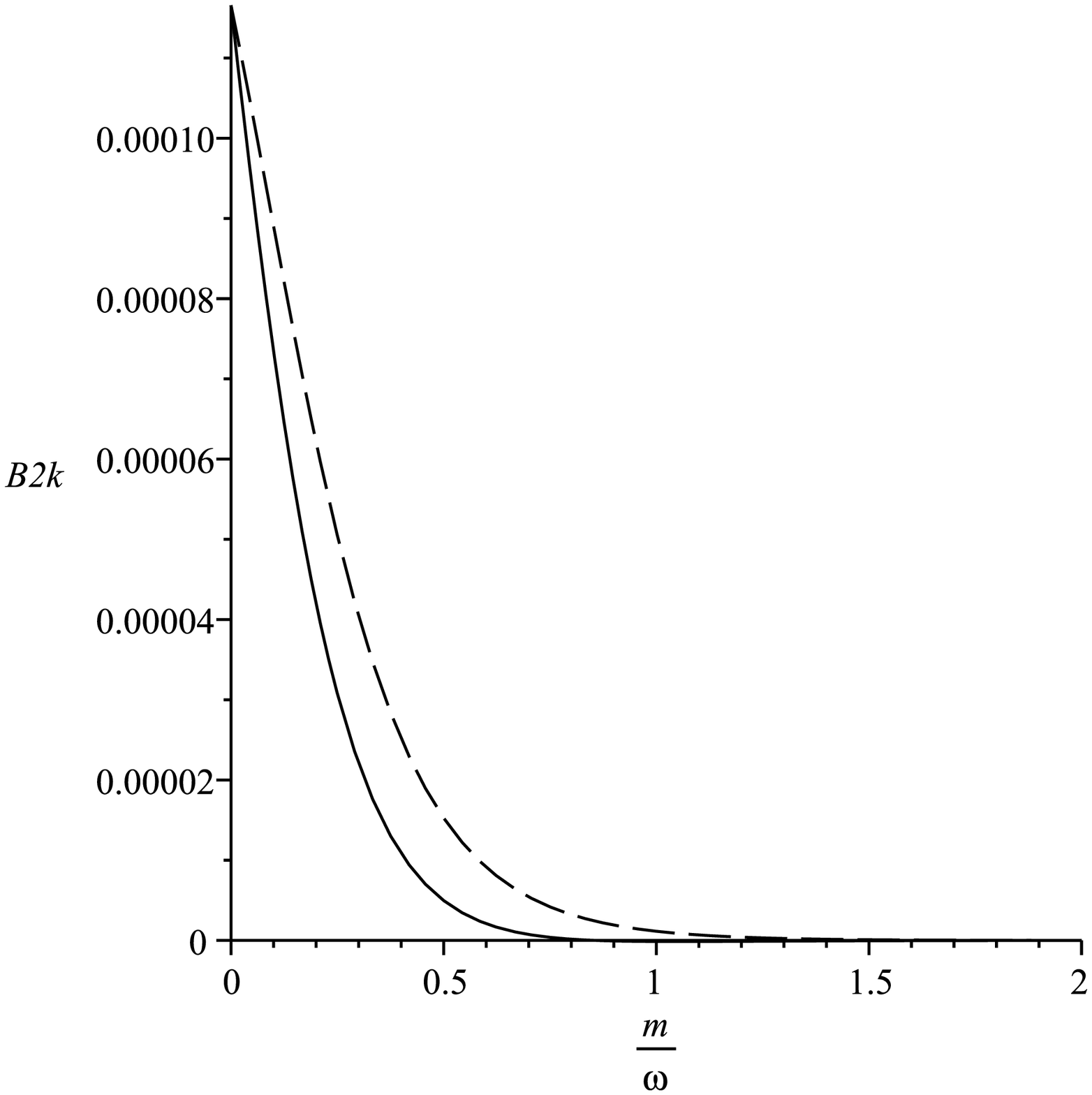}
\includegraphics[scale=0.35]{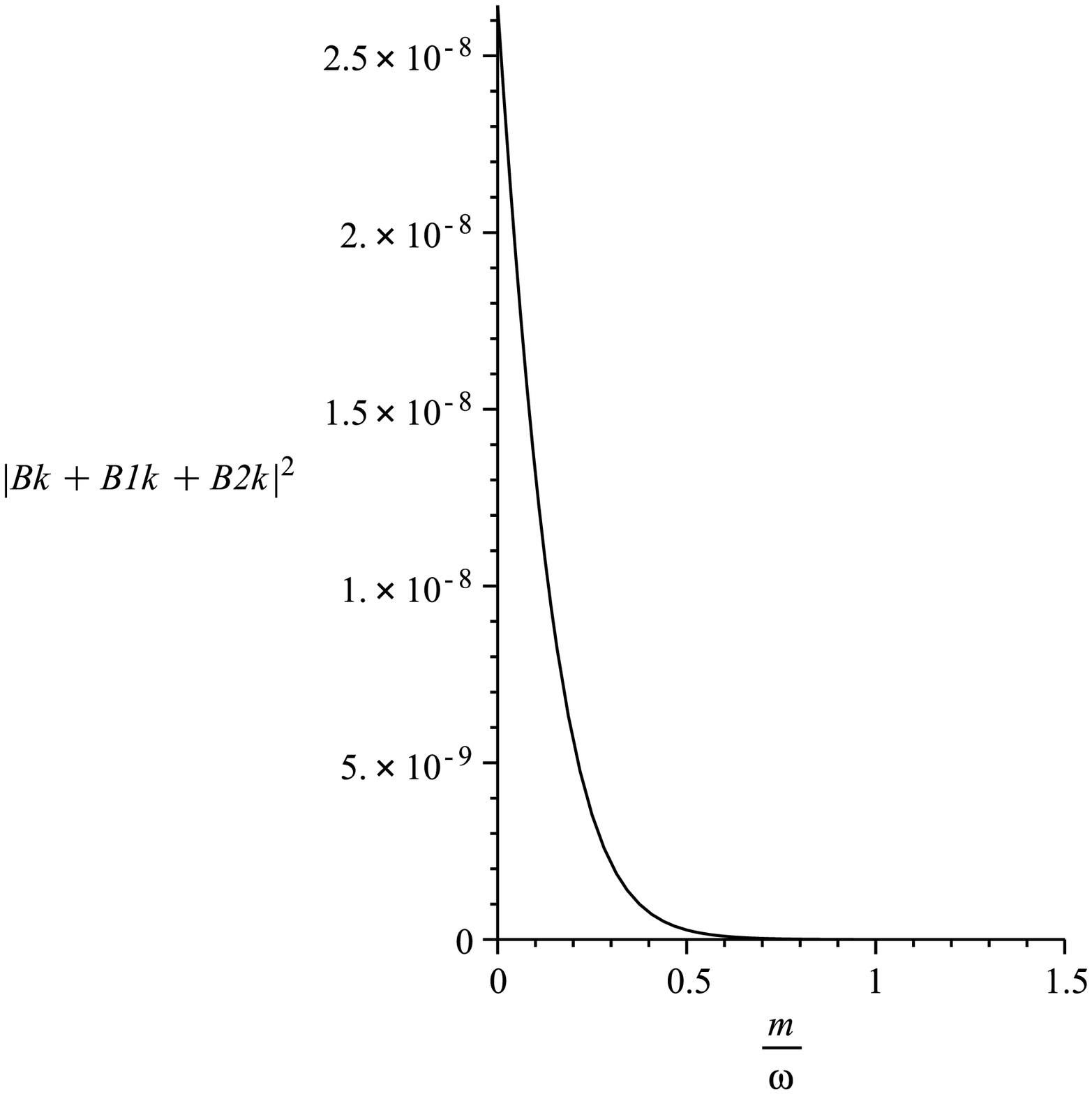}
\caption{The real and imaginary parts of the $B_{2Kk}$ functions, and the square modulus of the sum $B_{Kk}+B_{1Kk}+B_{2Kk}$ as functions of $m/\omega$ with $M_Z/\omega = 10$, $p = 0.1, p'=0.2, P=1$. The dotted lines represent the imaginary parts, and the solid lines the real parts.}
\centering
\end{figure}

\begin{figure}[h!t]
\includegraphics[scale=0.35]{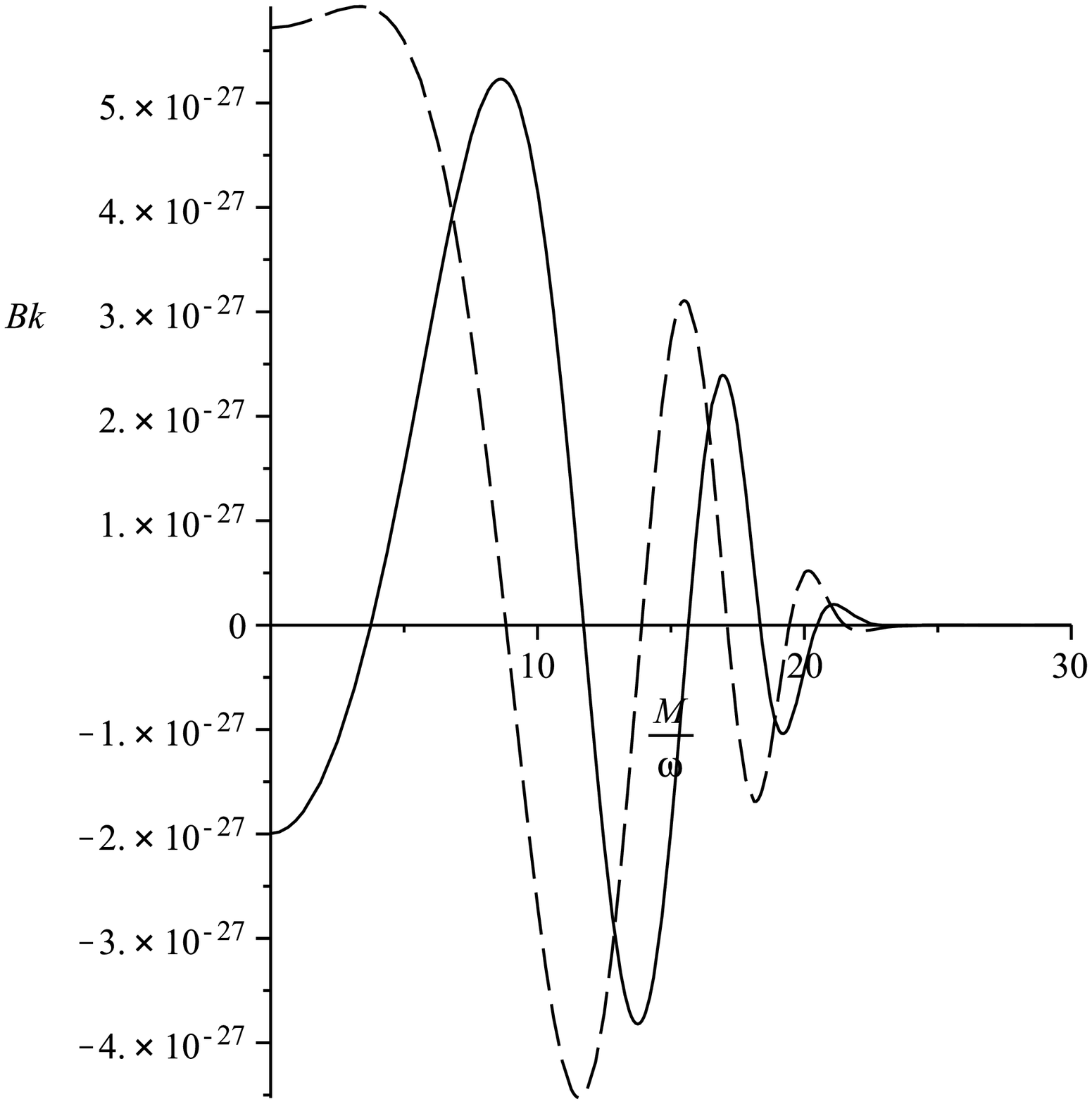}
\includegraphics[scale=0.35]{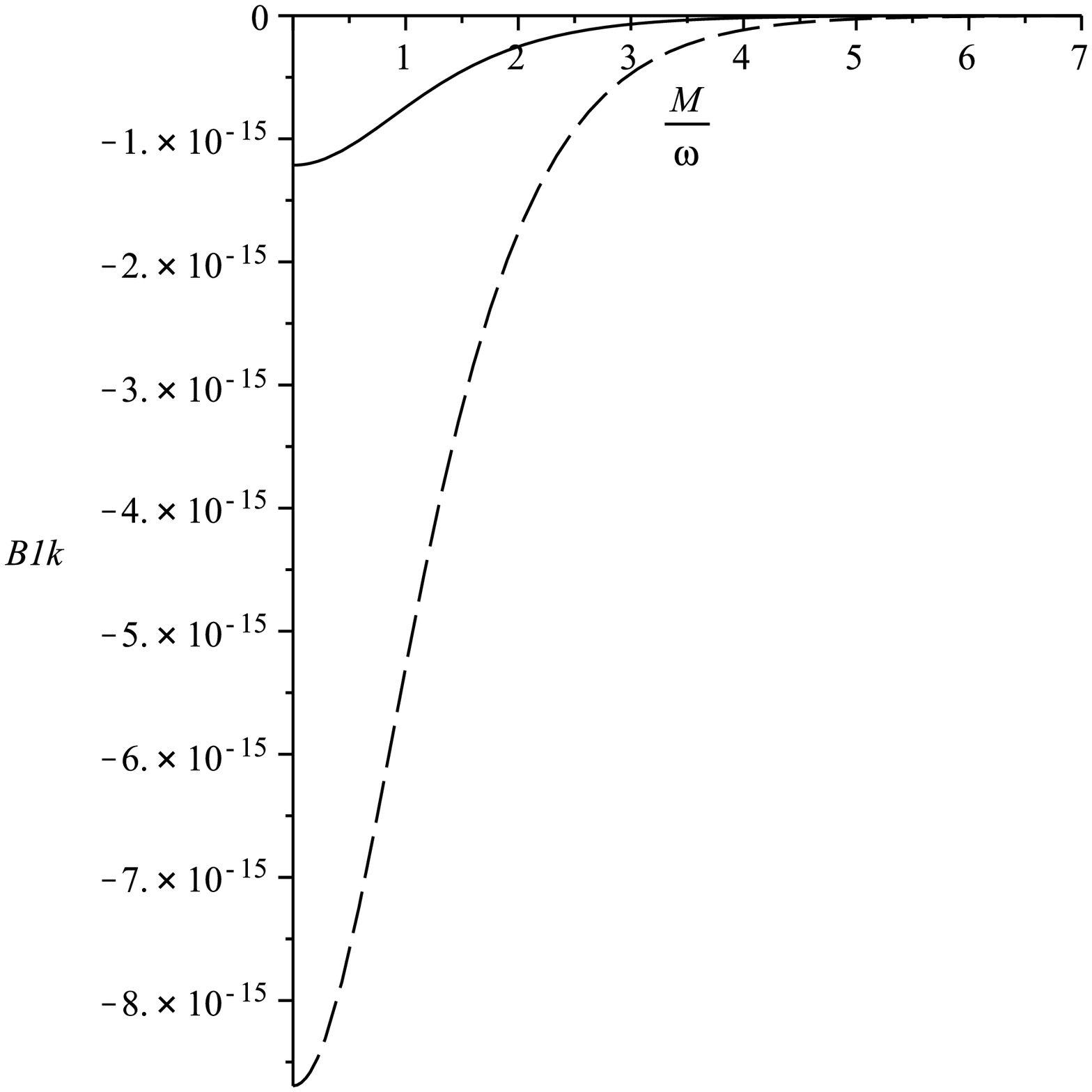}
\caption{The real and imaginary parts of the $B_{Kk}, B_{1Kk}$ as functions of $M_Z/\omega$ with $m/\omega = 10$, $p = 0.1, p'=0.2, P=1$. The dotted lines represent the imaginary parts, and the solid lines the real parts.}
\centering
\end{figure}
\begin{figure}[h!t]
\includegraphics[scale=0.35]{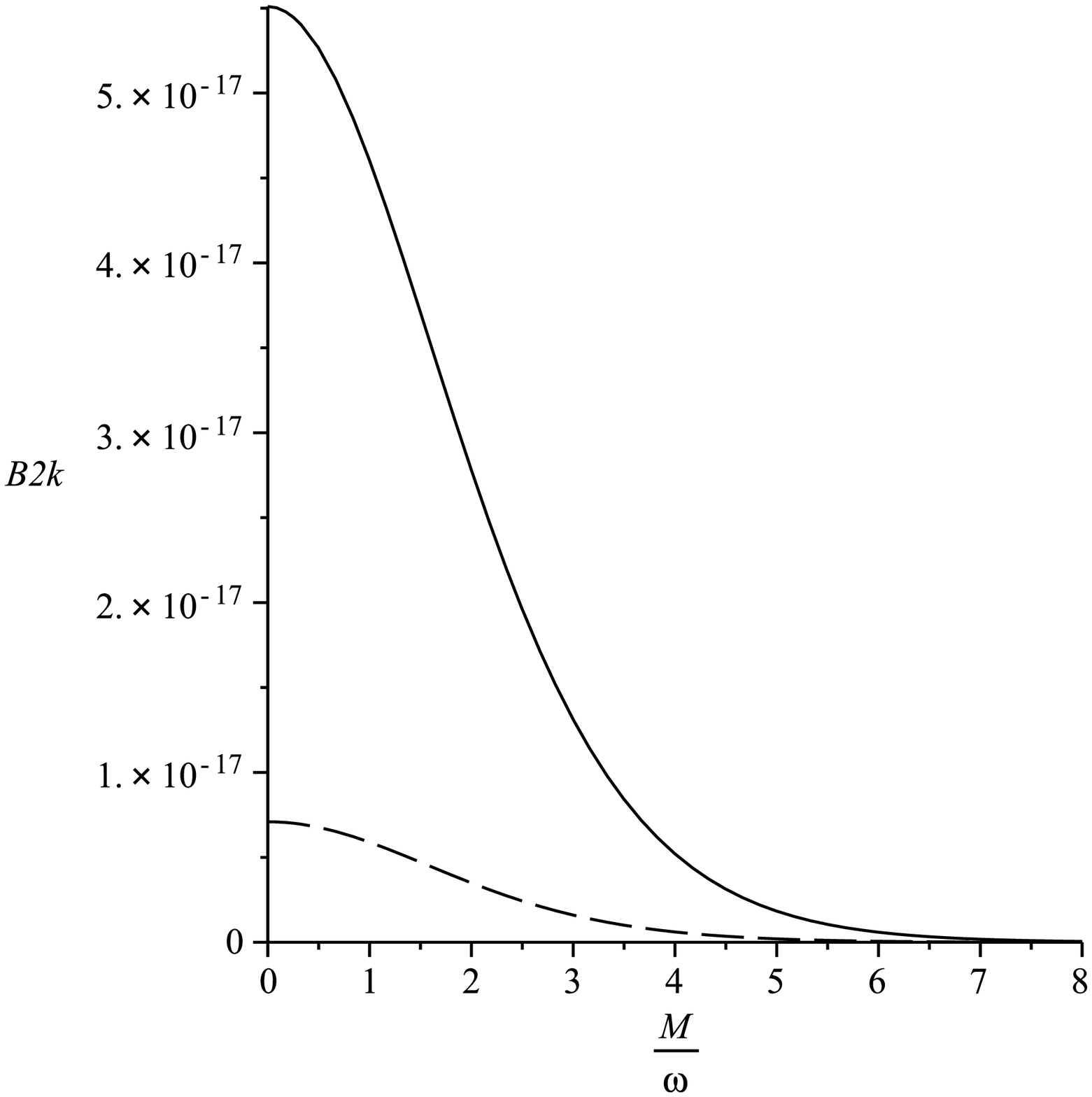}
\includegraphics[scale=0.35]{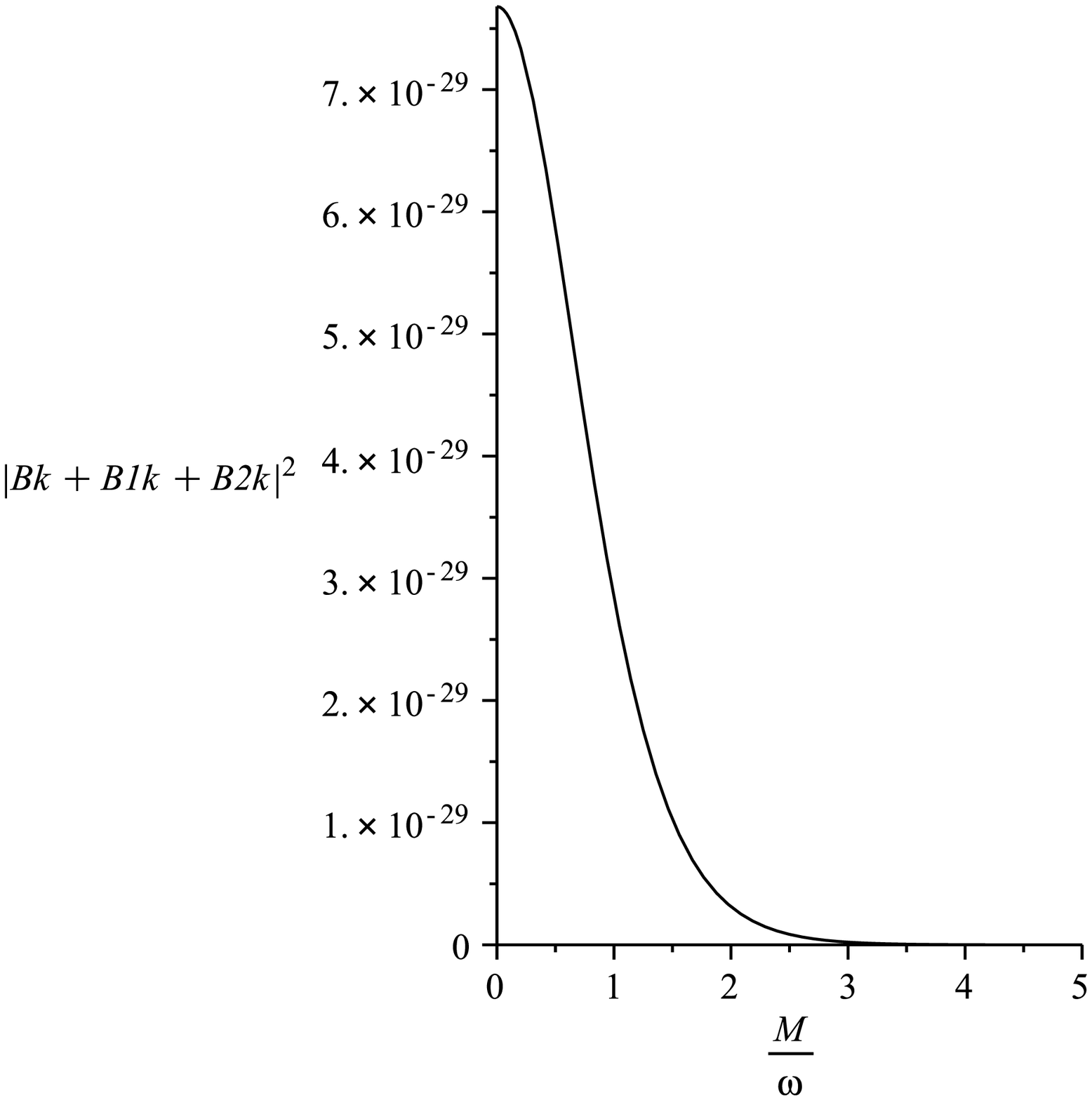}
\caption{The real and imaginary parts of the $ B_{2Kk}$ functions, and the square modulus of the sum $B_{Kk}+B_{1Kk}+B_{2Kk}$ as functions of $M_Z/\omega$ with $m/\omega = 10$, $p = 0.1, p'=0.2, P=1$. The dotted lines represent the imaginary parts, and the solid lines the real parts.}
\centering
\end{figure}
\,

\,

\,

\subsection{Total probability for $\frac{m}{\omega}=\frac{M_Z}{\omega} = 0$}
The limit where $m/\omega\sim M_Z/\omega \rightarrow0$ is important since in this situation the gravitational field is large, and this limit is interesting for the phenomena related to the early universe. In this limit the probability of transition in volume unit for the case $\lambda=\pm1$ is reduced to:
\begin{eqnarray}\label{prob00}
P(\lambda = \pm 1) =\frac{1}{8}\sum_{\sigma\sigma'\lambda}|A|^2=\frac{e^2}{64 \pi^3} \delta^3 (\vec{p}+\vec{p}\,' +\vec{P})\frac{1}{8}\sum_{\sigma\sigma'\lambda}|\xi_{\sigma}^+(\vec{p}\,)\,\vec{\sigma}\vec{\epsilon}\,^*\,(\vec{n}_{\mathcal{P}},\lambda=\pm1)\eta_{\sigma'}(\vec{p}\,')|^2\nonumber\\
\times \frac{1}{P(p+p'+P)^2}\left[\frac{\cos^2(2 \theta_W)}{\sin^2(2 \theta_W)} + \tan^2(\theta_W)-2\tan(\theta_W)\frac{\cos(2 \theta_W)}{\sin(2 \theta_W)}sgn(\sigma)sgn(\sigma')\right]
\end{eqnarray}
since $A_1(\lambda=\pm1)=A_2(\lambda=\pm1)$ in the limit  $m/\omega=M_Z/\omega = 0$.
The total probability is obtained by performing the integrals over the final momenta. We will consider the case when the electron and positron are emitted on the same direction that is the $z$ axis, but move in opposite senses in such a way that the angle between the momenta vectors is $\pi$ i.e $\vec p=p\,\vec e_3\,,\,\vec p\,'=-p'\,\vec e_3$. In this particular case the bispinor summation is reduced to a number since the helicity bispinors are reduced to a simple column matrix with elements $0,\,\pm 1$. The momentum of the Z boson is also considered on the third axis such that $\vec P=P\,\vec e_3$.

In a curved space the fact that the four momentum has constant length $g_{\mu\nu}p^{\mu}p^{\nu}=m^2$, we have as consequence that $g_{00}p^0=p_0-g_{0i}p^i$, with $p_0=\sqrt{g_{00}m^2+(g_{0i}g_{0j}-g_{00}g_{ij})p^ip^j}$. Then at a coordinate transformation the contravariant components of the momentum $p'^{\mu}=\frac{\partial x'^{\mu}}{\partial x^{\nu}}p^{\nu}$, the invariant element of volume in the space phase determined by space and momentum coordinates is \cite{42}:
\begin{equation}\label{invv}
\sqrt{-g}\frac{d^3 p}{p_0}
\end{equation}
This invariant element of volume was used to define the particle distribution function \cite{42} and we mention that it is the volume element defined with the contravariant components of momentum $d^3p=dp^1dp^2dp^3$. It is also important to mention that this invariant volume is always multiplied by a function dependent on coordinate such as the distribution functions in the theory of relativistic Boltzmann equation \cite{42} and all the equations can be expressed in terms of the momenta defined in a local frame. We adopt the definition for the volume elements as in equation (\ref{invv}), and express the momenta integrals that results from integrating equation (\ref{prob00}) in terms of physical momentum $\widehat{p}=e^{-\omega t}p$, up to some factors dependent on time that came out from the delta Dirac function and the factor $\frac{1}{P(p+p'+P)^2}$ which depends on conserved momentum, and we find that the integral is of the form:
\begin{equation}\label{dv12}
  \widehat{I} \sim \int \frac{d^3 \widehat{p}}{\sqrt{m^2+\widehat{p}^2}} \int \frac{d^3 \widehat{P}}{\sqrt{M_Z^2+\widehat{P}^2}} \int\frac{ d^3 \widehat{p'}}{\sqrt{m^2+\widehat{p'}^2}} \frac{\delta^3 (\widehat{\vec{p}}+\widehat{\vec{p}\,'} +\widehat{\vec{P}})}{(2\pi)^3\widehat{P}(\widehat{p}+\widehat{p'}+\widehat{P})^2}.
\end{equation}

To obtain the correct dimensionality in our total probability we will implement a similar definition for the volume element in momentum space, but defined with the conserved momentum so that we will add a factor of $\frac{1}{p_0}=\frac{1}{\sqrt{m^2+p^2}}$, with $p^2=g^{ij}p_ip_j$, to each volume element in momentum space. One can also think that $\frac{1}{p_0}$ factors are in fact similar to the density of the states function which in Minkowski theory has a dimension of $(energy \cdot volume)^{-1}$.

Then the total probability is defined as the following integral after the final momenta of the probability obtained in (\ref{prob00}):
\begin{equation}\label{pro}
 P_{tot}=\int \frac{d^3 p}{p_0} \int\frac{d^3 p'}{p'_0} \int \frac{d^3 P}{P_0} \, P(\lambda = \pm 1)
\end{equation}
This will led us to the correct dimension for the total probability and transition rate. Since the probability contains ultra-violet divergences, we will consider the case of ultra-relativistic momenta i.e. $p>>m$ for $p',P$ integrals, such that $p'_0=p',\,P_0=P$, while for the electron momenta we preserve the mass dependence i.e. $p_0=\sqrt{m^2+p^2}$. The momenta integrals that need to be computed in order to obtain the total probability are:
\begin{equation}\label{dv1}
  I = \int \frac{d^3 p}{\sqrt{m^2+p^2}} \int \frac{d^3 P}{P} \int\frac{ d^3 p'}{p'} \frac{\delta^3 (\vec{p}+\vec{p}\,' +\vec{P})}{(2\pi)^3P(p+p'+P)^2} = \int \frac{d^3 p}{\sqrt{m^2+p^2}} \int \frac{d^3 P}{P^2} \frac{1}{4(2\pi)^3(p+P)^3}
\end{equation}
where we use the setup with the particles moving on the same direction $|\vec{p}+\vec{P}|=p+P$.

Using equation (\ref{rdptot}) from Appendix we arrive at the final form for the $p$ integral:
\begin{eqnarray}\label{dv}
  I =2\pi \int \frac{d^3 p}{(2\pi)^3}\frac{1}{p^2\sqrt{p^2+ m^2}}.
\end{eqnarray}
The above integrals contain ultraviolet logarithmic divergences. The structure of the above integrals is not the same as in Minkowski field theory since the four momentum integral is missing. To obtain the result of the integral we apply the dimensional regularization proposed in \cite{WF,GHV,IT,GT}, which allows us to change the measure of momentum integration so that the dimension D in the integrals can be an arbitrary complex
number. We also specify that this method was used to study the propagators in de Sitter space-time \cite{PC,WT,PT}. Let us recall the well known result of the integral after $p$ in $D$ dimensions adapted to our situation:
\begin{equation}
\int \frac{d^3 p}{(2\pi)^3}\frac{1}{p^2\sqrt{p^2+ m^2}}\rightarrow I(D)=\int \frac{d^D p}{(2\pi)^D}\frac{1}{p^2\sqrt{p^2+ m^2}}=\frac{S_D}{(2\pi)^D}\int_0^{\infty}dp \frac{p^{D-1}}{p^2\sqrt{p^2+ m^2}},
\end{equation}
where $S_D$ is the result of the angular integrals in $D$ dimensions
\begin{equation}
S_D=\frac{2\pi^{D/2}}{\Gamma(D/2)}
\end{equation}
Then the variable is changed such that $p^2=ym^2$ and we arrive at the new integrals
\begin{equation}
I(D)=\frac{S_D\, m^{D-3}}{2(2\pi)^D}\int_0^{\infty}dy \frac{y^{D/2-2}}{(1+y)^{1/2}}.
\end{equation}
One can further use the integral of the Beta Euler function \cite{AS,21}
\begin{eqnarray}
B(a,b)=\frac{\Gamma(a)\Gamma(b)}{\Gamma(a+b)}=\int_0^{\infty}dy \frac{y^{a-1}}{(1+y)^{a+b}}
\end{eqnarray}
which in our case takes the values for $a=D/2-1\,,b=\frac{3-D}{2}$ and the final result reads \cite{WF,GHV,HGF}:
\begin{eqnarray}\label{dv2}
I(D)=\frac{2\pi^{D/2}m^{D-3}\Gamma(D/2-1)\Gamma(\frac{3-D}{2})}{2(2\pi)^{D}\Gamma(D/2)\Gamma(1/2)}
\end{eqnarray}
First is important to mention that the same result is obtained if we consider that the momenta of electron-positron pair is large and in equation (\ref{pro}) we take $p_0=p,\,p'_0=p',\,P_0=\sqrt{M_Z^2+P^2}$, and in all the above equations we make the substitution $m\rightarrow M_Z$. The above integral contains ultraviolet divergences for $D\geq3$ and infrared divergences for $D\leq0$, and these divergences are contained in the poles of the gamma Euler functions. In general these poles could be extracted by introducing an arbitrary mass parameter denoted by $\mu$ and an arbitrary coupling dimensionless constant denoted by $g$ so that our initial integral is replaced by:
\begin{equation}\label{dem}
I(D)_r=-\lambda\int\frac{d^D p}{(2\pi)^D}\frac{1}{p^2\sqrt{p^2+ m^2}}=-\lambda\frac{m^{D-3}\Gamma(D/2-1)\Gamma(\frac{3-D}{2})}{\sqrt{\pi}(4\pi)^{D/2}\Gamma(D/2)}
\end{equation}
where,
\begin{equation}
g=\lambda \mu^{D-3}=\lambda \mu^{-\varepsilon}
\end{equation}
and we write the result of the integral (\ref{dv2}) as a function of $g$ and $\varepsilon$ for $D=3-\varepsilon$ to adapt to our integral given in equation (\ref{dv}). Then the regularized integral gives:
\begin{equation}
I(D)_r=-\frac{g}{(4\pi)^{3/2}}\left(\frac{4\pi \mu^2}{m^2}\right)^{\varepsilon/2}\frac{\Gamma(\frac{1-\varepsilon}{2})\Gamma(\frac{\varepsilon}{2})}{\sqrt{\pi}\Gamma(\frac{3-\varepsilon}{2})}.
\end{equation}
The expansion in powers of $\varepsilon$
\begin{equation}\label{ep}
\left(\frac{4\pi \mu^2}{m^2}\right)^{\varepsilon/2}=1+\frac{\varepsilon}{2}\ln\left(\frac{4\pi \mu^2}{M_Z^2}\right)+o(\varepsilon^2)
\end{equation}
while the gamma Euler functions can be written as:
\begin{equation}
\frac{\Gamma(\frac{1-\varepsilon}{2})\Gamma(\frac{\varepsilon}{2})}{\Gamma(\frac{3-\varepsilon}{2})}=\frac{4}{\varepsilon}-2[\gamma+\psi(-1/2)+\psi(3/2)]+o(\varepsilon),
\end{equation}

The final result for $I(D)_r$ is written in terms of the Euler digamma function $\psi$ and we restrict it to the terms proportional with $\varepsilon$
\begin{eqnarray}\label{epsi}
I(D)_r=\frac{g}{(4\pi^2)}\left[\frac{2}{\varepsilon}+2-\gamma+\ln\left(\frac{4\pi \mu^2}{m^2}\right)+o(\varepsilon)\right]
\end{eqnarray}
The result obtained above is divergent for $\varepsilon=0$ and finite for arbitrary small values of $\varepsilon$, and we observe that the dimensional regularization does not removes all divergences from our integral when $D=3$.

In order to resolve this divergence in the total probability we will use the method of minimal substraction proposed in \cite{GT,SW}. Let us recall the result obtained in equation (\ref{dv2}), where we used the relation with gamma Euler functions $z\Gamma(z)=\Gamma(1+z)$ for the divergent gamma function $\Gamma \big(\frac{3-D}{2}\big)$ in $D=3$:

\begin{eqnarray}
  &&\bigg( \frac{3-D}{2} \bigg) \Gamma \bigg( \frac{3-D}{2} \bigg) =  \Gamma \bigg(\frac{5-D}{2} \bigg);\nonumber\\
 && \Gamma \bigg( \frac{3-D}{2} \bigg) = \frac{2 \Gamma \big( \frac{5-D}{2} \big)}{(3-D)}.
\end{eqnarray}
We rewrite the result for $I(D)$ in the form:
\begin{equation}
  I(D) = \frac{2 m^{D-3} \Gamma \big(\frac{D}{2} -1\big) \Gamma \big(\frac{5-D}{2} \big)}{\sqrt{\pi}\Gamma \big(\frac{D}{2} \big) (4\pi)^{D/2} (D-3)}.
\end{equation}
The above function has a pole in $D=3$ with residue:
\begin{equation}
R=\lim_{D\rightarrow 3}(D-3)I(D)=\frac{1}{2\pi^{2}}
\end{equation}

 Then we choose the a counter-term where we introduce the mass parameter $\mu$ of the form $\frac{\mu^{s}R}{D-3}$, where $s$ is taken such that this term has the same dimension as $ I(D)$. Then we define the renormalized integral as:
\begin{eqnarray}
  I(D)_{ren} &=& -I(D) + \frac{ \mu^{D-3}}{2\pi^2 (D-3) }\nonumber\\
 &=-& \frac{1}{ (D-3)} \Bigg[\frac{2m^{D-3}\Gamma \big( \frac{5-D}{2} \big)\Gamma\big( \frac{D}{2}-1 \big) }{\sqrt{\pi}(4\pi)^{\frac{D}{2}}\Gamma\big( \frac{D}{2}-1 \big)} - \frac{\mu^{D-3}}{2\pi^2} \Bigg].
\end{eqnarray}
The parenthesis from the above equation can be expanded around $D=3$
\begin{eqnarray}
 \frac{2m^{D-3}\Gamma \big( \frac{5-D}{2} \big)\Gamma\big( \frac{D}{2}-1 \big) }{\sqrt{\pi}(4\pi)^{\frac{D}{2}}\Gamma\big( \frac{D}{2}-1 \big)} - \frac{\mu^{D-3}}{2\pi^2}&=&\frac{D-3}{4\pi^2}\left[-\ln\left(\frac{4\pi \mu^2}{m^2}\right)-\psi(3/2)+\psi(1/2)\right] +o((D-3)^2)\nonumber\\
 &=&\frac{D-3}{4\pi^2}\left[-2+\gamma-\ln\left(\frac{4\pi \mu^2}{m^2}\right)\right]+o((D-3)^2)
\end{eqnarray}
The divergent term $D-3$ is canceled by the expansion around $D=3$, and the final result for the renormalised integral will be finite:
\begin{equation}\label{epvf}
I(D)_{ren} = \frac{1}{4\pi^2}\left[2-\gamma+\ln\left(\frac{4\pi \mu^2}{m^2}\right)\right],
\end{equation}
where we point out that parenthesis contain the exact finite terms obtained in equation (\ref{epsi}).

The total probability is obtained by collecting the results from equation (\ref{epvf}), with the observation that the summation after $\lambda,\sigma,\sigma'$ is included in our result:
\begin{eqnarray}\label{ptott}
P_{tot} &=&\frac{e^2}{64\cdot2\pi}\left[2-\gamma+\ln\left(\frac{4\pi \mu^2}{m^2}\right)\right]\left[\frac{\cos^2(2 \theta_W)}{\sin^2(2 \theta_W)} + \tan^2(\theta_W)\right]\nonumber\\
&=&\frac{M_W^2G_F \sin^2\theta_W}{16\sqrt{2}\pi}\left[2-\gamma+\ln\left(\frac{4\pi \mu^2}{m^2}\right)\right]\left[\frac{\cos^2(2 \theta_W)}{\sin^2(2 \theta_W)} + \tan^2(\theta_W)\right]
\nonumber\\
\end{eqnarray}
where in the second equality we introduce the Fermi constant $G_F$ i.e. $e^2=4\sqrt{2}M_W^2G_F \sin^2\theta_W$, with $M_W$ the mass of the W boson. We prove that the total probability is finite by using the dimensional regularization and the substraction method. Our integrals contain divergences of the form $\frac{1}{(D-3)^n}$ that can be removed by using the minimal substraction method \cite{GT,SW}.

The counter-term can also be cast in the lagrangean density, but we restrict to define the regularized total probability defined in equation (\ref{ptott}). However in a future work we want to study all the first order processes to obtain all counter-terms in order to propose a renormalization scheme based on dimensional regularization, in the limit of large expansion when $\omega>>m$.

\subsection{Amplitude of transition for $\lambda=0$}
\begin{figure}[h!t]
\includegraphics[scale=0.35]{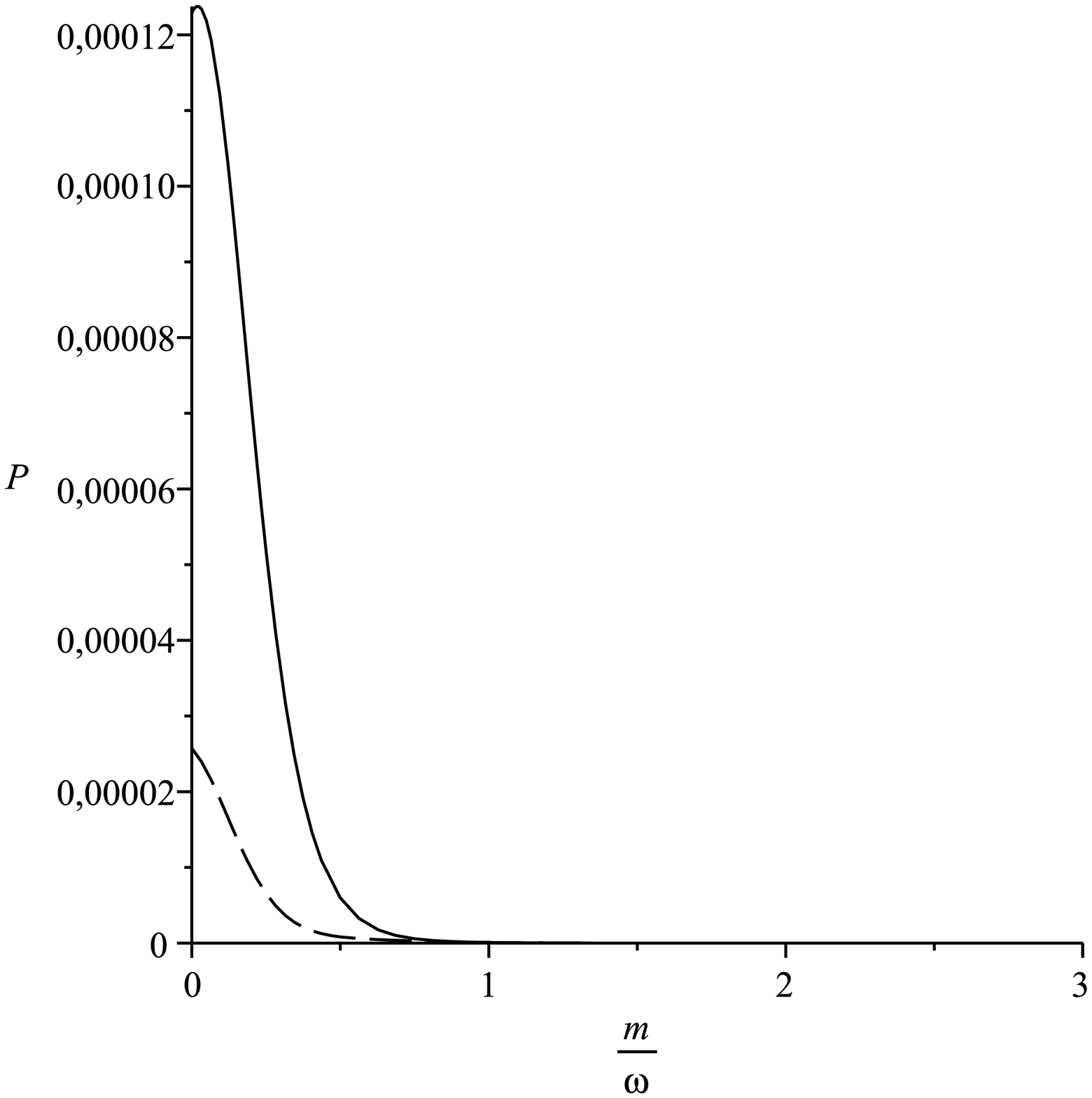}
\includegraphics[scale=0.35]{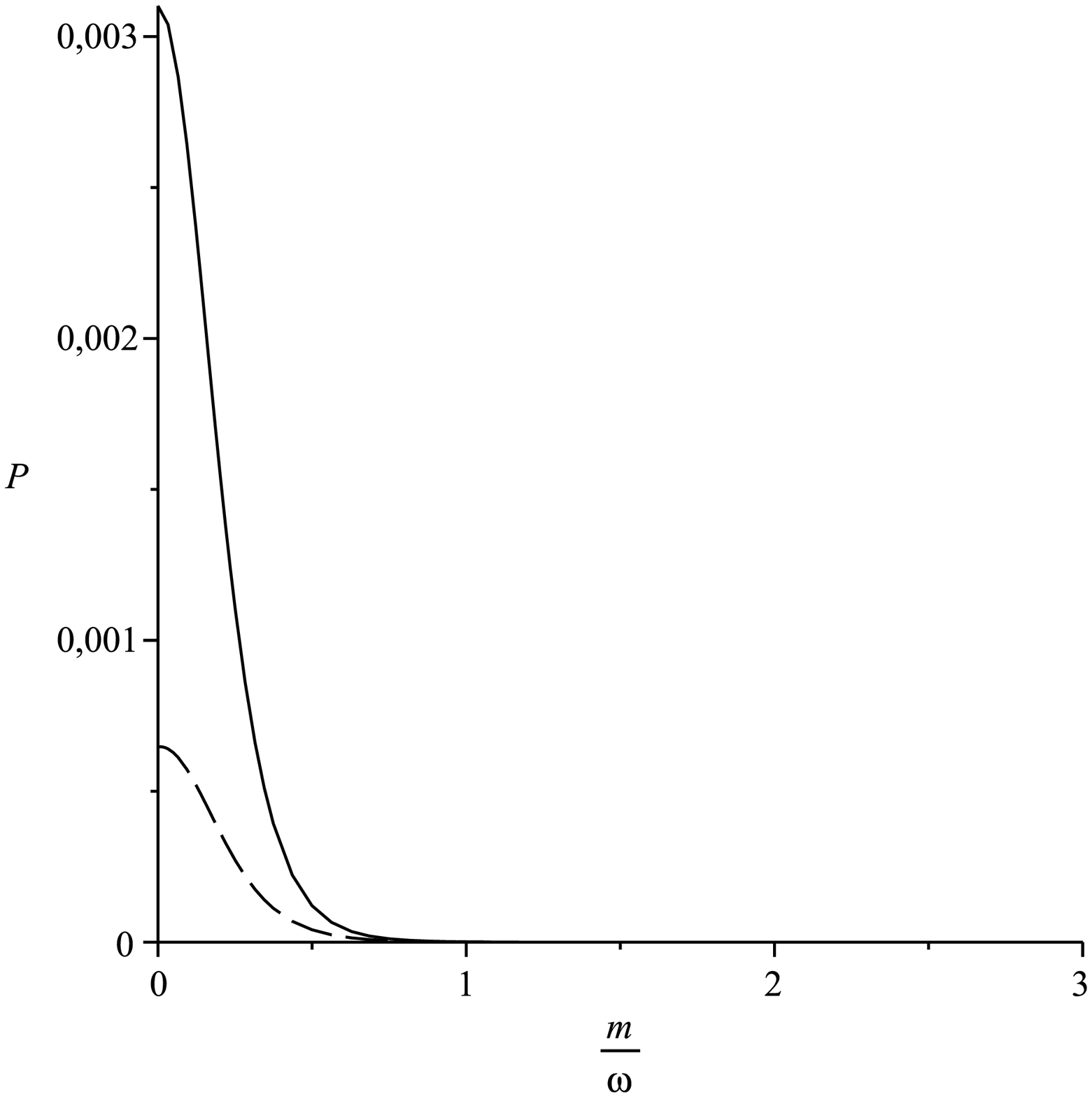}
\caption{Probability as a function of $m/\omega$ for $\lambda=0 $ with $M_Z/\omega = 0.9$, $\sigma$ and $\sigma'$ having the same sign in the left figure and opposite signs in the right figure. The dotted line is for $p/P = 0.1, p'/P=0.2$, and the solid line is for $p/P = 0.3, p'/P=0.4$ }
\centering
\label{f3}
\end{figure}

\begin{figure}[h!t]
\includegraphics[scale=0.35]{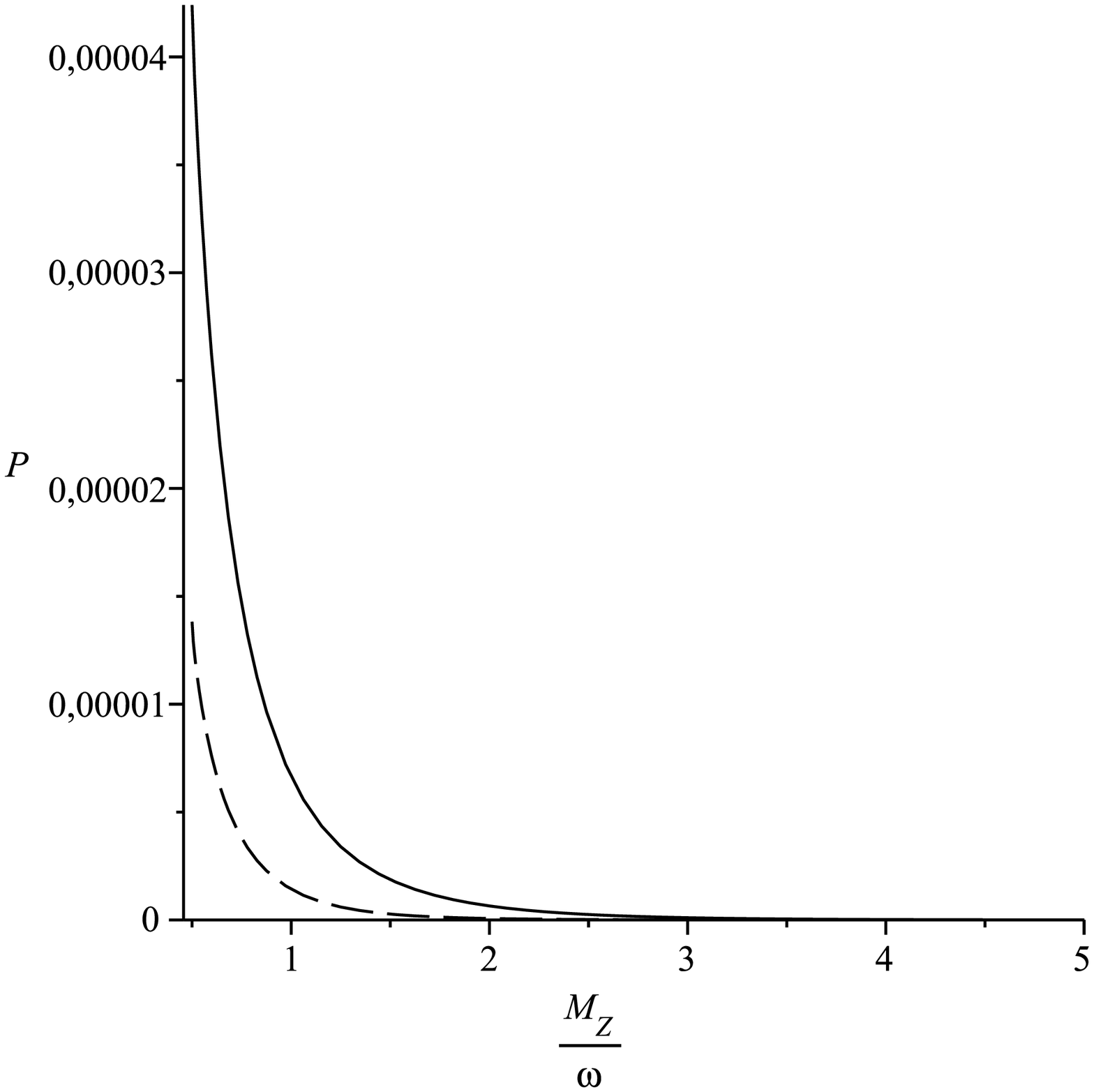}
\includegraphics[scale=0.35]{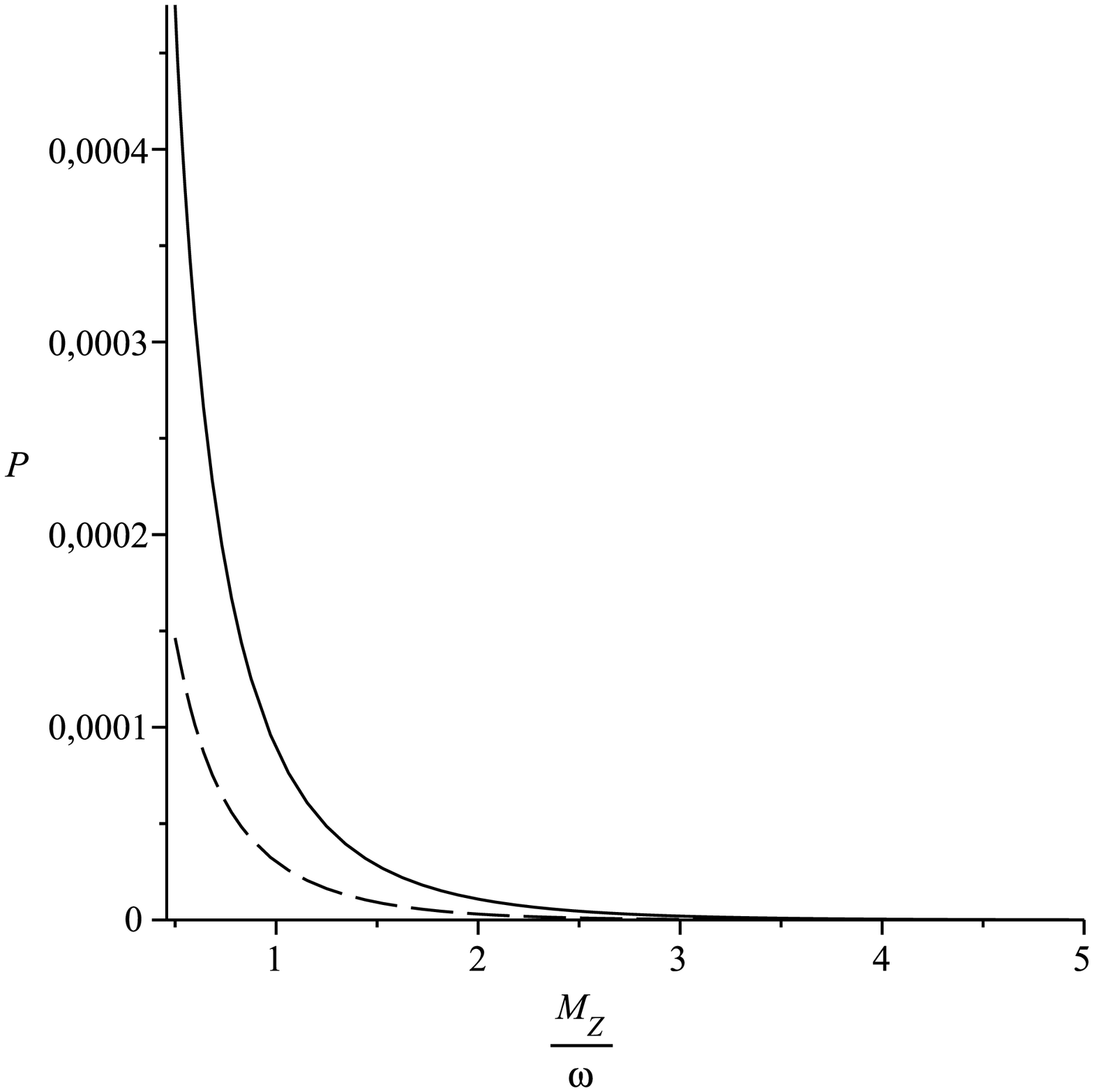}
\caption{Probability as a function of $M_Z/\omega$ for $\lambda=0$ with $m/\omega = 0.5$, $\sigma$ and $\sigma'$ having the same sign in the left figure and opposite signs in the right figure. The dotted line is for $p/P = 0.1, p'/P=0.2$, and the solid line is for $p/P = 0.3, p'/P=0.4$ }
\centering
\label{f4}
\end{figure}

In the case of production of Z bosons with longitudinal polarization the amplitude of transition contains both the spatial part of the solution and temporal parts of the amplitude reads:
\begin{eqnarray}
  A_{Ze\bar{e}} (\lambda = 0) &=& \int d^4 x \sqrt{-g} \Bigg\{ \bigg(\frac{e_0 \cos(2 \theta_W)}{\sin(2 \theta_W)} \bigg) \overline{U}_{\vec{p},\sigma}(x) \gamma^{\hat{0}} e^0_{\hat{0}} \bigg(\frac{1-\gamma^5}{2}\bigg) V_{\vec{p}\,',\sigma'}(x) f^*_{0 \vec{P},\lambda = 0}(x) \nonumber \\
   && - e_0 \tan(\theta_W) \overline{U}_{\vec{p},\sigma}(x) \gamma^{\hat{0}} e^0_{\hat{0}} \bigg(\frac{1+\gamma^5}{2}\bigg) V_{\vec{p}\,',\sigma'}(x) f^*_{0 \vec{P},\lambda = 0}(x) \Bigg\}   \nonumber\\
   && + \int d^4 x \sqrt{-g} \Bigg\{ \bigg(\frac{e_0 \cos(2 \theta_W)}{\sin(2 \theta_W)} \bigg) \overline{U}_{\vec{p},\sigma}(x) \gamma^{\hat{i}} e^j_{\hat{i}} \bigg(\frac{1-\gamma^5}{2}\bigg) V_{\vec{p}\,',\sigma'}(x) \vec{f}^*_{\vec{P},\lambda = 0}(x)   \nonumber\\
   && - e_0 \tan(\theta_W) \overline{U}_{\vec{p},\sigma}(x) \gamma^{\hat{i}} e^j_{\hat{i}} \bigg(\frac{1+\gamma^5}{2}\bigg) V_{\vec{p}\,',\sigma'}(x) \vec{f}^*_{\vec{P},\lambda = 0}(x) \Bigg\}
   = A_1(\lambda=0)+ A_2(\lambda=0).
   \nonumber\\
\end{eqnarray}
The integrals that define the $A_1(\lambda=0)\,,A_2(\lambda=0)$ are:
\begin{eqnarray}
  A_1(\lambda = 0)&=& \frac{\sqrt{pp'}}{8(2\pi)^{3/2}} \frac{\sqrt{\pi} \omega P e^{-\pi k/2}}{M_Z} \delta^3 (\vec{p}+\vec{p}\,' +\vec{P}) \int_{0}^{\infty} dz \cdot z^{5/2} \nonumber\\
  && \times \Bigg\{ -\frac{e_0 \cos(2 \theta_W)}{\sin(2 \theta_W)} sgn(\sigma') H^{(2)}_{\nu_+} (pz) H^{(2)}_{\nu_-} (p'z) H^{(2)}_{-ik} (Pz) \nonumber\\
  && + e_0 \tan (\theta_W) sgn(\sigma) H^{(2)}_{\nu_-} (pz) H^{(2)}_{\nu_+} (p'z) H^{(2)}_{-ik} (Pz) \Bigg\} \xi^+_{\sigma} (\vec{p}) \eta_{\sigma '} (\vec{p}\,');
  \nonumber\\
\end{eqnarray}
\begin{eqnarray}
  A_2 (\lambda = 0)&=& \frac{-i \sqrt{pp'} \omega P e^{-\pi k/2}}{8 M_Z (2\pi)^{3/2}} \delta^3 (\vec{p}+\vec{p}\,' +\vec{P}) \Bigg\{ \int_{0}^{\infty} dz \frac{e_0 \cos(2 \theta_W)}{\sin(2 \theta_W)} sgn(\sigma') \nonumber\\
  && \times \bigg[ \bigg(\frac{1}{2} - ik \bigg) \frac{z^{3/2}}{P} H^{(2)}_{\nu_+} (pz) H^{(2)}_{\nu_-} (p'z) H^{(2)}_{-ik} (Pz) - z^{/5/2} H^{(2)}_{\nu_+} (pz) H^{(2)}_{\nu_-} (p'z) H^{(2)}_{1-ik} (Pz) \bigg]  \nonumber\\
  && - \int_{0}^{\infty} dz \cdot e_0 \tan(\theta_W) sgn(\sigma) \bigg[ \bigg( \frac{1}{2} - ik \bigg) \frac{z^{3/2}}{P} H^{(2)}_{\nu_-} (pz) H^{(2)}_{\nu_+} (p'z) H^{(2)}_{-ik} (Pz)  \nonumber\\
  && - z^{5/2} H^{(2)}_{\nu_-} (pz) H^{(2)}_{\nu_+} (p'z) H^{(2)}_{1-ik} (Pz) \bigg] \Bigg\} \xi^+_{\sigma} (\vec{p}) \vec{\sigma} \cdot \vec{\epsilon}^* (\vec{n}_P , \lambda) \eta_{\sigma} (\vec{p}\,')
\end{eqnarray}
The integrals can be solved by using the same method as in the previous case, and the final result for obtaining the production of longitudinal modes is given by:
\begin{equation}
  A_1 (\lambda = 0)= \frac{i}{4 \pi^2} \frac{\omega P}{M_Z} \frac{\delta^3 (\vec{p}+\vec{p}\,' +\vec{P})}{\cosh^2 (\pi K)} \Bigg\{ - \frac{e_0 \cos(2 \theta_W)}{\sin(2 \theta_W)} sgn(\sigma') B_1 - e_0 \tan(\theta_W) sgn(\sigma) B_2 \Bigg\} \xi^+_{\sigma} (\vec{p}) \eta_{\sigma '} (\vec{p}\,')
\end{equation}

\begin{eqnarray}
  A_2 (\lambda = 0)&=& \frac{P}{4 \pi^{5/2}} \frac{\omega}{M_Z} \frac{\delta^3 (\vec{p}+\vec{p}\,' +\vec{P})}{\cosh^2 (\pi K)} \Bigg\{ \frac{e_0 \cos(2 \theta_W)}{\sin(2 \theta_W)} sgn(\sigma') \bigg[ \bigg(\frac{1}{2} - ik \bigg) B_1 + C_1 \bigg]   \nonumber\\
  && - e_0 \tan(\theta_W) sgn(\sigma) \bigg[ \bigg(\frac{1}{2} - ik \bigg) B_2 + C_2 \bigg] \Bigg\} \xi^+_{\sigma} (\vec{p}) \vec{\sigma} \cdot \vec{\epsilon}^* (\vec{n}_P , \lambda) \eta_{\sigma} (\vec{p}\,')
  \nonumber\\
\end{eqnarray}
where the functions $B\,,C$ are defined as:

\begin{eqnarray}
  B_1 &=&  p' B_{Kk} (pp'P) + p B_{-Kk} (pp'P) - A_{Kk} (pp'P) + C_{Kk} (pp'P) \\
  B_2 &=& p B_{Kk} (pp'P) + p' B_{-Kk} (pp'P) - A_{Kk} (pp'P) + C_{Kk} (pp'P) \\
  C_1 &=& p' D_{Kk} (pp'P) + p D_{-Kk} (pp'P) - F_{Kk} (pp'P) + G_{Kk} (pp'P) \\
  C_2 &=& p D_{Kk} (pp'P) + p' D_{-Kk} (pp'P) - F_{Kk} (pp'P) + G_{Kk} (pp'P).
\end{eqnarray}
Each function from the above relations is a combination of Appel hypergeometric function and gamma Euler functions as follows:
\begin{eqnarray}
  B_{Kk} (pp'P) = \frac{i (pp')^{-iK} e^{\pi K} (iP)^{-\frac{7}{2}+2iK}}{\Gamma \big( \frac{1}{2} - iK \big) \Gamma \big( \frac{3}{2} - iK \big) } \Gamma \bigg(\frac{7}{4} -iK - \frac{ik}{2} \bigg) \Gamma \bigg(\frac{7}{4} -iK + \frac{ik}{2} \bigg) \nonumber\\
  \times F_4 \bigg( \frac{7}{4} -iK - \frac{ik}{2}, \frac{7}{4} -iK + \frac{ik}{2}, \frac{1}{2}-iK, \frac{3}{2}-iK; \Big(\frac{p}{P}\Big)^2, \Big(\frac{p'}{P}\Big)^2 \bigg)
\end{eqnarray}

\begin{eqnarray}
  A_{Kk} (pp'P) = \frac{(p)^{-iK} (p')^{iK} (iP)^{-5/2}}{\Gamma \big( \frac{1}{2} - iK \big) \Gamma \big( \frac{1}{2} + iK \big) } \Gamma \bigg(\frac{5}{4} - \frac{ik}{2} \bigg) \Gamma \bigg(\frac{5}{4} + \frac{ik}{2} \bigg) \nonumber\\
  \times F_4 \bigg( \frac{5}{4} + \frac{ik}{2}, \frac{5}{4} - \frac{ik}{2}, \frac{1}{2}-iK, \frac{1}{2}+iK; \Big(\frac{p}{P}\Big)^2, \Big(\frac{p'}{P}\Big)^2 \bigg)
\end{eqnarray}

\begin{eqnarray}
  C_{Kk} (pp'P) = \frac{(p)^{1+iK} (p')^{1-iK} (iP)^{-9/2}}{\Gamma \big( \frac{3}{2} - iK \big) \Gamma \big( \frac{3}{2} + iK \big) } \Gamma \bigg(\frac{9}{4} - \frac{ik}{2} \bigg) \Gamma \bigg(\frac{9}{4} + \frac{ik}{2} \bigg) \nonumber\\
  \times F_4 \bigg( \frac{9}{4} + \frac{ik}{2}, \frac{9}{4} - \frac{ik}{2}, \frac{3}{2}-iK, \frac{3}{2}+iK; \Big(\frac{p}{P}\Big)^2, \Big(\frac{p'}{P}\Big)^2 \bigg)
\end{eqnarray}

\begin{eqnarray}
  D_{Kk} (pp'P) = \frac{i (pp')^{-iK} e^{\pi K} (iP)^{-\frac{7}{2}+2iK}}{\Gamma \big( \frac{1}{2} - iK \big) \Gamma \big( \frac{3}{2} - iK \big) } \Gamma \bigg(\frac{9}{4} -iK - \frac{ik}{2} \bigg) \Gamma \bigg(\frac{5}{4} -iK + \frac{ik}{2} \bigg) \nonumber\\
  \times F_4 \bigg( \frac{5}{4} -iK + \frac{ik}{2}, \frac{9}{4} -iK - \frac{ik}{2}, \frac{1}{2}-iK, \frac{3}{2}-iK; \Big(\frac{p}{P}\Big)^2, \Big(\frac{p'}{P}\Big)^2 \bigg)
\end{eqnarray}

\begin{eqnarray}
  F_{Kk} (pp'P) = \frac{(p)^{-iK} (p')^{iK} (iP)^{-5/2}}{\Gamma \big( \frac{1}{2} - iK \big) \Gamma \big( \frac{1}{2} + iK \big) } \Gamma \bigg(\frac{7}{4} - \frac{ik}{2} \bigg) \Gamma \bigg(\frac{3}{4} + \frac{ik}{2} \bigg) \nonumber\\
  \times F_4 \bigg( \frac{3}{4} + \frac{ik}{2}, \frac{7}{4} - \frac{ik}{2}, \frac{1}{2}-iK, \frac{1}{2}+iK; \Big(\frac{p}{P}\Big)^2, \Big(\frac{p'}{P}\Big)^2 \bigg)
\end{eqnarray}

\begin{eqnarray}
  G_{Kk} (pp'P) = \frac{(p)^{1+iK} (p')^{1-iK} (iP)^{-9/2}}{\Gamma \big( \frac{3}{2} - iK \big) \Gamma \big( \frac{3}{2} + iK \big) } \Gamma \bigg(\frac{11}{4} - \frac{ik}{2} \bigg) \Gamma \bigg(\frac{7}{4} + \frac{ik}{2} \bigg) \nonumber\\
  \times F_4 \bigg( \frac{7}{4} + \frac{ik}{2}, \frac{11}{4} - \frac{ik}{2}, \frac{3}{2}+iK, \frac{3}{2}-iK; \Big(\frac{p}{P}\Big)^2, \Big(\frac{p'}{P}\Big)^2 \bigg)
\end{eqnarray}
The probability of transition in volume unit is computed by summing after the helicities $\sigma\,,\sigma'$ and we obtain:
\begin{eqnarray}\label{po}
P(\lambda=0)=\frac{1}{4}\sum_{\sigma\sigma'}\left[|A_1 (\lambda = 0)|^2+|A_2 (\lambda = 0)|^2+A_1^* (\lambda = 0)A_2 (\lambda = 0)+A_1 (\lambda = 0)A_2^* (\lambda = 0)\right]
\nonumber\\
\end{eqnarray}
The probability from equation (\ref{po}) depends on the masses of the particles, expansion parameter and particles momenta, and the variation with the expansion parameter is given in Figs.(\ref{f3}),(\ref{f4})

The above results prove that the emission of longitudinal modes is possible for small values of the ratio between the masses and the expansion parameter. The probabilities are also vanishing in the Minkowski limit. For a clear picture of the ratios between masses and expansion parameter let us give the ratio $m c^{2}/\hbar\omega\sim5\cdot10^{37}$, (where $m$ is the electron mass). From this we observe that for large masses comparatively to the expansion factor, the probability of pair production at present is zero and our results refer to the early universe when the expansion parameter was bigger or around the same value of the particle mass.

\section{Transition rates}
In this paper we analyse the problem of generating Z bosons and massive fermions by using the perturbative transition amplitude generalized to a curved space-time. We use the electro-weak coupling with the neutral currents for analysing the spontaneous generation from vacuum of the massive Z bosons and massive fermions. The analytical and graphical results prove that the generation of massive Z bosons is a process that takes place along the entire inflation. This result is important in establishing the mechanisms that were involved in the generation of massive bosons in the early universe. A remarkable result that we have obtained is that in the Minkowski limit the probabilities are vanishing for both the transversal modes and longitudinal modes, a result that confirms the well established results from electro-weak theory where the process analysed here is forbidden by the energy conservation. In the non-stationary de Sitter metric this process is possible only at large expansion when the mass of the particle is close in value to the expansion parameter. We will compute the transition rate in de Sitter geometry for any values of the ratios between the particle masses and expansion parameter.
For defining the transition rate we work in the chart with conformal time where the definitions could be adapted from Minkowski space-time \cite{cpc}
\begin{equation}
R_{if}=\lim_{t_c\rightarrow 0}\frac{1}{2V}\frac{d}{dt_c}|A_{if}|^2=\lim_{t\rightarrow \infty}\frac{e^{\omega t}}{2V}\frac{d}{dt}| A_{if}|^2.
\end{equation}
Equation for transition amplitude in the case $\lambda=\pm1$ can be written in the form
\begin{equation}
A_{i\rightarrow f}(\lambda=\pm1)=\delta^3(\vec{p}+\vec{p}\,'+\vec{P})M_{i\rightarrow f}I_{i\rightarrow f}
\end{equation}

where $I_{i\rightarrow f}$ are the temporal integrals and $M_{i\rightarrow f}$ contain the constants and the bispinors. These functions are given bellow:
\begin{equation}
  M_{i\rightarrow f} = \frac{e_0 \pi^{3/2}\sqrt{pp'}}{8(2\pi)^{3/2}} \, \xi_{\sigma}^+(\vec{p}\,)\,\vec{\sigma} \cdot \vec{\epsilon}\,^*\,(\vec{n}_{P},\lambda=\pm1)\eta_{\sigma'}(\vec{p}\,')
\end{equation}

\begin{eqnarray}
  I_{i\rightarrow f} &=& \biggl\{ \frac{\cos(2\theta_{W})}{\sin(2\theta_{W})} \, sgn(\sigma')
  \int_0^{\infty} dz\left[
  z^{3/2}H^{(2)}_{\nu_{+}}(pz)H^{(2)}_{\nu_{-}}(p'z)K_{-ik}\left(iPz\right)\right] \nonumber \\
  && - \tan(\theta_{W})\,sgn(\sigma)\int_0^{\infty} dz\left[
  z^{3/2}H^{(2)}_{\nu_{-}}(pz)H^{(2)}_{\nu_{+}}(p'z)K_{-ik}\left(iPz\right)\right] \biggl\}.
\end{eqnarray}
The temporal integrals can be written in the form $I_{i\rightarrow f}=\int_0^{\infty}\mathcal{K}_{if}dz$, where we denote the integrand by $\mathcal{K}_{if}$
\begin{eqnarray}
  \mathcal{K}_{if} &=& \Big(\frac{2i}{\pi} \Big)\biggl[\frac{\cos(2\theta_{W})}{\sin(2\theta_{W})} \, sgn(\sigma') z^{3/2} H^{(2)}_{\nu_{+}}(pz) H^{(2)}_{\nu_{-}}(p'z) K_{-ik}\left(iPz\right) \\
   && - \tan(\theta_{W}) \, sgn(\sigma) z^{3/2} H^{(2)}_{\nu_{-}}(pz) H^{(2)}_{\nu_{+}}(p'z) K_{-ik}\left(iPz\right)\biggl]
\end{eqnarray}
Then the rate of transition for triplet generation from vacuum can be written in terms of the above quantities as:
\begin{equation}\label{rt}
R_{if}=\frac{1}{8} \sum_{\sigma \sigma'\lambda}\frac{1}{(2\pi)^3}\,\delta^3(\vec{p}+\vec{p}\,'+\vec{P})| M_{i\rightarrow f}|^2 | I_{i\rightarrow f}| \lim_{t\rightarrow \infty}| e^{\omega t}\mathcal{K}_{if}|,
\end{equation}
We consider that the transition from $in$ to $out$ state takes place after a sufficiently large time and we denote this time by $t_{\infty}$ and the limits from the rate equation (\ref{rt}) will be evaluated for this time.
For computing the limit we use the expansion of Hankel functions for small arguments since for $t\rightarrow \infty$ the argument $z=e^{-\omega t}$ become very small
\begin{equation}
H^{(1,2)}_\nu (z)=\mp i\left(\frac{2}{z}\right)^\nu\frac{\Gamma(\nu)}{\pi}.
\end{equation}
In the case of Z boson we use the approximation when $M_Z/\omega<<1/2$ and the index of Hankel functions becomes $-ik\rightarrow\frac{1}{2}$. The limit that defines our transition rate in the general case becomes:
\begin{eqnarray}
  &&\lim_{t \rightarrow t_{\infty}} |e^{\omega t} \mathcal{K}_{if}| = \Big(\frac{2}{\pi} \Big)\lim_{t \rightarrow +t_{\infty}} \bigg| \bigg( - \frac{\cos(2\theta_{W})}{\sin(2\theta_{W})} \frac{sgn(\sigma')}{\pi^2} \Big(\frac{e^{-\omega t}}{\omega} \Big)^{3/2} \Big(\frac{2}{p \, \frac{e^{-\omega t}}{\omega}} \Big)^{\frac{1}{2} +iK} \Big(\frac{2}{p' \, \frac{e^{-\omega t}}{\omega}} \Big)^{\frac{1}{2} -iK} \nonumber\\
  && \times \frac{\Gamma \big(\frac{1}{2} \big) \big| \Gamma \big( \frac{1}{2} - iK\big) \big|^2}{\sqrt{2} \big( \frac{iP}{\omega} e^{-\omega t}\big)^{1/2}} + \tan(\theta_{W}) \frac{sgn(\sigma')}{\pi^2} \Big(\frac{e^{-\omega t}}{\omega} \Big)^{3/2} \Big(\frac{2}{p \, \frac{e^{-\omega t}}{\omega}} \Big)^{\frac{1}{2} -iK} \Big(\frac{2}{p' \, \frac{e^{-\omega t}}{\omega}} \Big)^{\frac{1}{2} +iK} \nonumber  \\
  && \times \frac{\Gamma \big(\frac{1}{2} \big) \big| \Gamma \big( \frac{1}{2} - iK\big) \big|^2}{\sqrt{2} \big( \frac{iP}{\omega} e^{-\omega t}\big)^{1/2}} \bigg) \bigg|.
\end{eqnarray}
The final result for the limit is obtained after extracting the modulus from the imaginary quantity and this gives:
\begin{eqnarray}\label{lim}
  \lim_{t \rightarrow t_{\infty}} |e^{\omega t} \mathcal{K}_{if}| &=& \Big(\frac{2}{\pi} \Big)^{3/2} \frac{1}{\cosh(\pi K) \sqrt{pp'P}} \bigg[ \frac{\cos^2(2\theta_{W})}{\sin^2(2\theta_{W})} + \tan^2(\theta_{W}) \nonumber \\
  && - sgn(\sigma) sgn(\sigma ')  \frac{\cos(2\theta_{W})}{\sin(2\theta_{W})} \tan(\theta_{W}) 2\cos\Big(2K \ln\Big(\frac{p'}{p} \Big) \Big) \bigg]^{1/2}
  \nonumber\\
\end{eqnarray}

Collecting all the results we obtain for the transition rate
\begin{eqnarray}\label{ratafin}
  && R = \frac{1}{8} \sum_{\sigma \sigma'\lambda}\frac{\delta^3(\vec{p}+\vec{p}\,'+\vec{P})}{(2\pi)^3} \sqrt{\mathcal{I}_{i\rightarrow f} \cdot \mathcal{I}^*_{i\rightarrow f}} \sqrt{\frac{2}{\pi}} \, \frac{1}{\cosh(\pi K) \sqrt{pp'P}} \, \frac{e^2 \, pp'}{8\pi^2} \sum_{\sigma \sigma' \lambda} \big| \xi_{\sigma}^+(\vec{p}\,)\,\vec{\sigma} \cdot \vec{\epsilon}\,^*\,\eta_{\sigma'}(\vec{p}\,') \big|^2  \nonumber \\
  && \bigg[ \frac{\cos^2(2\theta_{W})}{\sin^2(2\theta_{W})} + \tan^2(\theta_{W}) - sgn(\sigma) sgn(\sigma ')  \frac{\cos(2\theta_{W})}{\sin(2\theta_{W})} \tan(\theta_{W}) \, 2\cos\Big(2K \ln\Big(\frac{p'}{p} \Big) \Big) \bigg]^{1/2}
\end{eqnarray}
where we introduce the notation $\mathcal{I}_{i\rightarrow f}=\frac{I_{i\rightarrow f}}{\sqrt{pp'}}$ which are the results of the temporal integrals as obtained in the amplitude equation (\ref{a1}):
\begin{eqnarray}
\mathcal{I}_{i\rightarrow f}=\frac{1}{\sqrt{pp'}}\biggl\{\frac{\cos(2\theta_{W})}{\sin(2\theta_{W})}\,sgn(\sigma')A_1(\lambda=\pm1)
-\tan(\theta_{W})\,sgn(\sigma)A_2(\lambda=\pm1) \biggl\},
\end{eqnarray}
with the specification that the functions $A_1(\lambda=\pm1),\,A_2(\lambda=\pm1)$ were defined in equation (\ref{aa3}).
\begin{figure}[h!t]
\includegraphics[scale=0.35]{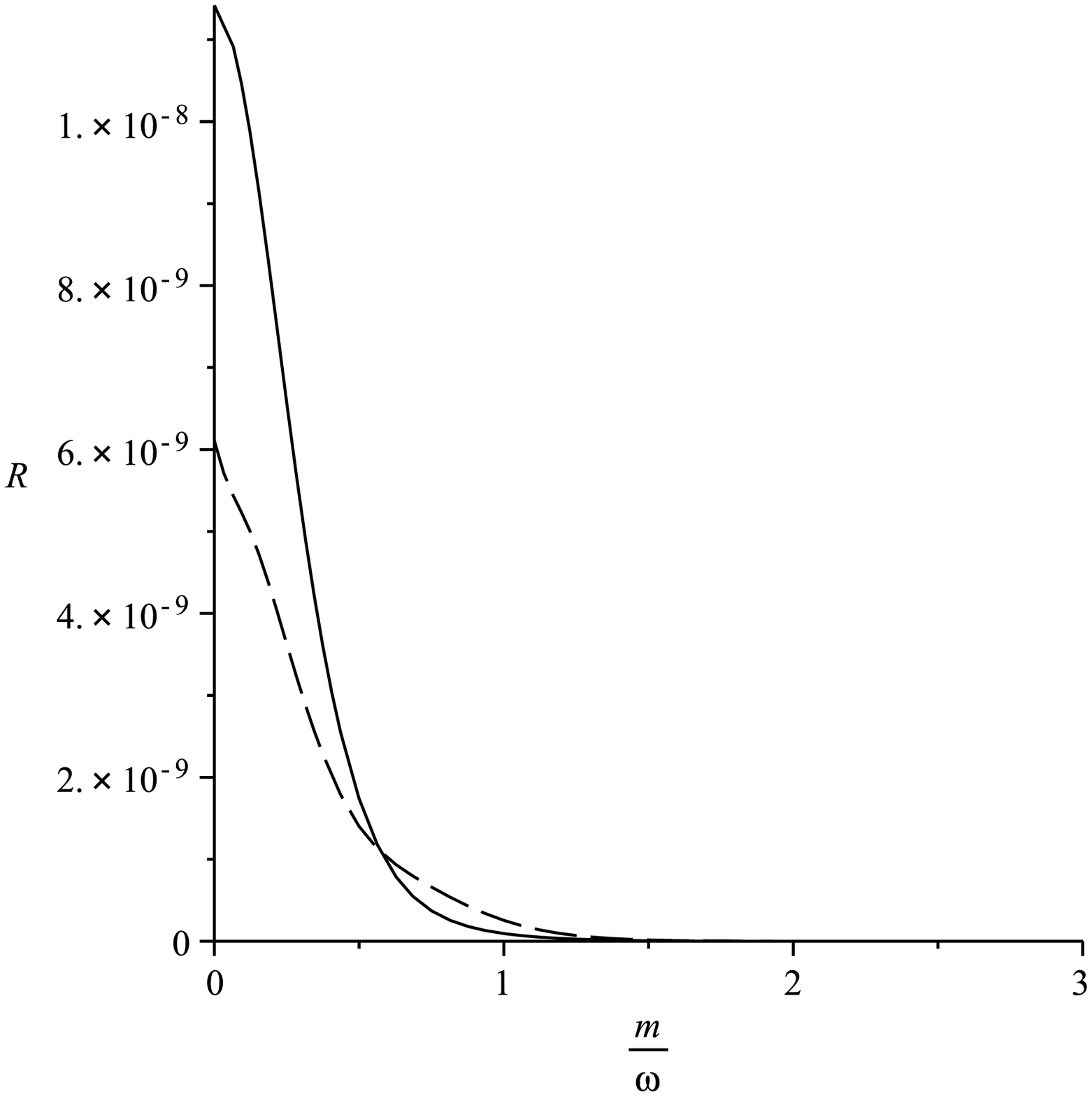}
\includegraphics[scale=0.35]{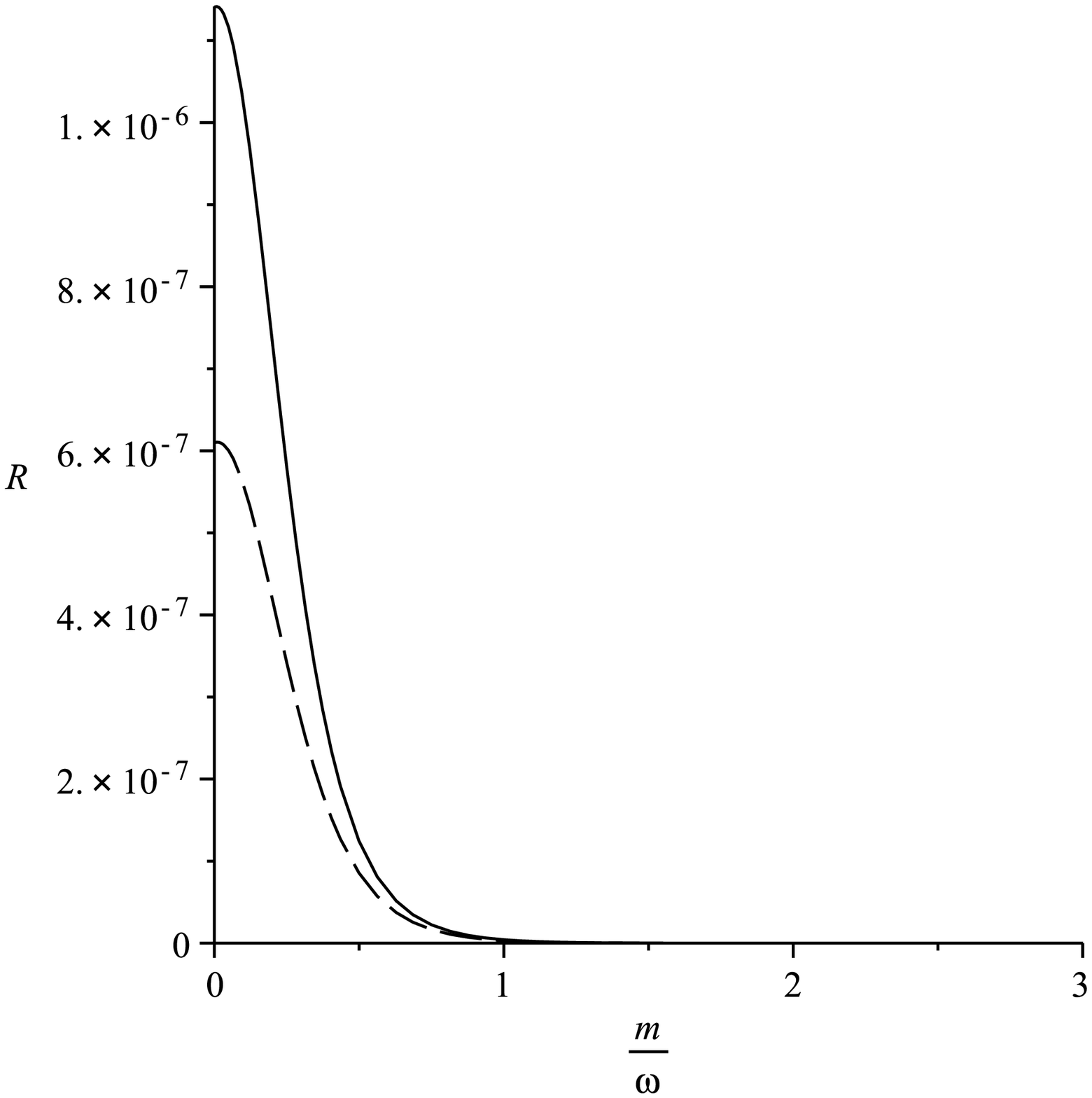}
\caption{Transition rate as a function of $m/\omega$ for $\lambda=\pm1 $, with $M_Z/\omega = 0.9$, $\sigma$ and $\sigma'$ having the same sign and opposite signs respectively. The dotted line is for $p/P = 0.1, p'/P=0.2$, and the solid line is for $p/P = 0.3, p'/P=0.4$ }
\centering
\end{figure}

\begin{figure}[h!t]
\includegraphics[scale=0.35]{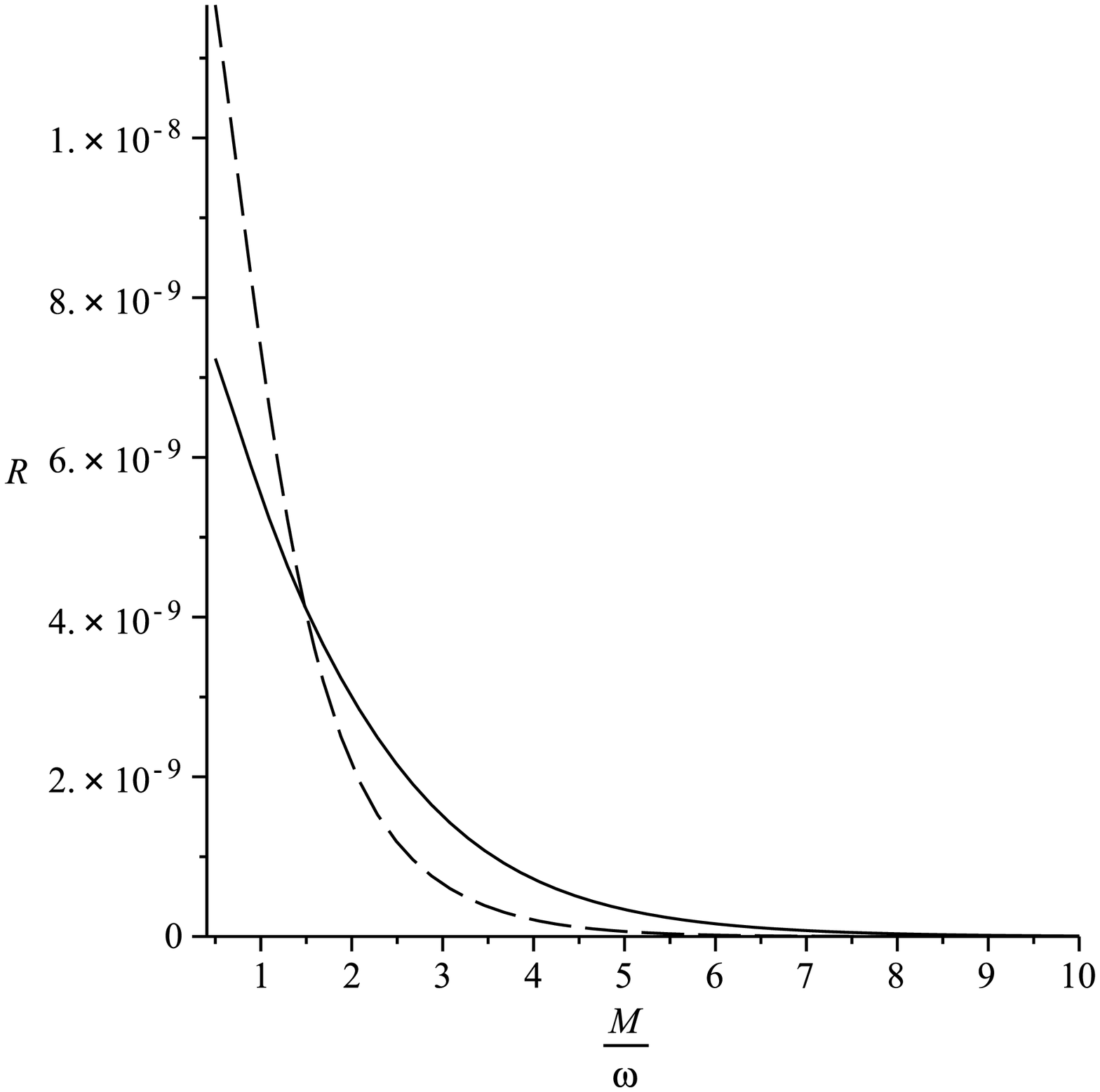}
\includegraphics[scale=0.35]{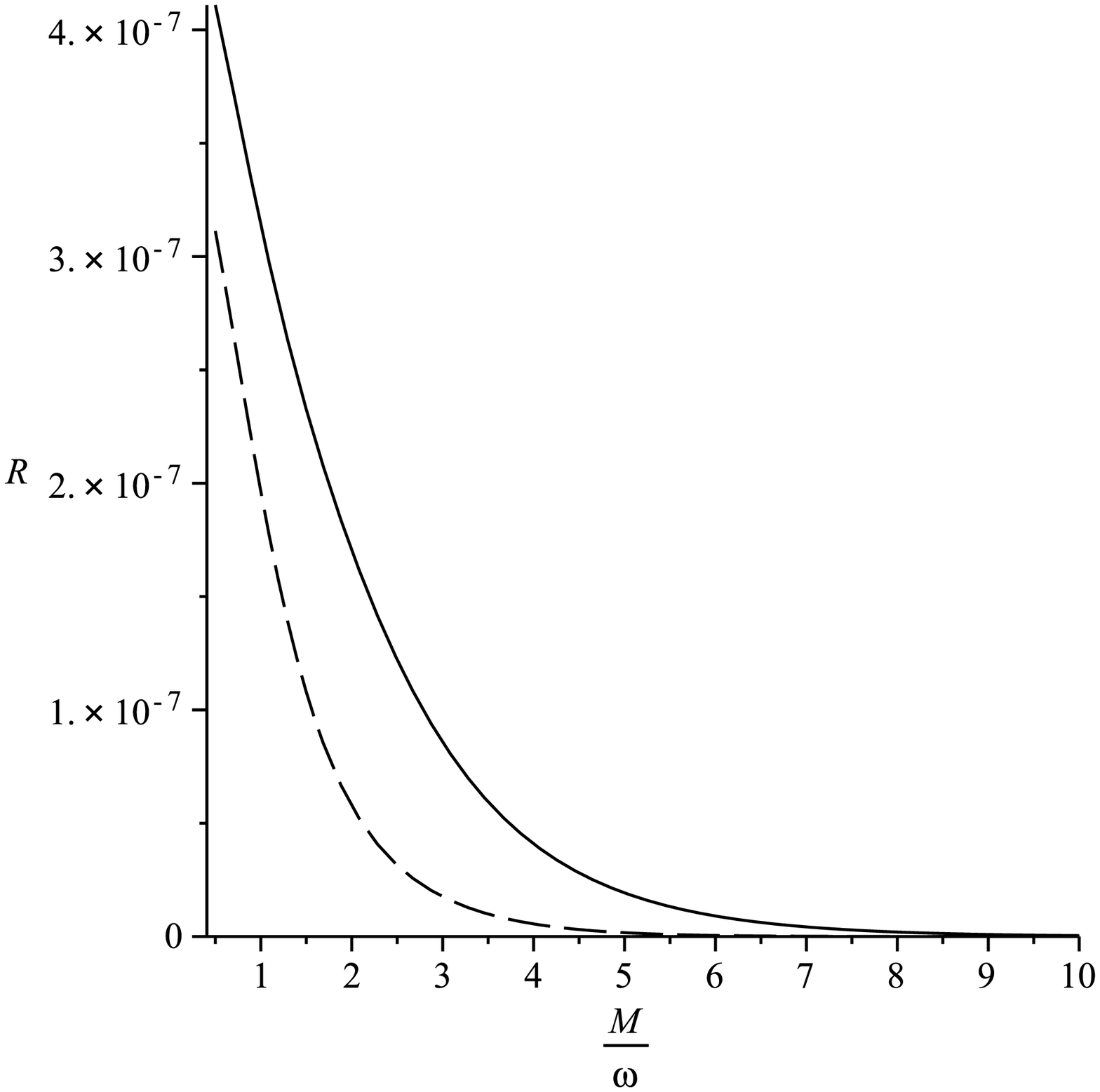}
\caption{Transition rate as a function of $M_Z/\omega$ for $\lambda=\pm1$, with $m/\omega = 0.5$, $\sigma$ and $\sigma'$ having the same sign and opposite signs respectively. The dotted line is for $p/P = 0.1, p'/P=0.2$, and the solid line is for $p/P = 0.3, p'/P=0.4$ }
\centering
\end{figure}
A graphical analysis of equation (\ref{ratafin}) in terms of parameters $\frac{m}{\omega},\,\frac{M_Z}{\omega}$ proves that the rates for the process of Z boson and fermion-antifermion generation from vacuum are important only in inflationary regime and are vanishing in the Minkowski limit when the expansion parameter becomes zero. The rates behaviour in terms of parameters $K,M_Z/\omega$ reproduce the probabilities behaviour proving that the production of gauge Z bosons was possible only in the large expansion conditions of early universe.

\subsection{Rates in the limit $\frac{m}{\omega}=\frac{M_Z}{\omega}=0$}
Since it is well known that the massive Z and W bosons could exist as stable particles only in the early universe there are some interesting observations that can be made to extend our result. The first is related to the fact that the transition rate in de Sitter case \cite{cpc} could in principle be computed for the case $\omega>>M_Z$ in which case the functions that define our amplitude could become simpler. For example in the case $\lambda=\pm1$ the amplitude is reduced in the limit $\frac{m}{\omega}\rightarrow0,\,\frac{M_Z}{\omega}\rightarrow0$ to a simpler form. The functions that define the amplitude can be written in the limit when the expansion parameter is much larger than the particle mass by using equation (\ref{a5}) from Appendix:
\begin{eqnarray}
&&B_{K=0k=0} (pp'P) =\frac{P^{-5/2}}{\pi}\left[\frac{1}{\left(1+\frac{p'}{P}\right)^2-\frac{p^2}{P^2}}+\frac{1}{\left(1-\frac{p'}{P}\right)^2-\frac{p^2}{P^2}}\right];\nonumber\\
&&B_{1K=0k=0} (pp'P) =\frac{(P)^{-3/2}}{\pi}\left[1+\frac{p^2+p'^2}{P^2}\right];\nonumber\\
&&B_{2K=0k=0} (pp'P) =\frac{2pp'(P)^{-7/2}}{\sqrt{\pi}}\left[1+2\frac{p^2+p'^2}{P^2}\right].
\end{eqnarray}
We observe that these much simpler functions are combinations that depend on the ratios of momenta and mention that for approximating the Appel hypergeometric functions in this limit we consider the ratio of momenta to be less than one. In this limit one obtains a polynomial expression in momenta so that these functions are integrable when computing the transition rate. However since we use a perturbative approach the ultraviolet divergences will appear in our computations. The dependence of the transition rates on the ratios between the particle masses and the expansion parameter will imply nontrivial computations since we will try to evaluate the limits from the definition of the total transition rates given bellow:
\begin{equation}
R_{if}=\int d^3p_{1f}...\int d^3p_{nf}\left[\lim_{t_c\rightarrow 0}\frac{1}{2V}\frac{d}{dt_c}|A_{Ze\bar{e}}|^2\right]=\int d^3p_{1f}...\int d^3p_{nf}\left[\lim_{t\rightarrow \infty}\frac{e^{\omega t}}{2V}\frac{d}{dt}| A_{Ze\bar{e}}|^2\right].
\end{equation}
where we integrate after the momenta of the final particles. In the general case the computations cannot be done but in the limit discussed above the transition rate could in principle be obtained thus determining the density number of particles. In what follows we will present the computations for the transition rate in the case of emission of massive fermions and Z bosons for $\omega>>m\,\,, \omega>>M_Z$. For our computations it will be convenient to use the amplitude written in terms of Hankel functions as:
\begin{eqnarray}
  A_{i\rightarrow f} (\lambda = \pm 1) &=& \frac{(\pi)^{3/2} \sqrt{pp'} e^{-\pi k/2}}{8 (2\pi)^{3/2}} \delta^3 (\vec{p}+\vec{p}\,' +\vec{P}) \Bigg\{ \int_{0}^{\infty} dz \cdot z^{3/2}  \nonumber \\
  &&\times \bigg(\frac{e_0 \cos(2 \theta_W)}{\sin(2 \theta_W)} \bigg) sgn(\sigma') H^{(2)}_{\nu_+} (pz) H^{(2)}_{\nu_-} (p'z) H^{(2)}_{-ik} (Pz) \nonumber  \\
  &&- \int_{0}^{\infty} dz \cdot z^{3/2}  e_0 \tan (\theta_W) sgn(\sigma) H^{(2)}_{\nu_-} (pz) H^{(2)}_{\nu_+} (p'z) H^{(2)}_{-ik} (Pz) \Bigg\}  \nonumber\\
  &&\times \xi^+_{\sigma} (\vec{p}) \vec{\sigma} \cdot \vec{\epsilon}^* (\vec{n}_P , \lambda) \eta_{\sigma} (\vec{p}\,')
\end{eqnarray}

Then in the limit for $m/\omega = 0$, $M_Z/\omega = 0$ the temporal integrals are of the type,
\begin{equation}
  I_{i\rightarrow f}(0) = \int_{0}^{\infty} dz \cdot z^{3/2} H^{(2)}_{\frac{1}{2}} (pz) H^{(2)}_{\frac{1}{2}} (p'z) H^{(2)}_{\frac{1}{2}} (Pz)\Bigg\{ \frac{\cos(2 \theta_W)}{\sin (2 \theta_W)} sgn(\sigma') - \tan (\theta_W) sgn(\sigma) \Bigg\}
\end{equation}
where we can use the relation
\begin{equation}
  H^{(2)}_{\frac{1}{2}} (z) = \sqrt{\frac{2}{\pi z}} \cdot i e^{-iz}.
\end{equation}
The final result for the temporal integrals now becomes:
\begin{equation}\label{is}
  I_{i\rightarrow f}(0) = \bigg(\frac{2}{\pi} \bigg)^{3/2} \frac{1}{\sqrt{pp' P}(p+p'+P)}\Bigg\{ \frac{\cos(2 \theta_W)}{\sin (2 \theta_W)} sgn(\sigma') - \tan (\theta_W) sgn(\sigma) \Bigg\}
\end{equation}
This way we arrive at the final result for the transition amplitude in the case $\lambda=\pm1$ for the case of large expansion where we can consider that $m/\omega = 0$, $M_Z/\omega = 0$
\begin{eqnarray}\label{azero}
  A_{i \rightarrow f} &=& \frac{\sqrt{2}}{2} (1-i) \frac{\delta^3 (\vec{p}+\vec{p}\,' +\vec{P})}{8 \pi^{3/2}} \frac{e_0}{\sqrt{P}(p+p'+P)} \Bigg\{ \frac{\cos
  (2 \theta_W)}{\sin (2 \theta_W)} sgn(\sigma') \nonumber \\
   && - \tan (\theta_W) sgn(\sigma) \Bigg\} \xi^+_{\sigma} (\vec{p}) \vec{\sigma} \cdot \vec{\epsilon}\,^* \eta_{\sigma'} (\vec{p}\,').
\end{eqnarray}
We specify that we evaluate the rate at $t_{\infty}$ as in the general case and the rate of transition for the emission from de Sitter vacuum of the triplet electron-positron and a Z boson is then \cite{cpc}:
\begin{equation}\label{rata}
  R = \frac{1}{8} \sum_{\sigma \sigma'\lambda}\frac{\delta^3 (\vec{p}+\vec{p}\,' +\vec{P})}{(2\pi)^3} \big|M_{if} \big|^2 \big|I_{if} (0)\big| \lim_{t \to t_{\infty}} \Big|e^{\omega t} \mathcal{K}_{if} \Big(\frac{m}{\omega} = \frac{M_Z}{\omega} = 0\Big) \Big|
\end{equation}
where we denote
\begin{equation}
  M_{i\rightarrow f} = \frac{e_0\pi^{3/2}\sqrt{pp'}}{8(2\pi)^{3/2}} \xi^+_{\sigma} (\vec{p}\,) \vec{\sigma} \cdot \vec{\epsilon}\,^* \eta_{\sigma'} (\vec{p}\,'),
\end{equation}
and we introduce the new notation
\begin{equation}
  S_{\sigma \sigma'} = \frac{\cos(2 \theta_W)}{\sin (2 \theta_W)} sgn(\sigma') - \tan (\theta_W) sgn(\sigma).
\end{equation}
The notation $I_{i\rightarrow f} (0)$ stands for the result of the temporal integral given in equation (\ref{is}) and $\mathcal{K}_{if}$ is the integrand from the temporal integral
\begin{equation}
  \mathcal{K}_{if} \Big(\frac{m}{\omega} = \frac{M_Z}{\omega} = 0\Big) = z^{3/2} H^{(2)}_{\frac{1}{2}} (pz) H^{(2)}_{\frac{1}{2}} (p'z) H^{(2)}_{\frac{1}{2}} (Pz) e^{-\omega t} S_{\sigma \sigma'}.
\end{equation}
The first step is to compute the $\lambda$ sum by using the relation :
\begin{equation}
\sum_{\lambda}\epsilon_i^*(\vec{n}_{\mathcal{P}},\,\lambda)\epsilon_j(\vec{n}_{\mathcal{P}},\,\lambda)=\delta_{ij},
\end{equation}
for $\lambda=0,\pm1$.
Computing the limit from the rate given in (\ref{rata}) we obtain:
\begin{equation}
  \lim_{t \to \infty} \Big|e^{\omega t} \mathcal{K}_{if} \Big(\frac{m}{\omega} = \frac{M_Z}{\omega} = 0\Big) \Big| = \bigg( \frac{2}{\pi}\bigg)^{3/2} \frac{1}{\sqrt{pp' P}}S_{\sigma \sigma'},
\end{equation}
remarkable is that this is the result from equation (\ref{lim}) when $\frac{m}{\omega} = \frac{M_Z}{\omega} = 0$.
Collecting all the above results we obtain the transition rate

\begin{equation}\label{rsff}
  R = \frac{1}{8} \sum_{\sigma \sigma'} \frac{\delta^3 (\vec{p}+\vec{p}\,' +\vec{P})}{64 (2\pi)^6} \frac{e_0^2}{P(p+p'+P)} \big| S_{\sigma \sigma'} \big|^2 \cdot \big| \xi^+_{\sigma} (\vec{p}\,) \sigma_i \eta_{\sigma'} (\vec{p}\,\,') \big|^2
\end{equation}
Because in this limit the dependence on momenta is far simpler we will compute the total transition rate by integrating the rate equation (\ref{rsff}) after the final momenta and we adopt the same definition of the integrals as used for the total probability:
\begin{equation}
  R_{tot} = \int \frac{d^3 p}{p_0} \int \frac{d^3 p'}{p'_0} \int \frac{d^3P}{P_0} \cdot R
\end{equation}
where the factor we take the same setup as in the case of total probability $p_0=\sqrt{m^2+p^2},\,p_0'=p',\,P_0=P$.

For facilitating our computations we choose the electron momentum and positron momentum on third axis such that $\vec{p} = p \cdot \vec{e_3},\,\vec p\,'=-p'\vec e_3$, and for the Z boson moneta $\vec P=P\vec e_3$, such that $|\vec{p}+\vec{P}|=p+P$.

The sum with the bispinors is simple, since the electron-positron pair is emitted on the third axis and the sum is reduced to a numerical factor. The helicity bispinors in this particular case are:

\begin{equation}
  \xi_{-\frac{1}{2}} (\vec{p}\,) = \begin{pmatrix}
                         0 \\
                         1
                       \end{pmatrix} ; \xi_{\frac{1}{2}} (\vec{p}\,) = \begin{pmatrix}
                                                             1 \\
                                                             0
                                                           \end{pmatrix}
\end{equation}
Solving the $p'$ integrals we obtain:
\begin{eqnarray}
&&I=\int \frac{d^3 p}{\sqrt{m^2+p^2}} \int \frac{d^3 P}{P} \frac{\int d^3 p'}{p'} \frac{\delta^3 (\vec{p}+\vec{p}\,' +\vec{P})}{(2\pi)^3P(p+p'+P)}\nonumber\\
&&=\int \frac{d^3 p}{\sqrt{m^2+p^2}} \int \frac{d^3 P}{2(2\pi)^3P^2(p+P)^2}=2\pi\int \frac{d^3 p}{(2\pi)^3p\sqrt{m^2+p^2}}
\end{eqnarray}

For solving the $p$ integral we will adopt the dimensional regularization \cite{WF,GHV,IT,GT,HGF}. The $D$ dimensional integral in our case is:
\begin{equation}
\int \frac{d^3 p}{(2\pi)^3p\sqrt{m^2+p^2}}\rightarrow \int \frac{d^D p}{(2\pi)^Dp\sqrt{m^2+p^2}}
\end{equation}
The $D$ dimensional integral in this case can be written as
\begin{equation}
I(D)=\int \frac{d^D p}{(2\pi)^Dp\sqrt{m^2+p^2}}=\frac{S_D}{(2\pi)^D}\int_0^{\infty}dp \frac{p^{D-1}}{p\sqrt{m^2+p^2}},
\end{equation}
and performing the substitution $p^2=ym^2$, we obtain the integral of the Beta Euler function
\begin{equation}
I(D)=\frac{S_D\,m^{D-2}}{2(2\pi)^D}\int_0^{\infty}dy\frac{y^{(D-3)/2}}{(1+y)^{1/2}}=\frac{2\pi^{D/2}m^{D-2}}{(2\pi)^D\Gamma\bigg(\frac{D}{2}\bigg)\Gamma\bigg(\frac{1}{2}\bigg)}\Gamma\bigg(\frac{D-1}{2}\bigg)\Gamma\bigg(1-\frac{D}{2}\bigg)
\end{equation}
The regularized integral is obtained if we introduce the dimensionless constant $g=\lambda\mu^{D-3}$ and write the result in terms of $g$ and $\varepsilon$ for $D=3-\varepsilon$
\begin{equation}
I(D)_r=-\lambda I(D)=-\frac{g m}{\sqrt{\pi}(4\pi)^{3/2}}\left(\frac{4\pi \mu^2}{m^2}\right)^{\varepsilon/2}\frac{\Gamma\bigg(1-\frac{\varepsilon}{2}\bigg)\Gamma\bigg(\frac{\varepsilon-1}{2}\bigg)}{\Gamma\bigg(\frac{3}{2}-\frac{\varepsilon}{2}\bigg)}
\end{equation}
Further we expand the term in powers of $\varepsilon$ as in equation (\ref{ep}) and use the equation (\ref{game}) from Appendix to expand the gamma function \cite{HGF}:
\begin{eqnarray}\label{game}
\frac{\Gamma\bigg(1-\frac{\varepsilon}{2}\bigg)\Gamma\bigg(\frac{\varepsilon-1}{2}\bigg)}{\Gamma\bigg(\frac{3}{2}-\frac{\varepsilon}{2}\bigg)}=-4-2\varepsilon[\gamma+\psi(-1/2)+\psi(3/2)]+o(\varepsilon^2)
\end{eqnarray}
The final result for the regularized integral is obtained by taking the limit $\varepsilon=0$ from the following relation:
\begin{eqnarray}\label{irfinal}
&&I(D)_r=-\frac{g m}{8\pi^{2}}\biggl\{-4-2\varepsilon\left[\gamma+\psi(-1/2)+\psi(3/2)+\ln\left(\frac{4\pi \mu^2}{m^2}\right)\right]+o(\varepsilon^2)\biggl\}
\nonumber\\
\end{eqnarray}
and one can observe that the integral is finite for $\varepsilon=0$. One may also try to use the Pauli-Villars regularization \cite{PV} to solve the divergent integrals from probability and transition rate, in which case counter-terms need to be added in order to obtain finite results. The momenta integrals for $\varepsilon=0$ become:
\begin{equation}
I=\frac{g m}{\pi}
\end{equation}

The final result for the total rate reads as:
\begin{eqnarray}\label{raa}
R_{tot} &=& \frac{e_0^2gm}{8\pi\cdot64(2\pi)^3 } \Bigg( \frac{\cos^2(2 \theta_W)}{\sin^2 (2 \theta_W)} + \tan^2 (\theta_W) \Bigg)\nonumber\\
&=&\frac{gmM_W^2G_F\sin^2(\theta_W)}{32\sqrt{2}(2\pi)^4 }\Bigg( \frac{\cos^2(2 \theta_W)}{\sin^2 (2 \theta_W)} + \tan^2 (\theta_W) \Bigg)
\end{eqnarray}
The rate is finite and can be used for defining the density number of particles in the large expansion conditions.
Then the density number of Z bosons obtained in the process of spontaneous generation from vacuum of a Z boson and an electron-positron pair is proportional with the ratio between the rate computed perturbatively $R_{vac\rightarrow Ze^-e+}$ and the rate of decay $R_{des}$ for the Z boson, that also must be computed in the de Sitter case. Thus one needs to study the decay rates of the Z boson in fermion-anti-fermion pair in the de Sitter geometry in order to take into account the influence of space expansion upon the decay processes. All perturbative processes give contribution to the density number of Z bosons including the processes in which the bosons can be emitted by fermions. In the case of Z emission by fermions the density number of Z bosons depends on the density number of fermions, the rate of the emission and the decay rate of Z boson.
\begin{equation}
n_{Z,emission}=\frac{R_{emission}}{R_{des}}n_{f},
\end{equation}
where $R_{emission}$ is the rate of Z emission by fermions and $n_f$ is the density number of fermions.

We do not approach the problem related to the origin of the massive bosons masses in the present paper. However in the early universe the mass of the Z boson is related to the expectation values of the Higgs field. In electroweak theory during inflation, the massive bosons masses can be generated by a large condensate of the Higgs field. In \cite{PT} it was proven that the generated vector mass scales as $m_W^2\sim \omega^2$, and that of the Higgs scalar remains perturbatively small, $m_{H}^2\sim g^2 \omega^2$, where $g$ is the gauge coupling.

This result should be considered alongside the results obtained by using other methods \cite{37}, where the authors consider the nonperturbative methods. Thus for a clear picture of the mechanisms that were involved in the problem of particle production in early universe we must take into account both methods as was discussed in \cite{17,18}.
This is because the space expansion could play a similar role in the perturbative approach as the thermal bath and the two effects would probably compete in the early universe.

Another observation related to the rate refers to the fact that the graphical results for the probabilities prove that the phenomenon of massive bosons generation is present during the entire inflation. This means that in principle the rate could be computed for all values of the ratios between the particles masses and Hubble parameter. We restricted in the present investigation only to the limit when $m/\omega=M_Z/\omega=0$, and this is due to the mathematical problems that we encounter. We hope to approach this interesting topic in a future study.

\section{Conclusions}
In this paper we developed the theory of interaction between Z bosons and massive fermions by using perturbative methods as in the theory of electro-weak interactions in flat space-time. The equations of interactions between massive neutral vector field and massive fermions were established and their solutions were written down with the help of the Green functions. This helps us introduce the $ in\ out $ fields and construct the formalism of reduction for the Proca field in de Sitter geometry. The definition of the transition amplitudes in the first order of perturbation theory was established and we mention that we work with Feynman rules in coordinates since the Proca propagators in de Sitter geometry need further study.

As an application of our formalism the problem of generating from de Sitter vacuum of the triplet Z boson and electron-positron was studied. We established the transition amplitudes for both the transversal and longitudinal modes and we defined the transition probabilities in volume unit. Our analytical and graphical results prove that the phenomenon of particle production is possible only for a large expansion factor that corresponds to the early universe. In the Minkowski limit we recovered the expected results that the process of particle production is no longer possible due to energy-momentum conservation. We computed the total probability and rate in the case of large expansion factor $\omega>>m,M_z$ and we adopted the dimensional regularization for computing the momenta integrals. By using the minimal substraction method we succed to remove the divergences from our total probability, proving that the theory we constructed here is renormable. The rate of transition was computed in the conditions of large expansion where the ratios $M_Z/\omega,\,m/\omega$ tend to zero and we determine the density number of Z bosons as the ratio between the production rate and decay rate.

For further study it will be of interest to take into account all possible processes that generate Z boson as a result of fields interactions including the processes of Z boson emission by electrons and positrons. The electro-weak interactions are less studied in curved backgrounds and we hope that our results will encourage others to study the Proca field and propagators in curved space-times. This could help us to understand the problems related to the mechanisms that were involved in generating massive bosons in early universe.

\section{Appendix}
For solving our integrals we use the relations that relate Hankel functions to Bessel functions
\begin{eqnarray}
H^{(1)}_{\mu}(z)=\frac{J_{-\mu}(z)-e^{-i\pi\mu}J_{\mu}(z)}{i\sin(\pi\mu)}\nonumber\\
H^{(2)}_{\mu}(z)=\frac{e^{i\pi\mu}J_{\mu}(z)-J_{-\mu}(z)}{i\sin(\pi\mu)}.
\end{eqnarray}
and the relations that transform Hankel function into modified Bessel K functions \cite{21}:
\begin{equation}\label{a2}
H^{(1,2)}_{\nu}(z)=\mp \left(\frac{2i}{\pi}\right)e^{\mp
i\pi\nu/2}K_{\nu}(\mp iz),
\end{equation}
The integrals that help compute our amplitudes are of the type \cite{21}:
\begin{eqnarray}\label{a3}
&&\int_0^{\infty} dz
z^{\lambda-1}J_{\mu}(az)J_{\nu}(bz)K_{\rho}(cz)=\frac{2^{\lambda-2}a^{\mu}b\,^{\nu}c^{-\lambda-\mu-\nu}}{\Gamma(1+\mu)\Gamma(1+\nu)}\Gamma\left(\frac{\lambda+\mu+\nu-\rho}{2}\right)\Gamma\left(\frac{\lambda+\mu+\nu+\rho}{2}\right)\nonumber\\
&&\times
\,F_{4}\left(\frac{\lambda+\mu+\nu-\rho}{2},\frac{\lambda+\mu+\nu+\rho}{2};1+\mu,1+\nu;-\frac{a^2}{c^2},-\frac{b^2}{c^2}\right),\nonumber\\
&&Re(\lambda+\mu+\nu)>Re(\rho)\,,Re(c)>|Im(a)|+|Im(b)|.
\end{eqnarray}
The definition of Appel hypergeometric function is given bellow \cite{21}:
\begin{equation}\label{a4}
  F_4 (a,b,c,d;x,y) = \sum_{m,n=0}^{\infty} \frac{\Gamma(a+m+n)\Gamma(b+m+n)\Gamma(c)\Gamma(d)x^m \cdot y^n}{\Gamma(a)\Gamma(b)\Gamma(c+m)\Gamma(d+n) m!\cdot n!} .
\end{equation}
In our computations we use the relation between the Appel hypergeometric functions and Gauss hypergeometric functions \cite{21}
\begin{eqnarray}\label{a5}
F_{4}\left(\alpha,\alpha+\frac{1}{2},\gamma,\frac{1}{2};x,y\right)&=&\frac{1}{2}(1+\sqrt y)^{-2\alpha}\, _2F_{1}\left(\alpha,\alpha+\frac{1}{2};\gamma;\frac{x}{(1+\sqrt y)^2}\right)\nonumber\\
&&+\frac{1}{2}(1-\sqrt y)^{-2\alpha}\, _2F_{1}\left(\alpha,\alpha+\frac{1}{2};\gamma;\frac{x}{(1-\sqrt y)^2}\right).
\end{eqnarray}
The $P$ integral from total probability has the result:
\begin{equation}\label{rdptot}
 \int \frac{d^3 P}{P^2}\frac{1}{(p+P)^3}=\frac{2\pi}{p^2}.
\end{equation}
The expansion for gamma Euler function for an integer $n$ is \cite{HGF}:
\begin{eqnarray}\label{game}
\Gamma(-n+\varepsilon)=\frac{(-1)^n}{n!}\biggl\{\frac{1}{\varepsilon}+\psi(n+1)+\frac{\varepsilon}{2}\biggl[\frac{\pi^2}{3}+\psi^2(n+1)-\psi'(n+1)\biggl]+o(\varepsilon^2)\biggl\}
\end{eqnarray}
where $\psi$ is the digamma Euler function and $\psi'$ its derivative.

\textbf{Acknowledgements}

Diana Dumitrele was supported by a grant of  the Romanian Ministry of Research and Innovation and West University of Timi\c soara, CCCDI-UEFISCDI, under project "VESS, 18PCCDI/2018",  within PNCDI III.

\end{document}